\documentclass[11pt,a4paper]{article}
\usepackage{epsfig}
\usepackage[T1]{fontenc}    
\usepackage{graphics}
\usepackage{graphicx}
\usepackage{pstricks,pst-coil,pst-fill,pst-plot}
\usepackage[fleqn]{amsmath}    
\usepackage{amssymb}    
\usepackage{amsfonts}   
\usepackage{verbatim}   
\usepackage{mathrsfs}   
\usepackage{dsfont}
\usepackage{euscript}
\usepackage{yfonts}
\usepackage{enumerate}     
\usepackage{txfonts}
\usepackage{marvosym}
\usepackage{vmargin}        
\usepackage{wasysym}		

\setmarginsrb{1.8cm}{2cm}{1.8cm}{2cm}{1cm}{1cm}{1cm}{1.6cm}
 \makeatletter
 \@addtoreset{equation}{section}
 \makeatother

\providecommand{\bysame}{\leavevmode\hbox to3em{\hrulefill}\thinspace}
\providecommand{\MR}{\relax\ifhmode\unskip\space\fi MR }

\providecommand{\href}[2]{#2}

\let\ua=\uparrow
\let\da=\downarrow
\let\tend=\rightarrow

\long\def\symbolfootnote[#1]#2{\begingroup%
\def\thefootnote{\fnsymbol{footnote}}\footnote[#1]{#2}\endgroup}

\newtheorem{theorem}{Theorem}[section]
\newtheorem{prop}[theorem]{Proposition}

\newtheorem{defin}[theorem]{Definition}

\newtheorem{lemme}{Lemma}[section]

\def\Proof{\medskip\noindent {\it Proof --- \ }}

\def\qed{\hfill\rule{2mm}{2mm}}

\newcommand\beq{\begin{equation}}
\newcommand\enq{\end{equation}}
\newcommand\bem{\begin{multline}}
\newcommand\enm{\end{multline}}

\def\beqa{\begin{eqnarray}}
\def\eeqa{\end{eqnarray}}
\def\ba{\begin{array}}
\def\ea{\end{array}}

\newcommand{\f}[2]{{\ensuremath{%
    \mathchoice%
    {\dfrac{#1}{#2}}
    {\dfrac{#1}{#2}}
    {\frac{#1}{#2}}
    {\frac{#1}{#2}}
}}}
\newcommand{\tf}[2]{\ensuremath{#1/#2}}
\newcommand{\pa}[1]{\ensuremath{\left(#1\right)}}

\def\a{\alpha}

\def\be{\beta}
\def\ga{\gamma}
\def\Ga{\Gamma}

\def\de{\delta}

\def\eps{\epsilon}
\def\veps{\varepsilon}
\def\la{\lambda}
\def\La{\Lambda}

\def\sg{\sigma}
\def\vsg{\varsigma}

\def\Om{\Omega}
\def\om{\omega}
\def\vp{\varphi}

\newcommand{\mc}[1]{\ensuremath{\mathcal{#1}}}
\newcommand{\mf}[1]{\ensuremath{\mathfrak{#1}}}
\newcommand{\msc}[1]{\ensuremath{\mathscr{#1}}}

\newcommand{\bs}[1]{\ensuremath{\boldsymbol{#1}}}

\DeclareFontFamily{OT1}{pzc}{}
\DeclareFontShape{OT1}{pzc}{m}{it}{<-> s * [1.10] pzcmi7t}{}
\DeclareMathAlphabet{\mathpzc}{OT1}{pzc}{m}{it}

\def \i{ \mathrm i}

\newcommand{\ov}[1]{\ensuremath{\overline{#1}}}
\newcommand{\wt}[1]{\ensuremath{\widetilde{#1}}}

\newcommand{\Int}[2]{\ensuremath{\int\limits_{#1}^{#2}}}

\newcommand{\sul}[2]{\ensuremath{\sum\limits_{#1}^{#2}}}
\newcommand{\pl}[2]{\ensuremath{\prod\limits_{#1}^{#2}}}

\newcommand{\R}{\ensuremath{\mathbb{R}}}
\newcommand{\Cx}{\ensuremath{\mathbb{C}}}

\newcommand{\Dp}[1]{\ensuremath{\partial_{#1}}}

\newcommand{\limit}[2]{\ensuremath{\underset{#1 \tend #2}{\longrightarrow} }}

\newcommand{\ex}[1]{\ensuremath{\e{e}^{#1}}}

\newcommand{\op}[1]{ \boldsymbol{ \texttt{#1} } }

\newcommand{\dd}{\mathrm{d}}
\newcommand{\e}[1]{\ensuremath{\mathrm{#1}}}

\newcommand{\intff}[2]{\ensuremath{ [  #1 \,; #2 ] }}

\newcommand{\intn}[2]{\ensuremath{[\![ \, #1 \,;\, #2 \,]\!]}}

\begin{document}

\begin{flushright}
DESY 18-088
\end{flushright}
\par \vskip .1in \noindent

\vspace{14pt}

\begin{center}
\begin{LARGE}
{\bf On the separation of variables for the modular XXZ magnet and the lattice Sinh-Gordon models}
\end{LARGE}

\vspace{1cm}

{\large Sergey \'E. Derkachov\footnote{e-mail: derkach@pdmi.ras.ru}}
\\[1ex]
Saint-Petersburg Department of Steklov Mathematical Institute
of Russian Academy of Sciences, Fontanka 27, 191023 St. Petersburg, Russia.\\[2.5ex]

{\large Karol K. Kozlowski\footnote{e-mail: karol.kozlowski@ens-lyon.fr} }
\\[1ex]
Univ Lyon, ENS de Lyon, Univ Claude Bernard Lyon 1, CNRS, Laboratoire de Physique, F-69342 Lyon, France. \\[2.5ex]

{\large Alexander N. Manashov\footnote{e-mail: alexander.manashov@desy.de}}%
\\[1ex]
Institut f\"{u}r Theoretische Physik, Universit\"{a}t Hamburg, D-22761 Hamburg, Germany. \\[1ex]

Institute for Theoretical Physics, University of Regensburg, D-93040 Regensburg, Germany.\\[1ex]

Saint-Petersburg Department of Steklov Mathematical Institute
of Russian Academy of Sciences, Fontanka 27, 191023 St. Petersburg, Russia.

\par

\par

\vspace{40pt}

\centerline{\bf Abstract} \vspace{1cm}
\parbox{12cm}{\small  We construct the generalised Eigenfunctions of the entries of the monodromy matrix of the $N$-site modular XXZ magnet and show, in each case, that these form a complete orthogonal system in $L^2(\R^N)$. 
In particular, we develop a new and simple technique, allowing one to prove the completeness of such systems. 
As a corollary of out analysis, we prove the Bystko-Teschner conjecture relative to the structure of the spectrum of the $\op{B}(\la)$-operator for the odd length lattice Sinh-Gordon model. }

\end{center}

\vspace{40pt}

\section{Introduction}

The quantum separation of variables has been developed by Sklyanin in \cite{SklyaninSoVFirstIntroTodaChain,SklyaninSoVGeneralOverviewFuncBA,SklyaninSoVGeneralOverviewAndConstrRecVectPofB}
and was applied, since then, to obtain the spectra and Eigenstates of numerous quantum integrable models having finite and infinite dimensional 
local spaces \cite{BytskoTeschnerSinhGordonFunctionalBA,DerkachovKorchemskyManashovXXXSoVandQopNewConstEigenfctsBOp,DerkachovKorchemskyManashovXXXreelSoVandQopABAcomparaison,
KharchevLebedevIntRepEigenfctsPeriodicTodaFromRecConstrofEigenFctOfB,KharchevLebedevSemenovTianShanskyRTCBEigenfunctions,NiccoliTeschnerSineGordonRevisited,NiccoliTerrasQuasiPEriodic8VertexBySOV}. 
The very essence of the method consists in mapping the original Hilbert space $\mf{h}_{\e{org}}$ where a quantum integrable model is formulated onto an auxiliary Hilbert space $\mf{h}_{\e{aux}}$.
This mapping is done in a very specific way that strongly simplifies the original spectral problem  for the model's transfer matrix. 
When formulated on the level of the original Hilbert space $\mf{h}_{\e{org}}$, the spectral problem for the transfer matrix is a genuine multi-dimensional and multi-parameter spectral problem. 
However, upon implementing the separation of variables transform, the latter is re-expressed in terms of a multi-parameter but \textit{one}-dimensional spectral problem
which takes the form of a scalar $T-Q$ equation. Thus, the method provides an outstanding simplification.

An important ingredient of the whole construction consists in establishing that the mapping $\mf{h}_{\e{org}} \tend \mf{h}_{\e{aux}}$ is unitary, a necessary step for guaranteeing the equivalence between the two Hilbert space descriptions
of the spectral problem. 
In the finite dimensional setting, one may establish this unitarity by an direct counting of dimension arguments. 
However, the situation is disproportionnally harder in the case of integrable models associated with infinite dimensional local Hilbert spaces. 
Indeed, even in the simplest possible case, the quantum $N$-particle Toda chain \cite{GutzwillerResolutionTodaChainSmallNPaper1}, 
establishing the unitarity of the transform is equivalent to proving the completeness and orthogonality of the system of Whittaker functions for $\mf{gl}_N$. 
This topic has a very long  history. The satisfactory and  full resolution of the problem was first achieved through rather   evolved tools of harmonic analysis on non-compact groups \cite{WallachRealReductiveGroupsII}.
These techniques, on top of their complexity, were completely unrelated to the algebraic structures usually dealt with in quantum integrable models. Hence their generalisation to 
more complex models was hardly imaginable. A first important progress towards a  quantum integrability based proof of unitarity was achieved in \cite{DerkachovKorchemskyManashovXXXSoVandQopNewConstEigenfctsBOp}. 
That paper proposed a method for constructing a pyramidal -the so called Gauss-Givental\cite{GiventalGaussGivIntRepObtainedForEFOfOpenToda} - representation for the integral kernel of the separation of variables transform
in the case of the non-compact XXX $\mf{sl}(2,\Cx)$ chain. In fact, the integral kernel appeared as a generalised system of Eigenfunctions of one of the entries of the model's  monodromy matrix. 
The work \cite{DerkachovKorchemskyManashovXXXSoVandQopNewConstEigenfctsBOp} developed  a technique allowing one to establish orthogonality of this system, which is "half" of the proof of unitarity. 
The mentioned technique was later applied to other models subordinate to the 
rational $R$-matrix: the non-compact XXX $\mf{sl}(2,\R)$ \cite{DerkachovKorchemskyManashovXXXreelSoVandQopABAcomparaison} or to the Toda chain \cite{SilantyevScalarProductFormulaTodaChain}. 
A quantum inverse scattering based method for proving completeness was developed in \cite{KozUnitarityofSoVTransform}, where all handlings leading 
to the proof of unitarity were set in a rigorous framework of distribution theory.  
Finally, one should mention the work \cite{DerkachovManashovInterativeConstrEigenFctsSL2C} where, on the example of the $\mf{sl}(2,\Cx)$ chain,
the technique of iterative construction of Eigenfunctions was perfected and set in a very natural to the quantum inverse scatterin method picture.

 The purpose of the present paper is to push these developments further.  On the one hand, we extend the method of constructing
Gauss-Givental  representations to the case of a non-compact, rank $1$, model associated with a trigonometric
$R$-matrix. One the other hand, we propose a  simple method for proving completeness. Indeed, as observed in
\cite{KozUnitarityofSoVTransform}, Mellin-Barnes integral representations are well-suited for proving completeness of the
unitarity transform's kernel. On a technical level, doing so demands to have at one's disposal a certain integral
representation for the symmetric delta function in $N$-variables. A technique for proving this integral representation was
developed in \cite{KozUnitarityofSoVTransform} and was based on a fine analysis of the consequences of the orthogonality
relations. While providing the desired results, this method is not based on a direct argument. In the present paper, we develop a direct and
systematic technique for establishing the integral identities of interest. We believe this to be an important contribution of
our work. In this work, we focus our study on the case of the modular $XXZ$ magnet. This model is closely connected with the lattice Sinh-Gordon
model. Thus, as a byproduct of our analysis, we prove the Bytsko-Teschner conjecture
\cite{BytskoTeschnerSinhGordonFunctionalBA} relative to the structure of the spectrum of the $\op{B}$-operator in this model.
Finally, at the time where this manuscript was beeing finalised, the work \cite{SchraderShapiroOrthogoAndCompletenessWhittakerFctsqToda} appeared on the ArXiV and dealt, 
by slightly different means with a related problem in the context of the $q$-Toda chain. It would be interesting to compare the methods.  

The paper is organised as follows.  
In Section \ref{Section modele et notations} we introduce the modular XXZ chain's Lax matrix and discuss some of the
fundamental objects related with this model. The Gauss-Givental integral representation for the system of  Eigenfunctions
associated with entries of the monodromy matrix is constructed in Section \ref{Section Gauss-Givental}. It immediately allows
us to establish the orthogonality of such a system. Section \ref{Section Mellin Barnes}  is devoted to  the proof of the
completeness of the Eigensystems. We derive  the  Mellin-Barnes integral representation for the Eigenfunctions and evaluate
some auxiliary integrals. Using these results  we prove the completeness of the orthogonal systems associated with entries of
the monodromy matrix. Then, in Section \ref{Section Complete and orthogonal system}, we discuss  properties of the complete
orthogonal system associated with  entries of the monodromy matrix.
 Finally, in Section \ref{Section Eigenfunctions for Sinh Gordon}, we apply our results to the case of the lattice Sinh-Gordon model. The paper contains two appendices. Appendix
\ref{Appendice notation} reviews the various notations introduced in the paper. Appendix \ref{Appendice fct speciale} reviews
the special functions used throughout the work.

\section{The modular XXZ magnet}
\label{Section modele et notations}

\subsection{The model}

The modular XXZ magnet's Hilbert space has the tensor product decomposition $\mf{h} \, = \, \otimes_{n=1}^{N} \mf{h}_{n} $ where the Hilbert space $\mf{h}_n$ associated with the $n^{\e{th}}$ site of the model 
is isomorphic to $L^2(\R,\dd x)$. The $n^{\e{th}}$-site Lax matrix takes the form 
\beq
\op{L}_n^{(\kappa)}(\la) \; = \;  2   \left( \ba{cc} - \i \sinh\Big[ \f{\pi}{\om_1} (\la - \op{p}_n) \Big] & \ex{-\pi \om_2 \op{x}_n } \cosh\Big[ \f{\pi}{\om_1} (\op{p}_n+\kappa) \Big] \ex{- \pi \om_2  \op{x}_n } \vspace{2mm}  \\ 
				   \ex{\pi \om_2 \op{x}_n } \cosh\Big[ \f{\pi}{\om_1} ( \op{p}_n -\kappa) \Big] \ex{\pi \om_2 \op{x}_n }  &     - \i \sinh\Big[ \f{\pi}{\om_1} (\la+\op{p}_n) \Big]   	    \ea  \right)\;. 
\enq
Here and in the following, $\om_1, \om_2$ and $\kappa$ are three real parameters such that $\om_a>0$ and $\kappa\not=0$. 
Further, $\op{x}_n$ and $\op{p}_n$ are operators on $\mf{h}_n$ satisfying to the canonical commutation relations $\big[ \op{x}_n, \op{p}_m \big]= \de_{n,m} \tfrac{\i }{2\pi}$, where $\de_{n,m}$ is the Kronecker symbol. 
In the following, we shall work in a representation where $\op{x}_n$ is the multiplication operator by the $n^{\e{th}}$ coordinate 
\beq
(\op{x}_n f)(\bs{x}_N)= x_n f(\bs{x}_N) \qquad \e{with} \qquad \bs{x}_N=(x_1,\dots, x_N)\quad \e{so} \; \e{that} \quad (\op{p}_n f)(\bs{x}_N)= -\tfrac{\i}{2\pi} \Dp{x_n} f(\bs{x}_N) \;. 
\enq
The model's monodromy matrix takes the form 
\beq
\op{T}_N(\la) \; = \; \op{L}_1^{(\kappa)}(\la)\cdots \op{L}_N^{(\kappa)}(\la) \; = \; \left(\ba{cc} \op{A}_N(\la)   & \op{B}_N(\la)   \\ 
								  \op{C}_N(\la)   &   \op{D}_N(\la) \ea  \right) \;. 
\label{ecriture matrice de monodromie XXZ}
\enq
The transfer matrix 
\beq
\op{t}(\la)=\e{tr}\big[ \op{T}(\la) \big] \; = \; (-\i)^N \ex{\frac{\pi}{\om_1}  N \la} \sul{k=0}{N} \Big(-\ex{ -\f{2\pi}{\om_1} \la } \Big)^k \op{T}_k
\enq
provides ones with a commutative algebra of positive self-adjoint operators $\Big\{ \op{T}_k \Big\}_1^N$ on $\mf{h}$, \textit{c.f.} \cite{BytskoTeschnerSinhGordonFunctionalBA} for more details.

\subsection{The elementary operator relations of the model}

The $\mf{R}$-operator for the modular XXZ magnet was first constructed in \cite{BytskoTeschnerRmatrixForModularDouble}. Its non-trivial part was given as
a special function of a positive self-adjoint operator coocked up from a representation of the coproduct of the Casimir of $ U_{q}(\mf{sl}_2)$.  
That charasterisation of $\mf{R}$ was rather implicit. The paper \cite{ChicherinDerkachovRmatrixForModularDouble} proposed an alternative way to construct this $\mf{R}$-operator. This allowed
for a much simpler and more explicit form of $\mf{R}$. 
This construction was based on the existence of an alternative factorisation of the model's Lax matrix
\beq
\op{L}_{n}^{(\kappa)}(\la) \, = \, -\i  \,  \op{M}_{u_2}(\op{x}_n) \cdot \op{H}(\op{p}_n) \cdot \op{N}_{u_1}(\op{x}_n)
\label{ecriture factorisation auxiliare de L XXZ}
\enq
where 
\beq
 \op{M}_{u_2}(\op{x}) \; = \; \left( \ba{cc}  U_2    & -U_2^{-1} \\ 
					 -U_2^{-1} \ex{2\pi \om_2 \op{x}}  & U_2 \ex{2\pi \om_2 \op{x}}  \ea \right)\;\; , \qquad 
 \op{N}_{u_1}(\op{x}) \; = \; \left( \ba{cc}  -U_1    &  U_1^{-1} \ex{- 2\pi \om_2 \op{x}}  \\ 
					 -U_1^{-1}  & U_1 \ex{-2\pi \om_2 \op{x}}  \ea \right)	
\enq
and $\op{H}(\op{p})=\ex{ -\f{\pi}{\om_1}(\op{p}-\i\f{\Om}{2}) \sg_3 }$. Here, $\sg_3=\e{diag}(1,-1)$ and we made use of the notation
\beq
U_a\, = \, \ex{ \f{\pi}{\om_1}u_a }\qquad \e{with} \qquad
\left\{ \ba{ccc} u_1 &= &\tfrac{1}{2}\cdot \big(\la +\kappa - \i \tfrac{\tau}{2}\big)   \vspace{2mm} \\
u_2 & = & \tfrac{1}{2}\cdot \big(\la - \kappa - \i \tfrac{\tau}{2}\big)   \ea \right. \;, 
\label{definition variables u1 et u2}
\enq
and, agree, from now on, to denote
\beq
\Om= \om_1+\om_2 \; , \qquad \tau \, = \, \om_2\, - \, \om_1 \; . 
\enq
The factorisation \eqref{ecriture factorisation auxiliare de L XXZ} allows one to interpret the Lax matrix as a function of the parameters $u_1$ and $u_2$, \textit{viz}. 
\beq
\op{L}_{n}^{(\kappa)}(\la) \, \equiv \, \op{L}_{n}(u_1,u_2)  \;. 
\enq
We shall adopt this notation in the following.  Note that changing $\kappa \hookrightarrow - \kappa$ produces and exchange of the parameters $u_1$ and $u_2$. 
It is well known that the operator $D_{-\kappa}(\op{p})$, the function $D_{\a}$ being defined in Appendix \ref{Appendice fct speciale} \textit{c.f.} \eqref{definition D alpha et pte conj complexe}, 
is an intertwining operator between the $\kappa$ and $-\kappa$ representations meaning that
\beq
D_{u_2-u_1}(\op{p}_1) \cdot  \op{L}_1(u_1,u_2) \; = \; \op{L}_1(u_2,u_1) \cdot D_{u_2-u_1}(\op{p}_1) \;. 
\label{equation entrelacement rep kappa et moins kappa}
\enq
The operator $D_{\a}(\op{p})$ acts as a multiplication operator on the spectrum of $\op{p}_1$ and can be represented as an integral operator on the spectrum of $\op{x}_1$ by means of the Fourier transform \eqref{ecriture TF de fct D}:
\beq
\big[ D_{\a}(\op{p}) \cdot f\big](x) \; = \; \sqrt{\om_1\om_2} \,  \mc{A}(\a) \Int{ \R }{}  D_{\a^{\star}}\big( \om_1\om_2 (x-x^{\prime}) \big) f\big( x^{\prime} \big) \cdot \dd x^{\prime}
\enq
where we have introduced
\beq
\a^{\star} = -\a -\i \f{ \Om }{2} \qquad \e{and} \qquad \mc{A}(\a) \; = \; \varpi\big( \a -\a^{\star}\big) \; = \; \varpi \Big( 2\a + \i \tfrac{ \Om }{2} \Big) \;. 
\label{definition star transform and A fct}
\enq
Also, $\varpi$ refers to the quantum dilogarithm whose definition is recalled in Appendix \ref{Appendice fct speciale} \textit{c.f.} \eqref{definition quantum dilog}. 

It was established in \cite{ChicherinDerkachovRmatrixForModularDouble}, on the basis of the factorisation \eqref{ecriture factorisation auxiliare de L XXZ}, that it holds
\beq
D_{u_1-v_2}\big( \om_1\om_2 \op{x}_{12} \big) \, \op{L}_1(u_1,u_2) \, \op{L}_2(v_1,v_2) \; = \; \op{L}_1(v_2,u_2) \,  \op{L}_2(v_1,u_1)  \, D_{u_1-v_2}\big( \om_1\om_2 \op{x}_{12} \big) \;. 
\enq
Above and in the following, for any quantities $\a_a,\a_b$ we agree to denote $\a_{ab}=\a_a-\a_b$, \textit{e.g.}  $\op{x}_{12}=\op{x}_1-\op{x}_2$. 

The two above intertwining relations thus ensure that the operator 
\beqa
\op{R}_{12}(u_1,u_2\mid v_2) & = & D_{u_2-u_1}\big( \om_1\om_2 \op{x}_{12} \big) \cdot  D_{u_2-v_2}(\op{p}_1) \cdot D_{u_1-v_2}\big( \om_1\om_2 \op{x}_{12} \big)  \\
& = & D_{u_1-v_2}\big(  \op{p}_1 \big)  \cdot D_{u_2-v_2}(\om_1\om_2 \op{x}_{12} ) \cdot  D_{u_2-u_1}\big(  \op{p}_1 \big) 
\label{ecriture expression explicite factorisation elementaire R12}
\eeqa
realises the intertwining 
\beq
\op{R}_{12}(u_1,u_2\mid v_2) \,   \op{L}_1(u_1,u_2) \, \op{L}_2(v_1,v_2) \; = \; \op{L}_1(u_1,v_2) \, \op{L}_2(v_1,u_2) \, \op{R}_{12}(u_1,u_2\mid v_2)  \;. 
\label{ecriture eqn intertwining}
\enq
The two factorisations of $\op{R}_{12}$ stem from the two possible ways of decomposing the permutation 
\beq
(u_1,u_2,v_1,v_2)\mapsto (u_1,v_2,v_1,u_2)
\enq
into $2$-cycles. Note that the equality between the two factorisations is, in fact, a consequence of the three term integral relation 
\eqref{ecriture identite itle 3 termes} satisfied by the $D$ functions which, in an operator form, is given in \eqref{ecriture identite etoile triangle}. 
Also, we stress that $\op{R}_{12}$ does \textit{not} correspond to the $\mf{R}$ operator of the XXZ-modular magnet. 
Nonetheless, it corresponds to half of its building block. We refer to \cite{ChicherinDerkachovRmatrixForModularDouble} for more details.

Finally, observe that one has the identities
\beq
  \bs{v}_{\eps}^{\bs{t}} \cdot \op{L}_n(u_1,u_2) \cdot \ex{ -2\i\pi \eps (v_2-u_2) \op{x}_n } \; = \;  \ex{ -2\i\pi \eps (v_2-u_2) \op{x}_n } \cdot \bs{v}_{\eps}^{\bs{t}} \cdot \op{L}_n(u_1,v_2) 
\label{ecriture echange us et v a gauche}
\enq
and
\beq
  \op{L}_n(u_1,v_2) \bs{v}_{\eps}  \cdot  \ex{ 2\i\pi \eps (u_1-v_1) \op{x}_n }  \; = \;  \ex{ 2\i\pi \eps (u_1-v_1) \op{x}_n }  \cdot \op{L}_n(v_1,v_2)    \bs{v}_{\eps}
\label{ecriture echange us et v a droite}
\enq
where we have introduced
\beq
\bs{v}_{+} \, = \, \left( \ba{c} 1 \\ 0 \ea \right)  \qquad \e{and} \qquad \bs{v}_{-} \, = \, \left( \ba{c} 0 \\ 1 \ea \right) \;.
\label{defintion vecteurs v pm}
\enq

\section{The Gauss-Givental integral representation and orthogonality}
\label{Section Gauss-Givental}

\subsection{The $\La$-operator}

We have now introduced enough notations so as to introduce the $\La_y^{(N)}$  operators and their adjoints. These operators play a crucial role in 
constructing the generalised Eigenfunctions of the operators $\op{A}_N$ and $\op{B}_N$. Below, we shall establish some of their exchange relations which are 
important for our further purposes. These should be understood in the weak sense, and when one considers sufficiently regular functions. See \cite{KozUnitarityofSoVTransform}
for more details.

\begin{defin}

Let $y_{\pm}$ stand for 
\beq
y_{\pm}\; = \; \f{1}{2} \big( y\pm \kappa -\i \f{ \Om }{2}  \big) \;. 
\label{definition parametres y pm}
\enq
Denote by $\La^{(N)}_{y,\eps}$ the below continuous operators $L^{\infty}(\R^{N-1}) \mapsto L^{\infty}(\R^N)$:
\beqa
\La^{(N)}_{y,\eps} & = &    \f{ \ex{2\i\pi y_- \op{x}_1}  }{ \Big(  \mc{A}(y_+) \sqrt{\om_1\om_2}   \Big)^{N-1} } \cdot \wt{\op{J}}_{N}\big( y_- - y_+\big) \cdot \op{U}_N(y_-,y_+) \cdot \ex{  2\i\pi \eps y_+^{\star} \op{x}_N} 1_N  \label{rep op LaN type droit}\\
& = &   \f{ \ex{2\i\pi y_- \op{x}_1}  }{ \Big(  \mc{A}(y_+) \sqrt{\om_1\om_2}   \Big)^{N-1} } \cdot \op{U}_N(y_+,y_-) \cdot \op{J}_{N}\big( y_- - y_+\big) \cdot \ex{  2\i\pi \eps y_+^{\star} \op{x}_N} 1_N  \label{rep op LaN type gauche}
\eeqa
with
\beq
\wt{\op{J}}_N(y) \, = \, \pl{a=1}{ N-1} D_{y}\big(\om_1\om_2 \op{x}_{a a+1} \big) \quad ,  \hspace{1.5cm}  \op{J}_N(y) \, = \, \pl{a=1}{ N-1} D_{y}\big( \op{p}_a  \big)
\label{definition operateurs J et J tilde N}
\enq
as well as 
\beq
 \op{U}_N(y_+,y_-) \; = \; \pl{a}{1 \curvearrowright N-1} \!\!\! \Big\{ D_{y_+}\big( \op{p}_a  \big)D_{y_-}\big(\om_1\om_2 \op{x}_{a a+1} \big) \Big\} \;. 
\enq
Finally, the operator $1_N$ stands for the constant function on the $N^{\e{th}}$ space and the formula is to be understood as the partial action of the chain of operators
on this function.

\end{defin}

Above and in the following we agree to denote the ordered products as
\beq
\pl{a}{1 \curvearrowright N} \op{O}_a \; = \; \op{O}_1\cdots \op{O}_N \qquad \e{and} \qquad \pl{a}{ N  \curvearrowright 1} \op{O}_a \; = \; \op{O}_N \cdots \op{O}_1 \;. 
\enq
Furthermore, the equality between \eqref{rep op LaN type droit} and \eqref{rep op LaN type gauche} follows from a multiple application of the relation \eqref{ecriture identite etoile triangle}.

Owing to the conjugation property of the $D_{\a}$ function \textit{c.f}. \eqref{definition D alpha et pte conj complexe}, and those of the quantum dilogarithm, the adjoint operators 
\beqa
\Big( \La^{(N)}_{y,\eps} \Big)^{\dagger} & = &   \bigg(  \f{ \mc{A} \big(  (-y_+^*)^{\star} \big)  }{ \sqrt{\om_1\om_2}  } \bigg)^{N-1}     \cdot 
1_N^{\dagger}  \ex{  -2\i\pi \eps \big(y_+^{\star}\big)^* \op{x}_N} \op{U}_N^{\dagger}(y_-,y_+)   \wt{\op{J}}_{N}^{-1}\big( y_- - y_+\big) \ex{ - 2\i\pi y_-^* \op{x}_1}  \\
&  = &   \bigg(  \f{ \mc{A} \big(  (-y_+^*)^{\star} \big)  }{ \sqrt{\om_1\om_2}  } \bigg)^{N-1}  \cdot  1_N^{\dagger}  \ex{  -2\i\pi \eps \big(y_+^{\star}\big)^* \op{x}_N} \op{J}_{N}^{-1}\big( y_- - y_+\big) \op{U}_N^{\dagger}(y_+,y_-) \ex{ - 2\i\pi y_-^* \op{x}_1}
\eeqa
give rise to continuous operators $L^{\infty}(\R^N) \tend L^{\infty}(\R^{N-1}) $. Here, ${}^*$ stands for the complex conjugation while ${}^{\star}$
for the transform \eqref{definition star transform and A fct}. Moreover, $1_N^{\dagger}$ represents the operation of integration \textit{versus} the $N^{\e{th}}$ coordinate. 
Finally, we have set 
\beq
\op{U}_N^{\dagger}(y_+,y_-) \; = \; \pl{a}{ N-1 \curvearrowright 1} D_{-y_-^*}\big(\om_1\om_2 \op{x}_{a a+1} \big) D_{-y_+^{*} }\big( \op{p}_a  \big)
\enq

We stress that the $\Lambda$-operators are in fact very closely related to the partial $\op{R}$-operators in that they can be expressed as
\beq
\La^{(N)}_{y, \eps}  \, = \,  \f{ \ex{2\i\pi y_-  \op{x}_1}  }{  \Big(  \mc{A}(y_+) \sqrt{\om_1\om_2}   \Big)^{N-1} }\cdot
\op{R}_{12}(u_1,u_2\mid v_2) \cdots \op{R}_{N-1N}(u_1,u_2\mid v_2)
\cdot \ex{   2\i\pi  \eps  y_+^{\star} \op{x}_N}  \, ,
\enq
where   $y_{\pm}$ are as defined in \eqref{definition parametres y pm}.

\vspace{2mm} 

The first interesting property of the operators $\La^{(N)}_{y,\pm}$ concerns their exchange relations with the operators $\op{A}_N,\op{B}_N$.   

\begin{lemme}
\label{Lemme action recursive AN et BN sur LambdaN}
The operators $\La^{(N)}_{y,\eps}$ satisfy 
\beqa
\op{A}_N(\la) \cdot \La^{(N)}_{y,-} & = & -2\i   \sinh\Big[\f{\pi}{\om_1} (\la-y) \Big]  \cdot \La^{(N)}_{y,-} \cdot \op{A}_{N-1}(\la)  \label{ecriture relation echange avec A}\\
\op{B}_N(\la) \cdot \La^{(N)}_{y,+} & = & -2\i   \sinh\Big[\f{\pi}{\om_1} (\la-y) \Big]  \cdot \La^{(N)}_{y,+} \cdot \op{B}_{N-1}(\la)  \label{ecriture relation echange avec B} \;. 
\eeqa

\end{lemme}

\Proof

Introduce the notation
\beq
 \op{T}_{N}(u_1,u_2) \equiv \op{T}_N(\la)  \; = \; \op{L}_{1}(u_1,u_2)\cdot \op{L}_{2}(u_1,u_2) \cdots \op{L}_{N}(u_1,u_2) \;, 
\enq
where $u_a$'s are given as in \eqref{definition variables u1 et u2}. 
Then, by virtue of \eqref{ecriture eqn intertwining}, \eqref {ecriture echange us et v a gauche}, \eqref{ecriture echange us et v a droite},  for arbitrary $v_1, v_2$, the operators
\beq
\op{O}_{\eps^{\prime},\eps}^{(N)}(u_1,u_2\mid v_1, v_2) \; = \; \ex{-2\i\pi \eps^{\prime} (v_2-u_2)  \op{x}_1} \cdot \op{R}_{12} (u_1,u_2\mid v_2) \cdots \op{R}_{N-1N} (u_1,u_2\mid v_2) 
\cdot \ex{2\i\pi  \eps (u_1-v_1) \op{x}_N}
\enq
satisfy to the exchange relation
\beq
\bs{v}_{\eps^{\prime}}^{\bs{t}} \cdot \op{T}_{N}(u_1,u_2 ) \, \bs{v}_{\eps} \cdot \op{O}_{\eps^{\prime},\eps}^{(N)}(u_1,u_2\mid v_2) \; = \; 
\op{O}_{\eps^{\prime},\eps}^{(N)}(u_1,u_2\mid v_1, v_2) \cdot \bs{v}_{\eps^{\prime}}^{\bs{t}} \cdot \op{T}_{N-1}(u_1,u_2) \cdot \op{L}_{N}(v_1,v_2) \, \bs{v}_{\eps}  \;. 
\label{ecritrue relation operateur O et T}
\enq
Upon adopting the parameterisation of $v_1,v_2$ in the form
\beq
v_1=\f{1}{2}\big( \mu+s-\i\f{\tau}{2}\big) \qquad \e{and} \qquad v_2 = \f{1}{2}\big( \mu - s - \i\f{\tau}{2}\big) \; , 
\enq
agreeing upon $\chi_{N;\eps}(x_N)=\ex{2\i\pi\eps x_N (s+\i\frac{\Om}{2} ) }$, it is readily checked that 
\bem
\Big( \bs{v}_{ + }^{\bs{t}}  \op{T}_{N-1}(u_1,u_2) \cdot \op{L}_{N}(v_1,v_2) \, \bs{v}_{\eps} \cdot \chi_{N;\eps}\Big)(x_N)  \\ 
\; = \; -2\i   \sinh\Big[ \f{\pi}{\om_1} (\mu - s - \i \tfrac{\Om }{ 2 } ) \Big] \,  \Big(  \op{A}_{N-1}(\la) \de_{\eps,1} \, + \,   \op{B}_{N-1}(\la) \de_{\eps,-1} \Big) \cdot \chi_{N;\eps}(x_N) \;. 
\label{ecriture action sur fct chi N}
\end{multline}
Finally, consider the re-parametrisation
\beq
v_2=\f{1}{2}\big(\la-y+\i\om_1 \big) \quad \e{and} \quad v_1=v_2+s\qquad \e{so} \; \e{that} \qquad 
\left\{ \ba{ccc} v_2 - u_2 & = & -y_-  \\
u_1-v_1+s+\i\f{\Om }{2} & = & -y_+^{\star} \ea \right.
\enq
where $y_{\pm}$ are as defined in \eqref{definition parametres y pm}. These then lead to the rewriting 
\bem
\op{R}_{12}\big( u_1, u_2 \mid v_2 \big) \; = \; D_{y_- - y_+}\big( \om_1 \om_2 \op{x}_{12} \big) \cdot D_{y_-}\big( \op{p}_1 \big) \cdot  D_{ y_+}\big( \om_1 \om_2 \op{x}_{12} \big) \\
\; = \;  D_{y_+}\big( \op{p}_1 \big) \cdot D_{y_- }\big( \om_1 \om_2 \op{x}_{12} \big) \cdot D_{y_--y_+}\big( \op{p}_1 \big) \;,
\end{multline}
which entails 
\beq
\op{O}_{+,\eps}^{(N)}(u_1,u_2\mid v_1, v_2)\cdot \chi_{N;\eps} \, = \,\Big(  \mc{A}(y_+) \sqrt{\om_1\om_2}   \Big)^{N-1} \cdot  \La^{(N)}_{y,-\eps} \, .
\enq
All together with \eqref{ecritrue relation operateur O et T}-\eqref{ecriture action sur fct chi N} this yields the two representations given in \eqref{ecriture relation echange avec A}-\eqref{ecriture relation echange avec B}

\qed

We now establish exchange relation between the $\La^{(N)}_{y,\pm}$ operators on the one hand and, on the other hand, between these operators and their duals. 
Below, we shall adopt hypergeometric-like notations for product of functions, as defined in \eqref{definition notation produit hypergeometrique}

\begin{prop}
\label{Proposition commutativite des Lambda et echange Lambda dagger Lambda}

Given $y,t \in \R$, let $y_{\pm}$, $t_{\pm}$ be defined according to \eqref{definition parametres y pm}.

For $y \not= t$, the operators $\Big( \La_{y,\eps}^{(N)} \Big)^{\dagger}$ and $\La_{t,\eps}^{(N)}$ satisfy to the exchange relations
\beq
\Big( \La_{y,\eps}^{(N)} \Big)^{\dagger} \cdot \La_{t,\eps}^{(N)} \; = \; \f{ 1 }{ \om_1 \om_2 } \cdot   \mc{A}\Big( t_+^{\star}-\big( y_+^{\star} \big)^* \, , \,  t_- - y_-^{*} \Big) \;  \La_{t,\eps}^{(N-1)} \cdot  \Big( \La_{y;\eps}^{(N-1)} \Big)^{\dagger} 
\label{ecriture relation echange lambda lambda dagger} 
\enq
and enjoy the pseudo-commutation relations
\beq
 \La_{y,\eps}^{(N)}  \, \La_{t,\eps}^{(N-1)}  \; = \;  \La_{t,\eps}^{(N)} \,  \La_{y,\eps}^{(N-1)} \;. 
\enq
\end{prop}

\Proof 

In order to establish the pseudo-commutativity, observe that 
\beq
 \La_{y,\eps}^{(N)}  \, \La_{t,\eps}^{(N-1)} \, = \; \f{ \Big(\mc{A}(t_+,y_+)\Big)^{2-N} }{  (\om_1\om_2)^{N-\frac{3}{2}} } \; \wt{\op{J}}_{N}\big( y_- - y_+\big)\cdot  \op{P}(y,t)\cdot \op{J}_{N-1}\big( t_- - t_+\big) \cdot 1_N\otimes 1_{N-1}
\enq
where 
\beq
 \op{P}(y,t)\, = \, \mc{A}(t_+) \ex{2\i\pi y_- \op{x}_1} \cdot \op{U}_N(y_-,y_+) \cdot \ex{  2\i\pi \eps y_+^{\star} \op{x}_N} \cdot 
 \ex{2\i\pi t_- \op{x}_1} \cdot \op{U}_{N-1}(t_+,t_-) \cdot \ex{  2\i\pi \eps t_+^{\star} \op{x}_{N-1}}  
\enq
Since $y_+-y_-=t_+-t_-$, it is enough to establish that $\op{P}(y,t)$ is symmetric under the exchange $t\leftrightarrow y$. 

One can recast $ \op{P}(y,t)$ in the below product form 
\bem
 \op{P}(y_{\pm},t_{\pm})\, = \, \bigg[  \ex{2\i\pi y_- \op{x}_1}  D_{y_-}\big(\op{p}_1\big) \cdot D_{y_+}\big(\om_1\om_2 \op{x}_{12}\big)  \ex{2\i\pi t_- \op{x}_1} D_{t_+}\big(\op{p}_1\big) \bigg] \\
\hspace{2cm} \times \, \pl{a}{2\curvearrowright N-2 } \bigg\{ D_{y_-}\big(\op{p}_a\big) \cdot D_{y_+}\big(\om_1\om_2 \op{x}_{aa+1}\big)   D_{t_-}\big(\om_1\om_2 \op{x}_{aa-1}\big)  D_{t_+}\big(\op{p}_a\big) \bigg\}  \\
 \hspace{1cm} \times \, \mc{A}(t_+)  \bigg[  D_{y_-}\big(\op{p}_{N-1}\big) \cdot D_{y_+}\big(\om_1\om_2 \op{x}_{N-1N}\big) D_{t_-}\big(\om_1\om_2 \op{x}_{N-2N-1}\big) \ex{2\i\pi \eps ( t_+^{\star}\op{x}_{N-1} +  y_+^{\star}\op{x}_{N} ) } \bigg] \;. 
\end{multline}
At this stage it remains to invoke the relations \eqref{ecriture identite echange 4 op D}, \eqref{ecriture identite echange 4 op D degeneree gauche} and \eqref{ecriture identite echange 4 op D degeneree droite}
so as to conclude that $  \op{P}(y_{\pm},t_{\pm}) $ is symmetric in $t \leftrightarrow y$.

The exchange relation \eqref{ecriture relation echange lambda lambda dagger}  for the adjoint can be established 
by means of the integral relations \eqref{ecriture identite integrale 3 fonction D et exposant}-\eqref{ecriture identite integrale 2 fonction D et exposant} for the $D_{\a}$ functions.
Also, one should use that provided $f$ is a regular function,
one has the integral representations 
\bem
\Big[ \ex{2\i\pi t_- \op{x}_1} \cdot \op{U}_N(t_-,t_+) \cdot \ex{  2\i\pi \eps t_+^{\star} \op{x}_N} 1_N \cdot f\big]\big( \bs{x}_N^{\prime} \big)  \\ 
\; = \;  \Int{\R^{N-1}}{}  \ex{2\i\pi t_- x_1^{\prime}} \cdot 
\pl{a=1}{N-1}\Big\{ \wt{\mc{A}}(t_-) \cdot D_{t_-^{\star}}\Big(\om_1\om_2(x_a^{\prime}-z_a)\Big) \cdot D_{t_+}\Big(\om_1\om_2(x_{a+1}^{\prime}-z_a)\Big)  \Big\} \cdot \ex{  2\i\pi \eps t_+^{\star} x_N^{\prime} }
f\big(\bs{z}_{N-1}\big) \cdot \dd^{N-1}z
\end{multline}
where $\wt{\mc{A}}(\a)=\sqrt{\om_1\om_2} \, \mc{A}(\a)$. Analogously, one has
\bem
\Big[  1_N^{\dagger} \ex{ -2\i\pi \eps \big(y_+^{\star}\big)^* \op{x}_N} \cdot \op{U}_N^{\dagger}(y_-,y_+) \cdot \ex{  -2\i\pi \eps y_-^{*} \op{x}_1}\cdot g\big]\big( \bs{x}_{N-1}^{\prime} \big)  \\ 
\; = \;  \Int{\R^{N}}{}  \ex{ -2\i\pi \eps \big(y_+^{\star}\big)^* z_N} \cdot 
\pl{a=1}{N-1}\Big\{ \wt{\mc{A}}(-y_-^*) \cdot D_{(-y_-^*)^{\star}}\Big(\om_1\om_2(x_a^{\prime}-z_a)\Big) \cdot D_{-y_+^{*}}\Big(\om_1\om_2(z_{a+1}-x_a^{\prime})\Big)  \Big\} \cdot \ex{  -2\i\pi  y_-^{*} z_1}
g\big(\bs{z}_{N-1}  \big) \cdot \dd^{N}z
\end{multline}
Above, we denote by $\bs{x}_k$ the $k$-dimensional vector $\bs{x}_k=(x_1,\dots,x_k)$. 
The claim then follows after a longish but straightforward calculation based on the integral identities \eqref{ecriture identite integrale 2 fonction D et exposant}, \eqref{ecriture identite integrale 3 fonction D et exposant}. 
\qed

\subsection{Further properties of the $\La$ operators}

\begin{prop}
\label{Proposition action operateur B} 
 It holds
\beq
\op{B}_N(y)  \cdot \La_{y,-}^{(N)} \, = \, \Big\{ b(y) \Big\}^N \cdot \La_{y+\i\om_2,-}^{(N)}  \quad with \quad b(\la) \; = \; -2\i  \sinh\Big[ \tfrac{\pi}{\om_1}\Big(\la+\kappa + \i \tfrac{\Om}{2}\Big)\Big]   
\enq
and
\beq
\op{C}_N(y)  \cdot \La_{y,-}^{(N)} \, = \, \Big\{ c(y) \Big\}^N \cdot \La_{y-\i\om_2,-}^{(N)}  \quad with \quad c(\la) \; = \; -2\i  \sinh\Big[ \tfrac{\pi}{\om_1}\Big(\la - \kappa - \i \tfrac{\Om}{2}\Big)\Big]  \;. 
\enq
 Furthermore, dual relations holds for the dual objects. 
 
\end{prop}

One possible way to establish the above relations is based on the use of the gauge transformation initially
suggested in the paper \cite{GaudinPasquierQOpConstructionForTodaChain} for the derivation of the Baxter $T-Q$ equation in the case
of the Toda chain and adapted to the case of XXX-spin chain in \cite{DerkachovQopForXXX}  
and later used in \cite{KozBabelonPasquierQOpTQEqnForqTodaAndToda2,DerkachovKorchemskyManashovXXXSoVandQopNewConstEigenfctsBOp,DerkachovKorchemskyManashovXXXreelSoVandQopABAcomparaison,DerkachovManashovInterativeConstrEigenFctsSL2C}. 
Here, however, but we shall present a proof which adapts, to the situation of interest, 
the reasoning introduced in \cite{ChicherinDerkachovKarakhanyanKirschnerBaxterOpForModularXXZandEllipticModels}. 
Although we do not provide these here, analogous relations can be obtained within the same technique for the action of the $\op{A}_N, \op{D}_N$ operators on the $\La_{y,+}^{(N)}$ operator.

\Proof 

The intertwining relation \eqref{ecriture eqn intertwining} may be recast in the form 
\beq
\op{R}_{12}(u_1,u_2\mid v_2) \,   \op{L}_1(u_1,u_2)    \; = \; \op{L}_1(u_1,v_2) \, \op{L}_2(u_1,u_2) \, \op{R}_{12}(u_1,u_2\mid v_2)  \cdot \op{L}_2^{-1}(u_1,v_2)  \;. 
\enq
Then, by using the elementary decomposition of the $\op{L}$ matrices \eqref{ecriture factorisation auxiliare de L XXZ}, one gets the identity
\beq
\op{R}_{12}(u_1,u_2\mid v_2) \,   \op{L}_1(u_1,u_2)    \; = \; -\i \, \op{M}_{v_2}(\op{x}_1) \cdot \op{G}   \cdot \op{M}_{v_2}^{-1}(\op{x}_2)  
\enq
with
\beq
 \op{G} \; = \; \op{H}(\op{p}_1) \cdot \op{N}_{u_1}(\op{x}_1) \cdot \op{M}_{u_2}(\op{x}_2) \cdot \op{H}(\op{p}_2) \cdot  \op{R}_{12}(u_1,u_2\mid v_2) \cdot \op{H}^{-1}(\op{p}_2) \;. 
\enq
Upon  multiplication  by the gauge matrices 
\beq
\op{Z}_k \; = \; \left(\ba{cc}  1 & 0 \\ \ex{2\pi \om_2 \op{x}_k} & 1  \ea \right) \;, 
\enq
the above relation takes the form 
\beq
\op{Z}_1 \cdot  \op{R}_{12}(u_1,u_2\mid v_2) \,   \op{L}_1(u_1,u_2)   \cdot  \op{Z}_2^{-1} \; = \;
 \left(\ba{cc}  \wt{\op{G}}_{11}  &  \wt{\op{G}}_{12} \cdot \big[ V_2-V_2^{-1} \big]^{-1}  \\ 
\big[ V_2-V_2^{-1} \big] \cdot  \wt{\op{G}}_{21}   &  \wt{\op{G}}_{22}  \ea \right) \;, 
\label{ecriture relation echange jaugee}
\enq
where $V_2$ is defined according to \eqref{definition variables u1 et u2} and we have set 
\beq
\wt{\op{G}} \; = \;  \f{ -\i }{  V_2  + V_2^{-1}  }  \left( \ba{cc} V_2    & - V_2^{-1} \\ 
					 - V_2^{-1} \ex{2\pi \om_2 \op{x}_1}  & V_2 \ex{2\pi \om_2 \op{x}_1}  \ea \right) \cdot \op{G} \cdot 
\left( \ba{cc}     & - V_2^{-1} \ex{-2\pi \om_2 \op{x}_2}  \\ 
					 -  1   & V_2 \ex{ -2\pi \om_2 \op{x}_2}  \ea \right) \;. 
\enq
The matrix elements of $\wt{\op{G}}$ are smooth in $V_2$ belonging to a vicinity of $1$. Furthemore the \textit{lhs} of \eqref{ecriture relation echange jaugee}
has a well-definied limit when $v_2 \tend 0$. Thus, the $12$-enrty of the \textit{rhs} of \eqref{ecriture relation echange jaugee} admits a $V_2\tend 1$
limit. Furthermore, the $21$-entry vanishes in this limit. Hence, one has the structure
\beq
\op{Z}_1 \cdot  \op{R}_{12}(u_1,u_2\mid 0) \,   \op{L}_1(u_1,u_2)   \cdot  \op{Z}_2^{-1} \; = \;
 \left(\ba{cc}  \wt{\op{G}}_{11}  &   *   \\ 
0    &  \wt{\op{G}}_{22}  \ea \right)_{\mid v_2=0} \;.  
\label{ecriture reduction jauge locale}
\enq
In order to compute the entries $\wt{\op{G}}_{11} $ and $\wt{\op{G}}_{22} $ it appears more convenient to slightly reorganise their expression.

Indeed, by substituing the second expression for $\op{R}_{12}$ given in \eqref{ecriture expression explicite factorisation elementaire R12}, using its independence on $\op{p}_2$
and then by simplifying factors issuing from the adjoint action of $\op{L}_{2}$ operators from its decomposition \eqref{ecriture factorisation auxiliare de L XXZ}
and finally, by moving the $D_{\a}(\op{p}_1)$ operators to the almost exteme left and right sides what can be done by using the intertwining relation \eqref{equation entrelacement rep kappa et moins kappa}, one gets 
\bem
\op{Z}_1 \cdot  \op{R}_{12}(u_1,u_2\mid 0) \,   \op{L}_1(u_1,u_2)   \cdot  \op{Z}_2^{-1}  \\ 
\; = \;\op{Z}_1 \cdot D_{u_1-v_2}(\op{p}_1) \cdot \op{L}_1( v_2, u_1 ) \cdot 
\op{M}_{u_2}(\op{x}_2 ) \cdot D_{u_2-v_2}\Big(\om_1\om_2 \op{x}_{12} -\i \tfrac{ \om_2 }{ 2 } \sg_3 \Big) \cdot  \op{M}_{v_2}^{-1}(\op{x}_2 )  \cdot  \op{Z}_2^{-1} \cdot D_{u_2-v_1}(\op{p}_1) \;. 
\end{multline}
In order to read out from that the expression for $ \wt{\op{G}}_{11}$ one should compose the \textit{rhs} with the vectore $\bs{v}_{+}^{\bs{t}}$ and $\bs{v}_{+}$
as defined in \eqref{defintion vecteurs v pm}. 
For doing so, it is useful to remark that 
\beq
\bs{v}_{+}^{\bs{t}} \, \op{Z}_1 \cdot D_{u_1-v_2}(\op{p}_1) \cdot \op{L}_1( v_2, u_1 ) \; = \; D_{u_1-v_2}(\op{p}_1) \,  \big( U_1 \;,  \,  - U_1^{-1} \big)
\quad \e{and} \quad 
\op{M}_{v_2}^{-1}(\op{x}_2 )  \cdot  \op{Z}_2^{-1}\bs{v}_{+} \; = \; \f{1}{V_2+V_2^{-1}} \left(\ba{c} 1 \\ -1 \ea \right) \;.
\enq
Also, one has 
\beq
\op{N}_{0}(\op{x}_1 ) \cdot  \op{M}_{u_2}(\op{x}_2 ) \cdot D_{u_2-v_2}\Big(\om_1\om_2 \op{x}_{12} -\i \tfrac{ \om_2 }{ 2 } \sg_3 \Big)  \left(\ba{c} 1  \\  -1 \ea \right)
 \; = \; - \left(\ba{c}  \mf{m}  \\  \mf{m} \ea \right)
\enq
with 
\beq
 \mf{m} \; = \; \Big( U_2 +  U_2^{-1} \ex{2\pi   \om_2 \op{x}_{21}} \Big) \cdot  D_{u_2}\Big(\om_1\om_2 \op{x}_{12} -\i \tfrac{ \om_2 }{ 2 }  \Big)
\; + \; \Big( U_2^{-1} +  U_2 \ex{2\pi   \om_2 \op{x}_{21}} \Big) \cdot D_{u_2}\Big(\om_1\om_2 \op{x}_{12} +\i \tfrac{ \om_2 }{ 2 }  \Big)  \;. 
\enq
By using the transmutation properties of $D_{\a}$ functions \eqref{ecriture ptes transmutation fct D alpha}, one recasts $\mf{m}$ in the form 
\beq
 \mf{m} \; = \; 2 \, \ex{ \pi  \om_2 \op{x}_{21} } \cdot D_{u_2 +\i \frac{ \om_2 }{ 2 } }\Big(\om_1\om_2 \op{x}_{12}   \Big)  \;. 
\enq
All of this leads to 
\beq
 \wt{\op{G}}_{11}   \mid_{v_2=0}  \, = \,    \i    D_{u_1}(\op{p}_1) \, \Big( U_1 \, , \, - U_1^{-1} \Big) \,  \ex{ -\f{\pi}{\om_1}(\op{p}_1-\i\f{\Om}{2}) \sg_3 }  \, \left( \ba{c }  1 \\  1 \ea \right)
 \, \ex{ \pi   \om_2 \op{x}_{21} } \cdot D_{u_2 +\i \frac{ \om_2 }{ 2 } } \big(\om_1\om_2 \op{x}_{12}   \big) \cdot D_{u_2-v_1}(\op{p}_1) \;. 
\enq
By taking the scalar products explicitly and then using, again, transmutation properties \eqref{ecriture ptes transmutation fct D alpha}, one arrives to 
\beq
 \wt{\op{G}}_{11} \mid_{v_2=0} \, = \, - \ex{- \pi   \om_2 \op{x}_{1} }  \op{R}_{12}\Big( u_1 +\i \tfrac{ \om_2 }{ 2 } , u_2 +\i \tfrac{ \om_2 }{ 2 } \mid 0 \Big) \cdot   \ex{ \pi   \om_2 \op{x}_{2} } \;. 
\enq
Quite similarly, one gets 
\beq
 \wt{\op{G}}_{11} \mid_{v_2=0} \, = \, 4 \sinh\Big[ \tfrac{2\pi}{\om_1} u_1\Big] \sinh\Big[ \tfrac{2\pi}{\om_1} u_2\Big] 
 \cdot \ex{  \pi   \om_2 \op{x}_{1} }  \op{R}_{12}\Big( u_1 -\i \tfrac{ \om_2 }{ 2 } , u_2 -\i \tfrac{ \om_2 }{ 2 } \mid 0 \Big) \cdot   \ex{ - \pi   \om_2 \op{x}_{2} } \;. 
\enq

The triangular structure in \eqref{ecriture reduction jauge locale} allows one to explicitly take products and leads to the relation 
\beq
\op{Z}_1 \cdot \Big( \op{R}_{12}\cdots \op{R}_{N0}\Big) (u_1,u_2\mid v_2) \,  \op{T}_N(u_1,u_2) \cdot  \op{Z}_0^{-1} 
\; = \; \ex{-\pi \om_2 \op{x}_1 \sg_3 }
\left( \ba{cc}   \mc{C}_{\ua}   & *  \\
      0    &    \mc{C}_{\da}  \ea \right) \ex{ \pi \om_2 \op{x}_0 \sg_3 } \;.  
\label{ecriture relation echange RT}
\enq
There, $\op{T}_N(u_1,u_2) =   \op{L}_1(u_1,u_2)  \cdots  \op{L}_N(u_1,u_2)  $ is the monodromy matrix and we agree upon 
\beq
\mc{C}_{\ua}\, = \,  (-1)^N \Big( \op{R}_{12}\cdots \op{R}_{N0}\Big) \Big( u_1 +\i \tfrac{ \om_2 }{ 2 } , u_2 +\i \tfrac{ \om_2 }{ 2 } \mid 0 \Big)
\enq
and
\beq
\mc{C}_{\da} \, = \, \Big\{ 4 \sinh\Big[ \tfrac{2\pi}{\om_1} u_1\Big] \sinh\Big[ \tfrac{2\pi}{\om_1} u_2\Big]  \Big\}^{N}
 \cdot\Big( \op{R}_{12}\cdots \op{R}_{N0}\Big) \Big( u_1 -\i \tfrac{ \om_2 }{ 2 } , u_2 -\i \tfrac{ \om_2 }{ 2 } \mid 0 \Big) \;. 
\enq
In order to deduce from \eqref{ecriture relation echange RT}  the action of the $\op{B}_N$ operator, one needs to exchange the position of the monodromy matrix 
and the string of partial $\op{R}$-operators. The intertwining relation \eqref{ecriture eqn intertwining}  leads to 
\bem
\Big( \op{R}_{12}\cdots \op{R}_{N0}\Big) (u_1,u_2\mid v_2) \,   \op{L}_1(u_1,u_2)  \cdots  \op{L}_N(u_1,u_2)  \op{L}_0(u_1,v_2)  \\ 
\; = \;   \op{L}_1(u_1,v_2) \, \op{L}_2(u_1,u_2)  \cdots  \op{L}_0(u_1,u_2)  \,\Big( \op{R}_{12}\cdots \op{R}_{N0}\Big) (u_1,u_2\mid v_2)  \;. 
\end{multline}
Then, upon using the relation \eqref{ecriture echange us et v a gauche} one obtains 
\bem
 \ex{ -2\i\pi \eps (v_2-u_2) \op{x}_1 } \cdot \Big( \op{R}_{12}\cdots \op{R}_{N0}\Big) (u_1,u_2\mid v_2) \, \cdot \bs{v}_{\eps}^{\bs{t}} \,  \op{T}_N(u_1,u_2) \, \op{L}_0(u_1,v_2) \\
\; = \; \bs{v}_{\eps}^{\bs{t}}  \, \op{T}_N(u_1,u_2) \, \op{L}_0(u_1,u_2)  \ex{ -2\i\pi \eps (v_2-u_2) \op{x}_1 }  \, \Big( \op{R}_{12}\cdots \op{R}_{N0}\Big) (u_1,u_2\mid v_2) \;. 
\end{multline}
Finally, upon using the independence on $\op{p}_0$ of $\op{R}_{N0}$ and inserting the explicit expression for the $\op{L}_0$ operators,  one arrives to 
\bem
 \ex{ -2\i\pi \eps (v_2-u_2) \op{x}_1 } \cdot \Big( \op{R}_{12}\cdots \op{R}_{N0}\Big) (u_1,u_2\mid v_2) \, \cdot \bs{v}_{\eps}^{\bs{t}} \,  \op{T}_N(u_1,u_2)   \\
\; = \; \bs{v}_{\eps}^{\bs{t}}  \, \op{T}_N(u_1,u_2)  \,   \ex{ -2\i\pi \eps (v_2-u_2) \op{x}_1 } \,  \Big( \op{R}_{12}\cdots \op{R}_{N-1N}\Big) (u_1,u_2\mid v_2)  \\
\times   D_{u_1-v_2}\big(  \op{p}_N \big)  \cdot  \op{M}_{u_2}(\op{x}_0)   \cdot D_{u_2-v_2}\Big(\om_1\om_2 \op{x}_{N0} -\i \tfrac{ \om_2 }{ 2 } \sg_3 \Big)  
  \cdot  \op{M}_{v_2}^{-1}(\op{x}_0) \cdot  D_{u_2-u_1}\big(  \op{p}_N \big) \;. 
\end{multline}
This last identity implies that one may recast \eqref{ecriture relation echange RT} in the form 
\bem
 \bs{v}_{+}^{\bs{t}}  \, \op{T}_N(u_1,u_2)  \,   \ex{  2\i\pi u_2 \op{x}_1 } \,  \Big( \op{R}_{12}\cdots \op{R}_{N-1N}\Big) (u_1,u_2\mid 0)    
\; = \; \ex{  \i\pi (2u_2 + \i \om_2) \op{x}_1 } 
\Big(    \mc{C}_{\ua}   \, , \,  *  \Big) \, \ex{ \pi \om_2 \op{x}_0 \sg_3 }   \\
 \times D_{u_1-u_2}\big(  \op{p}_N \big) \cdot \op{Z}_0 \cdot  \op{M}_{0}(\op{x}_0)   \cdot D_{-u_2}\Big(\om_1\om_2 \op{x}_{N0} -\i \tfrac{ \om_2 }{ 2 } \sg_3 \Big)  
  \cdot  \op{M}_{u_2}^{-1}(\op{x}_0) \cdot  D_{-u_1}\big(  \op{p}_N \big)
\;.
\end{multline}
Evaluating this expression on $\bs{v}_-$, acting with it on the function $1$ in the space $0$  and 
using that 
\beq
\lim_{x_0\tend +\infty} \Big\{ \ex{ - \pi \om_2 x_0  } \cdot D_{ u_2 + \i \frac{ \om_2 }{ 2 }  }\Big(\om_1\om_2 x_{N0}  \Big)   \cdot D_{-u_2}\Big(\om_1\om_2 x_{N0} -\i \tfrac{ \om_2 }{ 2 } \sg_3 \Big)   \Big\}
\; = \; \ex{-\pi \om_2 x_N} \, U_2^{-\sg_3}  \;, 
\enq
leads to 
\beq
\op{B}_N(y-\i\om_1) \cdot \ga_N(u_1,u_2) \; = \; (-1)^{N+1} \ga_N \Big( u_1 +\i \tfrac{ \om_2 }{ 2 } , u_2 +\i \tfrac{ \om_2 }{ 2 }  \Big)\cdot 
  D_{ u_1+ \i \frac{ \om_2 }{ 2 }}\big(  \op{p}_N \big) \ex{-\pi \om_2 \op{x}_N} D_{-u_1}\big(  \op{p}_N \big)
\enq
where we used that the condition $v_2=0$ is equivalent to $\la = y - \i\om_1$ and we have set 
\beq
\ga_N(u_1,u_2) \; = \; \ex{  2\i\pi u_2 \op{x}_1 }  \Big( \op{R}_{12}\cdots \op{R}_{N-1N}\Big) (u_1,u_2\mid 0)  \;. 
\enq
Finally, acting on the function  $\ex{-2\i\pi y_+^{\star}x_N}$ on the $N^{\e{th}}$ space produces the $\La$-operator on the \textit{lhs} and one gets, by recombining the factors
\beq
\op{B}_N(y-\i\om_1) \cdot \La^{(N)}_{y,-} \; = \; (-1)^{N+1}  \Bigg(  \f{ \mc{A}\big( (y +\i \om_2)_+ \big)  }{ \mc{A}(y_+) } \Bigg)^{N-1}   
D_{ u_1+ \i \frac{ \om_2 }{ 2 }}\Big(   y_+^{\star} -+ \i \frac{ \om_2 }{ 2 } \Big) \, D_{-u_1}\big(   y_+^{\star} \big) \cdot \La^{(N)}_{y+\i \om_2,-}  \;. 
\enq
Upon observing that one has $u_1=y_+$, a straightforward calculation leads to 
\beq
\op{B}_N(y-\i\om_1) \cdot \La^{(N)}_{y,-} \; = \; (-1)^{N-1} b(y)  \La^{(N)}_{y+\i \om_2,-}  
\enq
hence yielding the claim upon observing that $\op{B}_N(y-\i\om_1)=(-1)^{N-1}\op{B}_N(y)$. \qed

For the statement of the next result, it appears convenient to keep track of the $\kappa$ dependence of the operator $\La^{(N)}_{y,-}$
and denote it as $\La^{(N)}_{y,\kappa,-}$. 

\begin{lemme}
 \label{Proposition action operateur spin flip sur operateur Lambda}
One has the exchange relations 
\beq
 \op{J}_{N+1}(y_- - y_+) \cdot \La^{(N)}_{y,\kappa,-}  \, = \, D_{-\kappa}^N(y) \cdot  \La^{(N)}_{y, - \kappa,-}  \cdot  \op{W}_{N}(y_- - y_+) \cdot \op{J}_{N}(y_- - y_+)
\enq
where we have introduced
\beq
\op{J}_{N+1}( \a ) \, = \, \pl{a=1}{N} D_{\a}(\op{p}_a) \qquad \e{and} \qquad   \op{W}_{N+1}( \a )  \; = \;
\op{J}_{N+1}( \a ) \ex{2\i\pi \a \op{x}_N } \cdot  \pl{a=1}{N} D_{\a}\Big( \om_1 \om_2 \op{x}_{a \, a+1} \Big) \cdot  \ex{2\i\pi \a \op{x}_1 }\;. 
\enq
Also, it holds 
\beq
 \op{W}_{N+1}(y_- - y_+)  \cdot \La^{(N)}_{y,-\kappa,-}  \, = \, D_{-\kappa}(y) \cdot  \La^{(N)}_{y, - \kappa,-} \cdot   \op{W}_{N}(y_- - y_+)  \;. 
\enq

 \end{lemme}
Note that, in the above definition, it is understood that 
\beq
 \op{W}_{1}( \a )  \; = \;  \op{J}_{1}( \a )  \, = \, D_{\a}(\op{p}_1) \;. 
\enq

\Proof 
This is a direct consequence of a multiple application of the star-triangle relation and of the interchange relation of $D$s and exponentials \eqref{ecriture identite integrale 2 fonction D et exposant}. \qed

\subsection{Elementary properties of a basic complete orthogonal system}

In this subsection we discuss the  complete orthogonal system of generalised Eigenfunctions of the one-site operator $\op{B}_1(\la)$  and establish its properties under specific
  action of $\La^{(2)}_{y,-}$ operators. 

\begin{lemme}
\label{Lemme fct propre B un site}

The functions 
\beq
\phi_{y}(x) \; = \; \Int{ \R }{} \varpi(\kappa+t) \ex{ 2i\pi t (x+y) }\cdot \dd t 
\enq
are self-dual under the exchange $\om_1\leftrightarrow \om_2$ and satisfy to 
\beq
\op{B}_1(\la) \cdot \phi_y(x) \, = \,  \ex{2\pi \om_2 y} \, \phi_y(x) \, . 
\enq
The family $\{ \phi_{y}(x) \}$ forms a complete orthogonal system on $L^2(\R)$, \textit{viz}. 
\beq
\Int{ \R }{}  \phi_{y^{\prime}}^{*}(x)\cdot \phi_y(x) \cdot \dd x \; = \; \de\big( y^{\prime}-y\big) \qquad and \qquad
\Int{ \R }{}  \phi_{y}^{*}(x^{\prime})\cdot \phi_y(x) \cdot \dd y \; = \; \de\big( x^{\prime}-x\big) \;. 
\enq
\end{lemme}

\Proof 
The generalised spectral problem for the  operator $\op{B}_1(\la)$ associated with the generalised Eigenvalue $ \ex{2\pi \om_2 y}$
can be recast in the form of the below finite-difference equation
\beq
2\i \sinh\Big[\f{\pi}{\om_1}\big(\kappa+\op{p}+\i\f{\tau}{2} \big) \Big]\cdot \phi_{y}(x)\; = \; \ex{2\pi \om_2(x+y) }  \phi_{y}(x)  \;. 
\enq
   One possible solution is given by $\phi_{y}(x)=\phi_{0}(x+y)$. Thus we focus on $\phi_{0}$. Passing to the Fourier space, one gets the below 
finite difference equation satisfied by the Fourier transform
\beq
\mc{F}[\phi_0](t+ \i \om_2 ) \; = \; 2\i \sinh\Big[\f{\pi}{\om_1}\big(\kappa + t + \i\f{\tau}{2} \big) \Big] \mc{F}[\phi_0](t) \;. 
\enq
This equation, along with its dual, is easily solved in terms of the dilogarithm leading, all-in-all, to the claim. \textit{c.f.} \eqref{ecriture eqn diff finite dilog}. The completeness and orthogonality follows
from straightforward handlings. In particular,  these entail that each generalised Eigenvalue of $\op{B}_1(\la)$  appears with multiplicity one. \qed

\begin{lemme}
\label{Lemme Action  operateurs Lambda dag Lambda deux sur psi y}
For $y, t, t_0 \in \R$, it holds 
\beq
\Big( \La^{(2)}_{y,+} \Big)^{\dagger} \cdot  \La^{(2)}_{t,+}  \cdot \phi_{t_0} \; = \; \de(y-t) \;  \f{ \ex{-2\pi \Om t_0 }  }{ \big(\om_1 \om_2\big)^{\frac{3}{2}} }\, \phi_{t_0} \;. 
\enq
\end{lemme}

\Proof

One has the representation
\bem
\Big( \La^{(2)}_{y,+} \Big)^{\dagger} \cdot  \La^{(2)}_{t,+}   \; = \; \f{ 1 }{ \om_1 \om_2 } \mc{A}\bigg( \ba{c} \big( -y_+^{*}\big)^{\star}  \\ t_+ \ea \bigg) \\
 \times \e{tr}_2\bigg[
\ex{- 2\i\pi  \big( y_+^{\star}\big)^{*} \op{x}_2 }  D_{-y_+^*}\big(\om_1\om_2\op{x}_{12}\big) D_{-y_-^{*}}\big(\op{p}_1\big) \cdot \ex{2\i\pi \big( t_- - y_-^*\big) \op{x}_1} 
\cdot  D_{t_-}\big(\op{p}_1\big)  \cdot D_{t_+}\big(\om_1\om_2 \op{x}_{12}\big)  \ex{ 2\i\pi   t_+^{\star} \op{x}_2 }  \Big]
\end{multline}
where the trace runs through the second quantum space. Here, we remind that we use the hypergeometric-like notation \eqref{definition notation produit hypergeometrique}.
In virtue of \eqref{ecriture degenerescence exponentielle etoile triangle} the expression can be recast as
\bem
\Big( \La^{(2)}_{y,+} \Big)^{\dagger} \cdot  \La^{(2)}_{t,+}  \cdot   \mc{A}\bigg( \ba{c} t_+  \\ \big( -y_+^{*}\big)^{\star}  \ea \bigg) \cdot \om_1 \om_2  \\
= \; \e{tr}_2\bigg[  \ex{ 2\i\pi \big[t_-\op{x}_1-  \big( y_+^{\star}\big)^{*} \op{x}_2\big] }  D_{-y_+^*}\big(\om_1\om_2\op{x}_{12}\big) D_{t_- - y_-^{*} }\big(\op{p}_1\big) 
\cdot D_{t_+}\big(\om_1\om_2 \op{x}_{12}\big)  \ex{ 2\i\pi  \big[ t_+^{\star} \op{x}_2 -y_-^*\op{x}_1 \big] }  \bigg] \;. 
\end{multline}
Then, invoking the star-triangle relation \eqref{ecriture identite etoile triangle} one obtains
\bem
\Big( \La^{(2)}_{y,+} \Big)^{\dagger} \cdot  \La^{(2)}_{t,+}     \cdot   \mc{A}\bigg( \ba{c} t_+  \\ \big( -y_+^{*}\big)^{\star}  \ea \bigg)  \cdot \om_1 \om_2  \\
\hspace{1cm} \; = \; \e{tr}_2\bigg[
\ex{ 2\i\pi \big[t_-\op{x}_1-  \big( y_+^{\star}\big)^{*} \op{x}_2\big] }  D_{t_+}\big(\op{p}_1\big) D_{t_- - y_-^{*} }\big(\om_1\om_2\op{x}_{12}\big) 
\cdot D_{-y_+^*}\big(\op{p}_1\big)  \ex{ 2\i\pi  \big( t_+^{\star} \op{x}_2 -y_-^*\op{x}_1 \big) }  \bigg] \\
\; = \; \ex{ 2\i\pi t_-\op{x}_1 } \cdot D_{t_+}\big(\op{p}_1\big)  \cdot  \ex{ - 2\i\pi  \big( y_+^{\star}\big)^{*} \op{x}_1 }  \cdot 
\e{tr}_2\bigg[ \ex{ - 2\i\pi  \big[ t_+^{\star} - \big( y_+^{\star}\big)^{*} \big] \op{x}_{12} }     D_{t_- - y_-^{*} }\big(\om_1\om_2\op{x}_{12}\big) \bigg]
\cdot \ex{ 2\i\pi   t_+^{\star} \op{x}_1 } \cdot D_{-y_+^*}\big(\op{p}_1\big) \cdot  \ex{ - 2\i\pi y_-^*\op{x}_1  } \;. 
\label{equation reecriture PS Lambda a deux sites}
\end{multline}
At this stage one can already take the trace over the second space. However, for this purpose, one needs to regularise the integral appropriately. 
Indeed, one ends up with the below integral
\beq
\e{tr}_2\bigg[ \ex{ - 2\i\pi  \big[ t_+^{\star} - \big( y_+^{\star}\big)^{*} \big] \op{x}_{12} }     D_{t_- - y_-^{*} }\big(\om_1\om_2\op{x}_{12}\big) \bigg] \; = \; 
\lim_{\eta \tend 0^+} \Int{ \R }{}  \underbrace{ \ex{  2\i\pi  \big[ t_+^{\star} - \big( y_+^{\star}\big)^{*} +\i \f{3}{2} \eta \big] x }     D_{t_- - y_-^{*} + \i\f{\eta}{2} }\big(\om_1\om_2x \big) }_{= f_{\eta}(x)} \cdot \dd x \;. 
\label{ecriture rep int trace espace 2 pour PS 2 sites fct propres B}
\enq
The above regularisation is such that $f_{\eta}\in L^{1}(\R)$. Indeed, one has 
\beq
\left\{ \ba{ccc} t_+^{\star} - \big( y_+^{\star}\big)^{*} & = &  \tfrac{1}{2}(y-t)-\i \tfrac{ \Om }{2} \vspace{2mm} \\
t_--y_-^* & =  & \tfrac{1}{2}(t-y)-\i\tfrac{ \Om }{2}   \ea \right. \quad viz. \quad \Big[ t_- - y_-^* + \i\tfrac{\eta}{2} \Big]^{\star} \, =  \,\tfrac{y-t}{2} - \i\tfrac{\eta}{2}
\enq
and hence
\beq
f_{\eta}(x)  \underset{ x\tend +\infty }{ \sim } \ex{2 \i \pi \big[ y-t+ \i \eta \big] x }  \qquad 
f_{\eta}(x)  \underset{ x\tend -\infty }{ \sim } \ex{2 \i \pi \big[-\i \Om + 2 \i \eta \big] x }  \;. 
\enq
The integral in the \textit{rhs} of \eqref{ecriture rep int trace espace 2 pour PS 2 sites fct propres B} can be taken in closed form by means of \eqref{ecriture TF de fct D}. This leads, after some re-organisations, to 
\beq
\e{tr}_2\bigg[ \ex{ - 2\i\pi  \big[ t_+^{\star} - \big( y_+^{\star}\big)^{*} \big] \op{x}_{12} }     D_{t_- - y_-^{*} }\big(\om_1\om_2\op{x}_{12}\big) \bigg] \; = \; 
\f{ 1 }{ \sqrt{\om_1\om_2} } \cdot \lim_{\eta \tend 0^+} \Big\{ \mc{A}(-\i\eta) \cdot D_{ (-i\eta)^{\star} }\big( y - t \big) \Big\} \; = \; \f{  \de(t-y)  }{ \sqrt{\om_1\om_2} }   \;, 
\enq
where we used \eqref{ecriture realisation fct delta via fct D alpha}. Then, inserting this result into \eqref{equation reecriture PS Lambda a deux sites} and observing that 
\beq
t_+^{\star}-\big( y_+^{\star} \big)^{*}\, = \, \f{y-t}{2} - \i \f{\Om}{2} \quad , \quad t_+ - y_+^{*}\, = \, \f{t-y}{2} - \i \f{\Om}{2} 
\enq
one concludes that two quantities coincide when $t=y$. Further $\big( -y_+^{*} \big)^{\star}=y_+$ what ensures that the $\mf{A}$ factors in \eqref{equation reecriture PS Lambda a deux sites} 
cancel out. All-in-all, one gets that 
\beqa
\Big( \La^{(2)}_{y,+} \Big)^{\dagger} \cdot  \La^{(2)}_{t,+}   & = &  \f{  \de(t-y)  }{ \big(\om_1 \om_2\big)^{\frac{3}{2}}  } \cdot \ex{ 2\i\pi t_-\op{x}_1 } \cdot D_{t_+}\big(\op{p}_1\big)  \cdot  \ex{  2\i\pi  \big(  t_+ - y_+^{*} \big) \op{x}_1 }  \cdot 
 D_{-y_+^*}\big(\op{p}_1\big) \cdot  \ex{ - 2\i\pi y_-^*\op{x}_1  } \nonumber \\ 
 & = &    \f{  \de(t-y)  }{  \big(\om_1 \om_2\big)^{\frac{3}{2}}  } \cdot \ex{ 2\i\pi (t_--y_+^*)\op{x}_1 } \cdot D_{t_+ - y_+^{*} }\big(\op{p}_1\big)  \cdot   \ex{  2\i\pi(t_+- y_-^*)\op{x}_1  } \nonumber  \\
 & = &   \f{  \de(t-y)  }{ \big(\om_1 \om_2\big)^{\frac{3}{2}}  }\cdot \ex{ 2\i\pi \big( t_- + t_+ - y_-^* - y_+^* \big) \op{x}_1 } \cdot D_{t_+ - y_+^{*} }\big(\op{p}_1 + t_+- y_-^*\big)  \nonumber  \\
 & = &  \f{  \de(t-y)  }{ \big(\om_1 \om_2\big)^{\frac{3}{2}}  } \cdot \ex{ - 2\i\pi \i \Om  \op{x}_1 } \cdot D_{-\i \f{ \Om }{ 2 }  }\big(\op{p}_1  + \kappa -\i \f{ \Om }{2} \big) \nonumber \\
 & = &  \f{  \de(t-y)  }{ \big(\om_1 \om_2\big)^{\frac{3}{2}} }\cdot \ex{ - 2\i\pi \i \Om  \op{x}_1 } \varpi \bigg( \ba{c} \kappa -\i \Om + \op{p}_1  \\   \kappa  + \op{p}_1  \ea \bigg) \;. 
\eeqa
It then solely remains to represent $\phi_{t_0}$ as
\beq
\phi_{t_0}(x) \; = \; \Int{\R}{}  \varpi\big( \kappa + y +\i \Om \big) \ex{2\i \pi (y+\i \Om)(x+t_0)  } \cdot \dd y \;. 
\enq
Then by moving the action of $ \Big( \La^{(2)}_{y,+} \Big)^{\dagger} \cdot  \La^{(2)}_{t,+}  $ under the integral sign one gets the claim. \qed

\subsection{The orthogonal sets of Eigenfunctions of $\op{A}_N(\la)$, $\op{B}_N(\la)$}

We are now in position to present a set of generalised Eigenfunctions associated with the operators $\op{A}_N(\la)$ and  $\op{B}_N(\la)$. 
Then, we establish that these Eigenfunctions form, in each case, an orthogonal system.

\begin{prop}

The functions 
\beq
\vp_{ \bs{y}_N }^{ (-) }( \bs{x}_N ) \; = \;   \Big( \La_{y_N,-}^{(N)} \cdots \La_{y_1,-}^{(1)} \Big) (\bs{x}_N)
\enq
and
\beq
\vp_{ y_0 , \bs{y}_{N-1} }^{(+)}( \bs{x}_N ) \; = \;   \ex{ (N-1) \pi \Om y_0 } \cdot \Big(\om_1 \om_2 \Big)^{ \frac{3}{4} (N-1) } \cdot \Big( \La_{y_{N-1},+}^{(N)}\cdots \La_{y_1,+}^{(2)} \cdot \phi_{y_0}  \Big) (\bs{x}_N)
\enq
are, respectively, symmetric functions of $\bs{y}_N$ and $\bs{y}_{N-1}$ and satisfy to 
\beq
\op{A}_N(\la) \cdot \vp_{\bs{y}_N}^{(-)}(\bs{x}_N) \; = \;   \pl{a=1}{N} \Big\{ -2\i \sinh\Big[\f{\pi}{\om_1} \big( \la-y_a \big) \Big] \Big\} \cdot \vp_{\bs{y}_N}^{(-)}(\bs{x}_N)
\enq
and 
\beq
\op{B}_N(\la) \cdot \vp_{y_0,\bs{y}_{N-1}}^{(+)}(\bs{x}_N) \; = \;  \ex{2\pi \om_2 y_0 }  
		      \pl{a=1}{N-1} \Big\{ -2\i \sinh\Big[\f{\pi}{\om_1} \big( \la-y_a \big) \Big] \Big\} \cdot \vp_{y_0,\bs{y}_{N-1}}^{(+)}(\bs{x}_N) \;. 
\enq

\end{prop}

\Proof 

The symmetry property in the $y$-variables follows from commutativity of the $\La$ operators established in Proposition \ref{Proposition commutativite des Lambda et echange Lambda dagger Lambda}. 
The form of the action of $\op{B}_N$ and $\op{A}_N$ operators is a consequence of Lemmatae \ref{Lemme fct propre B un site} and \ref{Lemme action recursive AN et BN sur LambdaN}
and the fact that 
\beq
\Big(\op{A}_{1}(\la) \cdot \La^{(1)}_{y,-}\Big)(x_1) \; = \; -2\i    \sinh\Big[\f{\pi}{\om_1} \big( \la-y \big) \Big] \, \La^{(1)}_{y,-}(x_1) 
\qquad \e{where} \qquad \La^{(1)}_{y,-}(x)=\ex{2\i\pi y x} \;. 
\enq

\qed

\begin{prop}

The families 
\beq
\big\{ \vp_{\bs{y}_N}^{(-)}(\bs{x}_N) \big\}_{\bs{y}_N \in \R^N} \quad  and  \quad  \big\{ \vp_{y_0,\bs{y}_{N-1}}^{(+)}(\bs{x}_N)\big\}_{(y_0, \bs{y}_{N-1}) \in \R^N}
\enq
form an orthogonal system in $L^2(\R^N,\dd^N x)$, \textit{viz}.
\beq
\Int{\R^N}{}  \Big( \vp_{\bs{y}_N^{\prime}}^{(-)}(\bs{x}_N) \Big)^*\cdot \vp_{\bs{y}_N}^{(-)}(\bs{x}_N) \cdot \dd^N x \; = \; 
\f{ 1 }{ \mu_{N}\big(\bs{y}_N\big) } \cdot \de_{\e{sym}}\big( \bs{y}_N - \bs{y}_N^{\prime}  \big)
\enq
and 
\beq
\Int{\R^N}{}  \Big( \vp_{y_0^{\prime},\bs{y}_{N-1}^{\prime}}^{(+)}(\bs{x}_N) \Big)^*\cdot \vp_{y_0,\bs{y}_{N-1}}^{(+)}(\bs{x}_N) \cdot \dd^N x \; = \; 
\f{ 1 }{ \mu_{N-1}\big(\bs{y}_{N-1}\big) } \cdot \de\big( y_0 - y_0^{\prime} \big) \cdot \de_{\e{sym}}\big( \bs{y}_{N-1} - \bs{y}_{N-1}^{\prime}  \big)
\enq
where 
\beq
\mu_{N}\big(\bs{y}_N\big) \; = \; \pl{a<b}{N}\Big\{ 4 \om_1 \om_2   \sinh\big[\f{\pi}{\om_1}(y_a-y_b) \big] \cdot \sinh\big[\f{\pi}{\om_2}(y_a-y_b) \big]  \Big\} 
\label{definition densite mesure Harish-Chandra}
\enq
and 
\beq 
\de_{\e{sym}}\big( \bs{y}_N - \bs{y}_N^{\prime}  \big) \; = \; \sul{\sg \in \mf{S}_N }{} \pl{a=1}{N} \de\big(y_a-y_{\sg(a)}^{\prime}\big) \;. 
\enq
\end{prop}

\Proof 
We start  by discussing the family $\vp_{\bs{y}_N}^{(-)}$ which is slightly easier to deal with. The orthogonality integral, which is to be understood in the sense of distributions, 
can be recast as 
\beq
\Int{\R^N}{}  \Big( \vp_{\bs{t}_N}^{(-)}(\bs{x}_N) \Big)^*\cdot \vp_{\bs{y}_N}^{(-)}(\bs{x}_N) \cdot \dd^N x \; = \; 
\Big( \La^{(1)}_{t_1;-} \Big)^{\dagger} \cdots \Big( \La^{(N)}_{t_N;-} \Big)^{\dagger} \cdot \La^{(N)}_{y_N;-}  \cdots  \La^{(1)}_{y_1;-} \;. 
\enq
One now moves the Hermitian conjugate operators through the string of $\La$ operators 
by means of the exchange relations \eqref{ecriture relation echange lambda lambda dagger}. However, in order to exchange the operator 
$ \Big( \La^{(N)}_{t;-} \Big)^{\dagger}$ with $\La^{(N)}_{y;-} $ one needs to ensure that $y \not= t$. This can be achieved 
by first, introducing the compactly supported smooth function $\varrho_{\eps}$ such that 
\beq
\varrho_{\eps}=1 \quad \e{on} \quad \intff{-\eps}{\eps} \quad  \e{and} \quad \varrho_{\eps}=0 \quad \e{on} \quad \R\setminus \intff{-2\eps}{2\eps} \;.  
\enq
Then, one considers 
\beq
g\big(\bs{y}_N,\bs{t}_N\big) \, = \, \sul{ \sg \in \mf{S}_N }{} \pl{a \not= b }{ N } \Big( 1-\varrho_{\eps}\big(y_a-t_{\sg(b)}\big) \Big)
\enq
which is uniformly away from $0$ on $\mc{D}_{\eps}\times \mc{D}_{\eps}$ with $\mc{D}_{\eps}\, = \, \Big\{ \bs{y}_N \in \R^N \, : \, \e{min}_{a \not= b} |y_a-y_b| \geq 7 \eps \Big\}$. 
Thus, one gets a partition of the unity on $\mc{D}_{\eps}\times \mc{D}_{\eps}$:
\beq
\vsg_{\sg}(\bs{y}_N,\bs{t}_N) \, = \, \f{ 1 }{g\big(\bs{y}_N,\bs{t}_N\big) }\pl{a \not= b }{ N } \Big( 1-\varrho_{\eps}\big(y_a-t_{\sg(b)}\big) \Big) \qquad \e{so} \, \e{that} \qquad 
 \sul{ \sg \in \mf{S}_N }{}\vsg_{\sg}(\bs{y}_N,\bs{t}_N)  \; = \; 1 \;. 
\enq
Thus, one gets on $\mc{D}_{\eps}\times \mc{D}_{\eps}$:
\beq
\Big( \La^{(1)}_{t_1;-} \Big)^{\dagger} \cdots \Big( \La^{(N)}_{t_N;-} \Big)^{\dagger} \cdot \La^{(N)}_{y_N;-}  \cdots  \La^{(1)}_{y_1;-} \; = \; 
 \sul{ \sg \in \mf{S}_N }{}\vsg_{\e{id}}\big( \bs{y}_N,\bs{t}_N^{(\sg)} \big)  \Big( \La^{(1)}_{t_{\sg(N)};-} \Big)^{\dagger} \cdots \Big( \La^{(N)}_{t_{\sg(1)};-} \Big)^{\dagger} \cdot \La^{(N)}_{y_N;-}  \cdots  \La^{(1)}_{y_1;-}
\enq
where $\bs{t}_N^{(\sg)} \, = \, \big(t_{\sg(1)}, \cdots , t_{\sg(N)} \big)$. 
One may already apply the exchange relation on the level of each summand  \eqref{ecriture relation echange lambda lambda dagger} since the presence of $\vsg_{\e{id}}\big( \bs{y}_N,\bs{t}_N^{(\sg)} \big)$
does ensure an appropriate distinctiveness of the variables, \textit{c.f.} \cite{KozUnitarityofSoVTransform} for a rigorous treatment of this feature. 
This leads, in the $\eps \tend 0^+$ limit, to 
\bem
\Int{\R^N}{}  \Big( \vp_{\bs{t}_N}^{(-)}(\bs{x}_N) \Big)^*\cdot \vp_{\bs{y}_N}^{(-)}(\bs{x}_N) \cdot \dd^N x \\ \; = \;  
\sul{\sg \in \mf{S}_N }{}  \pl{a<b}{N}\bigg\{  \f{1}{\om_1 \om_2 } \mc{A}\Big( (t_{\sg(a)})_+^{\star} -\big( (y_b)_+^{\star}\big)^*, (t_{\sg(a)})_- - \big((y_b)_-\big)^* \Big) \bigg\}
\pl{a=1}{N}\Big( \La_{t_{\sg(a)},-}^{(1)} \Big)^{\dagger} \cdot \La_{y_a,-}^{(1)} \;.
\end{multline}
At this stage it solely remains to observe that for real $t,y$ one has 
\beq
\mc{A}\Big( t_+^{\star} - ( y_+^{\star})^*, t_- - y_-^* \Big)  \; = \; \mc{A}\Big( \f{y-t}{2} - \i \f{ \Om }{2} \, , \, \f{t-y}{2} - \i \f{ \Om }{2}  \Big)
\; = \; \bigg\{  4 \sinh\big[\f{\pi}{\om_1}(y-t) \big] \cdot \sinh\big[\f{\pi}{\om_2}(y-t) \big]  \bigg\}^{-1}
\enq
and also that 
\beq
\Big( \La_{t,-}^{(1)} \Big)^{\dagger} \cdot  \La_{y,-}^{(1)} \; = \;  \de(y-t) \;. 
\enq
In the above handlings, we have omitted all technical details that need to be discussed so as to deal rigorously with the various steps of the calculations in the sense of distributions. 
The appropriate theory for that has been developed in \cite{KozUnitarityofSoVTransform} and we refer to this work for all the details leading to a rigorous treatment. 

In the case of $\vp_{y_0,\bs{y}_{N-1}}^{(+)}$, the same kind of operations leads to 
\bem
\Int{\R^N}{}  \Big( \vp_{t_0,\bs{t}_{N-1}}^{(+)}(\bs{x}_N) \Big)^*\cdot \vp_{y_0,\bs{y}_{N-1}}^{(+)}(\bs{x}_N) \cdot \dd^N x  \; = \;  \ex{(N-1)\pi \Om(t_0+y_0) } \cdot \Big( \om_1 \om_2 \Big)^{ \frac{3}{2}(N-1) } \\
 \times \sul{\sg \in \mf{S}_{N-1} }{}  \pl{a<b}{N-1} \bigg\{  \f{1}{\om_1 \om_2 }  \mc{A}\Big( (t_{\sg(a)})_+^{\star} -\big( (y_b)_+^{\star}\big)^*, (t_{\sg(a)})_- - \big((y_b)_-\big)^* \Big) \bigg\} \cdot 
\Big( \phi_{t_0} \, , \, \Big( \La_{t_{\sg(N-1)},-}^{(2)} \Big)^{\dagger} \La_{y_{N-1},-}^{(2)} \cdots \Big( \La_{t_{\sg(1)},-}^{(2)} \Big)^{\dagger} \La_{y_{1},-}^{(2)} \cdot \phi_{y_0}\Big)_{L^2(\R) } \; . 
\end{multline}
Then, by trivial induction based on Lemma \ref{Lemme Action  operateurs Lambda dag Lambda deux sur psi y}, one gets 
\bem
\Big( \phi_{t_0} \, , \, \Big( \La_{t_{N-1},-}^{(2)} \Big)^{\dagger} \La_{y_{N-1},-}^{(2)} \cdots \Big( \La_{t_{1},-}^{(2)} \Big)^{\dagger} \La_{y_{1},-}^{(2)} \cdot \phi_{y_0}\Big)_{L^2(\R) } \\
\; = \;  \f{ \ex{-(N-1)2\pi \Om y_0 } }{  \big( \om_1 \om_2 \big)^{ \frac{3}{2}(N-1) }  } \cdot \pl{a=1}{N-1} \de\big( t_a - y_a \big) \; \cdot\;  \big( \phi_{t_0}, \phi_{y_0} \big)_{L^2(\R) }  
\; = \; \f{ \ex{-(N-1)\pi \Om(y_0+t_0) } }{   \big( \om_1 \om_2 \big)^{ \frac{3}{2}(N-1) }   } \pl{a=0}{N-1} \de\big( t_a - y_a \big) 
\end{multline}
by virtue of the orthogonality of the system of functions $\phi_{y_0}$, \textit{cf}. Lemma \ref{Lemme fct propre B un site}.  \qed

\section{The Mellin-Barnes representation and completeness}
\label{Section Mellin Barnes}
In this section, we construct the Mellin-Barnes representation for a family $ \big\{ \psi^{(-)}_{ \bs{y}_N } \big\}_{\bs{y}_N \in \R^N}$, resp. $\big\{ \psi^{(+)}_{ y_0,\bs{y}_{N-1} } \big\}_{y_0 \in \R, \bs{y}_{N-1} \in  \R^{N-1} }$,
of generalised Eigenfunctions of the operator $\op{A}_{N}(\la)$, resp. $\op{B}_{N}(\la)$.  
We then use this integral representation to show that $\big\{ \psi^{(-)}_{ \bs{y}_N } \big\}_{\bs{y}_N \in \R^N}$, resp. $\big\{ \psi^{(+)}_{ y_0,\bs{y}_{N-1} } \big\}_{y_0 \in \R, \bs{y}_{N-1} \in  \R^{N-1} }$,
forms a complete system on $L^2( \R^{N}, \dd^Nx)$ in respect to an  $L^2_{\e{sym}}\Big( \R^{N}, \dd \mu_N(\bs{y}_N)  \Big) $, resp. 
 $L^2_{\cdot \times \e{sym}}\Big( \R\times \R^{N-1}, \dd y_0 \otimes \dd \mu_{N-1}(\bs{y}_{N-1})  \Big) $, integration of the $y$-variables.  
Here,  the subscript $\e{sym}$ refers to functions invariant under the natural action of the symmetric group $\mf{S}_N$, resp. $\mf{S}_{N-1}$.

\subsection{The Mellin-Barnes integral representation} 

Given an $k$-dimensional vector $\bs{v}_k$, we agree upon the shorthand notation $\ov{\bs{v}}_k=\sum_{a=1}^{k} v_a$. 
Also, we shall denote by $\bs{e}_k$ the canonical unit base vector with $1$ in the $k^{\e{th}}$ position and $0$ everywhere else. 
Depending on the context, these vectors will belong to $\R^N$ or $\R^{N-1}$. 

\begin{prop}
\label{Proposition rep MellinBarnes}

Let $\bs{y}_N\in \R^N$. Define the functions $\psi_{\bs{y}_N}^{(-)}(\bs{x}_N ) $   inductively by 
\beq
\psi_{\bs{y}_N}^{(-)}(\bs{x}_N ) \; = \; \Int{ \mc{C}^{N-1}_{\bs{y}_N}  }{} \psi_{\bs{w}_{N-1}}^{(-)}(\bs{x}_{N-1} ) \psi_{\ov{\bs{y}}_N-\ov{\bs{w}}_{N-1}}^{(-)}(x_N ) \Phi\big(\bs{w}_{N-1}\mid \bs{y}_N \big) \cdot \f{ \dd^{N-1} w }{ (N-1)!} \; \;
\label{definition recurrence MB}
\enq
and
\beq
\psi^{(-)}_y(x)= \ex{2\i\pi x y } \;. 
\enq
Here
\beq
\Phi\big(\bs{w}_{N-1}\mid \bs{y}_N \big) \; = \; \Big( \f{1}{\om_1\om_2} \Big)^{N-1}
\pl{a=1}{N}\f{ \varpi(y_a-\kappa) }{ \varpi^{N-1}(y_a+\kappa) } \cdot \f{  \pl{a=1}{N-1}\varpi^{N-1}(w_a+\kappa)  }{ \varpi\big( \ov{\bs{y}}_N-\ov{\bs{w}}_{N-1}-\kappa\big)  }
\cdot \f{ \pl{a=1}{N-1}\pl{b=1}{N} \varpi\big(  y_b-w_a -\i\tfrac{\Om }{2} \big)  }{ \pl{a\not=b}{N-1} \varpi\big(  w_b-w_a -\i\tfrac{\Om }{2} \big)   }  
\enq
and $\mc{C}_{\bs{y}_N}$ is a curve corresponding to small deformation of the real axis such that all the simple poles for each $w$-variable at 
\beq
 w \, = \, y_b+ \i \ell \om_1  + \i k \om_2\; , \quad with \; \; (\ell,k) \in \mathbb{N}^2 \;\; and \;\;   b\in \intn{1}{N} 
\enq
are located above it. 

Then, the functions $\psi_{\bs{y}_N}^{(-)}$ are well-defined, self-dual under $\om_1\leftrightarrow \om_2$ and satisfy
\beq
\op{A}_N(\la) \cdot \psi_{\bs{y}_N}^{(-)}(\bs{x}_N ) \; = \; a\big( \la \mid \bs{y}_N \big)  \cdot \psi_{\bs{y}_N}^{(-)}(\bs{x}_N ) \qquad with \qquad 
a\big( \la \mid \bs{y}_N \big)  \; = \; \pl{b=1}{N} \bigg\{-2\i  \sinh\Big[ \f{\pi}{\om_1}(\la-y_b)\Big]  \bigg\}
\label{ecriture action A}
\enq
as well as
\beqa
\op{B}_N(\la) \cdot \psi_{\bs{y}_N}^{(-)}( \bs{x}_N ) & = & \sul{k=1}{N} \Big\{ b(y_k) \Big\}^N \pl{ \substack{ p=1  \\ \not= k  }}{ N } \f{  \sinh\Big[ \f{\pi}{\om_1}(\la-y_p)\Big] }{ \sinh\Big[ \f{\pi}{\om_1}(y_k-y_p)\Big] }
\cdot \psi_{\bs{y}_N+\i\om_2 \bs{e}_k }^{(-)}( \bs{x}_N ) \label{ecriture action B}\\
\op{C}_N(\la) \cdot \psi_{\bs{y}_N}^{(-)}( \bs{x}_N ) & = & \sul{k=1}{N} \Big\{ c(y_k) \Big\}^N \pl{ \substack{ p=1  \\ \not= k  }}{ N } \f{  \sinh\Big[ \f{\pi}{\om_1}(\la-y_p)\Big] }{ \sinh\Big[ \f{\pi}{\om_1}(y_k-y_p)\Big] }
\cdot \psi_{\bs{y}_N-\i\om_2 \bs{e}_k }^{(-)}( \bs{x}_N )  \label{ecriture action C}
\eeqa
where
\beq
b(\la) \; = \; -2\i  \sinh\Big[ \tfrac{\pi}{\om_1}\Big(\la+\kappa + \i \tfrac{\Om}{2}\Big)\Big] \qquad and \qquad 
c(\la) \; = \; 2\i  \sinh\Big[ \tfrac{\pi}{\om_1}\Big(\la-\kappa - \i \tfrac{\Om}{2}\Big)\Big]  \;. 
\enq

Further, given $(y_0,\bs{y}_{N-1})\in \R^N$,   the functions $\psi_{y_0,\bs{y}_{N-1}}^{(+)}(\bs{x}_N ) $   defined as 
\beq
\psi_{y_0,\bs{y}_{N-1}}^{(+)}(\bs{x}_N ) \; = \; \Int{ \mc{C}^{N-1}_{\bs{y}_N} \times \mc{C}_{-\bs{y}_N}  }{} 
\psi_{\bs{w}_{N-1}}^{(-)}(\bs{x}_{N-1} ) \psi_{w_{N}}^{(-)}(x_N ) \Psi\big(\bs{w}_{N}\mid y_0,\bs{y}_{N-1}  \big) \cdot \f{ \dd^{N-1} w \otimes \dd w_N}{ (N-1)!} \; \;
\enq
in which 
\bem
 \Psi\big(\bs{w}_{N}\mid y_0,\bs{y}_{N-1}  \big) \, = \, \Big( \f{1}{ \om_1\om_2 } \Big)^{N-1} \cdot   \f{  \ex{ 2\i \pi y_0 \ov{\bs{w}}_N } \,   \varpi(w_N+\kappa) } {  \pl{a=1}{N-1}\varpi^{N-1}(w_a+\kappa)  }
\cdot \pl{a=1}{N-1}\f{ \varpi(y_a-\kappa) }{ \varpi^{N-1}(y_a+\kappa) }  \\ 
\times \pl{a=1}{N-1} \f{ \varpi \big( w_a + w_N + \i\tfrac{\Om }{2} \big)  }{   \varpi \big( y_a + w_N + \i\tfrac{\Om }{2} \big)   }
\cdot \f{ \pl{a=1}{N-1}\pl{b=1}{N-1} \varpi\big(  y_b-w_a -\i\tfrac{\Om }{2} \big)  }{ \pl{a\not=b}{N-1} \varpi\big(  w_b-w_a -\i\tfrac{\Om }{2} \big)   }   \;, 
\end{multline}
satisfy to 
\beq
\op{B}_N(\la) \cdot \psi_{y_0,\bs{y}_{N-1}}^{(+)}(\bs{x}_N ) \; = \; \ex{2\pi \om_2 y_0} \,  a\big( \la \mid \bs{y}_{N-1} \big)  \cdot \psi_{y_0,\bs{y}_{N-1}}^{(+)}(\bs{x}_N )  
\enq
and
\bem
\op{A}_N(\la) \cdot \psi_{y_0,\bs{y}_{N-1}}^{(+)}(\bs{x}_N ) \; = \;  -\i a\big( \la \mid \bs{y}_{N-1} \big)  \cdot
\bigg\{  \ex{ \frac{\pi}{\om_1}(\la + \ov{\bs{y}}_{N-1})  }   \psi_{y_0+\frac{\i}{2\om_1}    ,\bs{y}_{N-1} }^{(+)}( \bs{x}_N )  
\; - \;  \ex{ -\frac{\pi}{\om_1}(\la + \ov{\bs{y}}_{N-1})  }   \psi_{y_0-\frac{\i}{2\om_1}   ,\bs{y}_{N-1} }^{(+)}( \bs{x}_N )  \bigg\} \\
\; + \; \sul{k=1}{N-1} \Big\{ b(y_k) \Big\}^N \pl{ \substack{ p=1  \\ \not= k  }}{ N-1 } \f{  \sinh\Big[ \f{\pi}{\om_1}(\la-y_p)\Big] }{ \sinh\Big[ \f{\pi}{\om_1}(y_k-y_p)\Big] }
\cdot \psi_{y_0,\bs{y}_{N-1}+\i\om_2 \bs{e}_k }^{(+)}( \bs{x}_N )  
\end{multline}

\end{prop}

\Proof  The proof goes by induction on $N$. One builds $\psi_{\bs{y}_N}^{(-)}(\bs{x}_N ) $ as in \eqref{definition recurrence MB}
and assumes that the properties \eqref{ecriture action A}, \eqref{ecriture action B} and \eqref{ecriture action C} hold up to $N-1$. 
Then decomposing the monodromy matrix $\op{T}_N(\la)=\op{T}_{N-1}(\la)L_N^{(\kappa)}(\la)$ allows one to compute the action of the 
operator $\op{A}_N,\op{B}_N$ and $\op{C}_N$ on $\psi_{\bs{y}_N}^{(-)}(\bs{x}_N ) $. The system of first order difference equations 
satisfied by $\Phi\big(\bs{w}_{N-1}\mid \bs{y}_N \big)$ then allows one to conclude that, indeed, the claimed form of the actions hold. 
See \cite{SklyaninSoVGeneralOverviewAndConstrRecVectPofB} or, in a closer setting to the present one, \cite{KharchevLebedevMellinBarnesIntRepForWhittakerGLN,KharchevLebedevSemenovTianShanskyRTCBEigenfunctions}.
Self-duality of $\psi_{\bs{y}_N}^{(-)}(\bs{x}_N ) $ is manifest. 
The convergence of the integral can be established along the lines discussed in \cite{GerasimovKharchevLebedevRepThandQISM} for the analogous representation arising in the 
case of the quantum Toda chain. \qed

\subsection{An auxiliary integral}

Prior to proving completeness, as already discussed in the introduction, we need to establish an auxiliary integral representation for the symmetric $\de$-function in many variables. This identity 
appears as a central tool allowing to prove, in a simple and direct manner, the completeness of a system of functions defined through a Mellin-Barnes like representation. 
The technique for establishing this identity that we develop below thus emerges as an important tool for studying quantum integrable models with non-compact local spaces and solvable by the 
 quantum separation of variables method. 

\vspace{3mm} 

Given $\bs{v}_{N-1}, \bs{w}_{N-1} \in \R^{N-1}$ and $\bs{\veps}_N \in \R^N$ such that $\veps_a>0$ for all $a$ and is small enough. 
Define $\msc{C}$ to be a curve obtained from $\R$ by a small deformation of $\R$ such that, for any $a=1,\dots, N-1$ and $b=1,\dots, N$: 
\begin{itemize}
\item the lattice $v_a  - \i \veps_b + \i \om_1 \mathbb{N} + \i \om_2 \mathbb{N} $ lies entirely above $\msc{C}$; 
\item the lattice $w_a + \i \veps_b  - \i \om_1 \mathbb{N} - \i \om_2 \mathbb{N} $ lies entirely below of $\msc{C}$. 
\end{itemize}
Such a curve always exists provided that $||\bs{\veps}_N||$ is small enough and the $\veps_a$'s are generic.

A very useful object occurring in the proof of completeness is given by the below family of integrals parameterised by $\bs{\veps}_N$ as above: 
\beq
\msc{J}^{(N;m)}_{\bs{\veps}_N}\big( x\mid \bs{w}_{N-1}, \bs{v}_{N-1} \big) \; = \; 
\Int{ \msc{C}^m }{} \dd^m y \Int{ \R^{N-m} }{} \pl{a=m+1}{N} \dd y_a \, \cdot \; \mc{I}_{N}^{(m)}\Big(\bs{\veps}_N\mid x,\bs{y}_N ; \bs{w}_{N-1}, \bs{v}_{N-1}  \Big) \;, 
\label{definition integrale mathscr J N m-}
\enq
with 
\beq
\mc{I}_{N}^{(m)}\Big(\bs{\veps}_N\mid x,\bs{y}_N ; \bs{w}_{N-1}, \bs{v}_{N-1}  \Big) \, = \,  \f{ \ex{2\i\pi \ov{\bs{y}}_N x} }{ N! \, (\om_1 \om_2)^{N-1} }
 \f{ \pl{a=1}{N-1}\pl{b=1}{N} \varpi\Big(  y_b-w_a -\i\frac{\Om }{2} +\i \zeta^{(m)}_{b}\, , \,  v_a -y_b - \i\frac{\Om }{2}    +\i \zeta^{(m)}_{b} \Big)  }
{ \pl{a<b}{N} \varpi\Big(  y_{ba} -\i\frac{\Om }{2} +\i \zeta^{(m)}_{b} + \i \zeta^{(m)}_{a}, y_{ab} - \i\frac{\Om }{2} - \i \zeta^{(m)}_{b} + \i \zeta^{(m)}_{a} \Big)  } \;. 
\label{ecriture expression explicite integrande modele}
\enq
The sequence $\zeta_a^{(m)}$ is defined as 
\beq
\zeta_a^{(m)} \, = \, - \veps_a \;\; , \;\; a=1,\dots, m \quad \e{and} \quad 
\zeta_a^{(m)} \, = \,  \big(\wt{  \bs{\veps}_N }\big)_a^{(m)} \, = \, \veps_{N+m+1-a} \; \; , \;  \; a=m+1,\dots, N \;. 
\label{ecriture regulateur zetam}
\enq
Also, we remind that $y_{ab}=y_a-y_b$ and we used hypergeometric-like notation for products of $\varpi$ factors , \textit{c.f.} \eqref{definition notation produit hypergeometrique}.

\begin{prop}
\label{Proposition integrale multiple pour delta sym}
For any $N\geq 1$ and $m\in \intn{0}{N}$, it holds, in the sense of distributions, 

\bem
\lim_{\bs{\veps}_N\tend 0^+} \Big\{ \msc{J}^{(N;m)}_{\bs{\veps}_N}\big( x\mid \bs{w}_{N-1}, \bs{v}_{N-1} \big)   \Big\}  \; = \; 
\pl{a\not=b }{N} \varpi\Big( w_a-w_b - \i \frac{\Om}{2} \Big) \cdot  \de(x)   \\
				  \times \de_{\e{sym}}\Big(\bs{w}_{N-1}-\bs{v}_{N-1} \Big) \cdot 
							    \left\{ \ba{cc} 1   & m=0 \vspace{2mm} \\ 
									  \frac{1}{N}  & m=1 \vspace{2mm} \\ 
									0     & m \in \intn{2}{N} 	  
\ea \right.   \;. 
\end{multline}

\end{prop}

\Proof 

The result is established by  a triangular induction. 

\vspace{2mm}

{\bf $ \bullet$ Initialisation}

\vspace{2mm}

It is obvious that 
\beq
\lim_{\veps_1\tend 0^+} \Big\{ \msc{J}^{(1;a)}_{ \veps_1 }\big( x\mid \emptyset, \emptyset \big)   \Big\} \; = \; \de(x) \;, \quad \e{for} \;\; a \in \{0,1\}. 
\enq
Next, assume that the claim holds up to $N-1$ and for any $m\in \intn{0}{N-1}$.

\vspace{2mm}

{\bf $\bullet $ Initialisation step at $N$: the integral $\msc{J}^{(N;N)}_{\bs{\veps}_N}$ }

\vspace{2mm}

First of all, observe that the appropriate choice of signs of the regulators $\zeta_a^{(N)}$ in \eqref{ecriture expression explicite integrande modele}-\eqref{ecriture regulateur zetam}
makes the integrand 
\beq
\mc{I}_{N}^{(N)}\Big(\bs{\veps}_N\mid x,\bs{y}_N ; \bs{w}_{N-1}, \bs{v}_{N-1}  \Big) 
\enq
algebraically decreasing at $\infty$ along the lines  $\big( \R \pm  \i r \big)^N$, with $r>0$. This property is enough so as to allow for computing the integral by taking residues in the upper or lower complex plane -relatively to any variables-
and conclude that, in the sense of distributions, the contours at $\infty$ do not contribute to the integral. Thus, in the sense of distributions,  the integral can be taken by computing the residues of all  
the poles lying above $\msc{C}$ -if $x\geq 0$-  or below $\msc{C}$ -if $x \leq 0$-. Here, we only discuss the case when $x \geq 0$. In respect to each variable $y_a$, the integrand has simple poles at 
\beq
y_a \, = \,  v_b - \i \eps_a +\i \ell_a \om_1 + \i k_a \om_2  \qquad \e{with} \qquad \ell_a, k_a \in \mathbb{N} \quad \e{and} \quad b \in \intn{1}{N} \;.   
\enq
Thus, one gets that, in the sense of distributions,  
\bem
 \msc{J}^{(N;N)}_{\bs{\veps}_N}\big( x\mid \bs{w}_{N-1}, \bs{v}_{N-1} \big) \\
 \; = \; \sul{ \a \in \mc{M}_N }{} \sul{ \substack{ \bs{\ell}_N, \bs{k}_{N} \\ \in \mathbb{N}^N } }{ }
\e{Res}\Big(\mc{I}_{N}^{(N)}\big(\bs{\veps}_N\mid x,\bs{y}_N ; \bs{w}_{N-1}, \bs{v}_{N-1}  \big)\cdot \dd^N y , \bs{y}_N=\bs{v}^{(\a)}+\i  \bs{\ell}_N \om_1 + \i  \bs{k}_{N} \om_2 -\i \bs{\veps}_N  \Big)  \; . 
\label{ecriture dvpment integrale NN sur les résidus}
\end{multline}
The sums in \eqref{ecriture dvpment integrale NN sur les résidus} run through all maps $\a\in \mc{M}_N$, where $\mc{M}_N=\Big\{ \a\; : \; \intn{1}{N} \tend \intn{1}{N-1}\Big\}$.
Furthermore, given a map $\a \; : \; \intn{1}{N} \tend \intn{1}{N-1}$, we denote 
\beq
\bs{v}^{(\a)} \in \R^N \quad  \e{the}\, \e{vector} \, \e{given} \, \e{by} \;   \big(\bs{v}^{(\a)}\big)_a = v_{\a(a)}  \quad \e{for} \quad  a=1,\dots,N \; .
\enq

Once that the residues are computed, one may send $\veps_a \tend 0^+$ for any $a=1,\dots, N$. The expression for the residues of the dilogarithm \eqref{ecriture formule residu general du dilogarithme} 
leads to  
\beq
\e{Res}\Big(\mc{I}_{N}^{(N)}\big(\bs{0}\mid x,\bs{y}_N ; \bs{w}_{N-1}, \bs{v}_{N-1}  \big)\cdot \dd^N y \,  , \,  \bs{y}_N \, = \, \bs{v}^{(\a)}+\i  \bs{\ell}_N \om_1 + \i  \bs{k}_{N} \om_2   \Big)   
\, = \, P^{(\a)}_{\bs{\ell}_n;\bs{k}_N} \cdot  G^{(\a)}_{\bs{\ell}_n;\bs{k}_N}  \cdot \ga^{(\a)}_{\bs{\ell}_n;\bs{k}_N}  \;, 
\enq
in which 
\beq
 \ga^{(\a)}_{\bs{\ell}_n;\bs{k}_N} \; = \; \f{ 2^{N(N-1)}   }{ N! \cdot \big( \om_1 \om_2 \big)^{ \frac{N}{2} - 1 } } 
 \pl{a=1}{N} \Bigg\{ (-1)^{k_a\ell_a} \, \big(  2\i \big)^{\ell_a + k_a}  \cdot 
\pl{p=1}{k_a}   \sinh\Big[ \i p \pi \tfrac{ \om_2 }{\om_1} \Big]   \cdot \pl{ p = 1 }{ \ell_a }  \sinh\Big[ \i p \pi \tfrac{\om_1 }{\om_2} \Big] \Bigg\}^{-1} \;, 
\enq
\beq
P^{(\a)}_{\bs{\ell}_n;\bs{k}_N} \, = \,  \ex{ 2\i\pi x \big( \ov{\bs{v}^{(\a)}} + \i \ov{\bs{\ell}}_N \om_1 + \i \ov{\bs{k}}_N \om_2    \big) }
\cdot \pl{a=1}{N-1}\pl{b=1}{N} \varpi\Big(  v_{\a(b)}-w_a -\i\frac{\Om }{2} + \i \ell_b \om_1 + \i k_b \om_2 \Big) \;. 
\enq
Note that above, we made use of the notation \eqref{definition moyenne coord vecteur}. Finally, one has 
\bem
  G^{(\a)}_{\bs{\ell}_n;\bs{k}_N} \; = \;  \pl{b=1}{N} \pl{ \substack{ a=1  \\ a \not= \a(b)}  }{N-1} \varpi\Big(  v_a -  v_{\a(b)} - \i\frac{\Om }{2}   - \i \ell_b \om_1 - \i k_b \om_2  \Big)   \\
\times \pl{a<b}{N} \bigg\{ \sinh\Big[ \f{\pi}{\om_1} \big( v_{\a(a)\a(b)} + \i \ell_{ab} \om_1 + \i k_{ab} \om_2 \big) \Big] \cdot \sinh\Big[ \f{\pi}{\om_2} \big( v_{\a(a)\a(b)} + \i \ell_{ab} \om_1 + \i k_{ab} \om_2 \big) \Big] \bigg\} \;. 
\end{multline}

Let $\a \in \mc{M}_N $ be given. Then, there exists $a\not= b$ such that $\big(\bs{w}^{(\a)}\big)_a=\big(\bs{w}^{(\a)}\big)_b$. 
A longish but straightforward calculation then shows that one has the factorisation
\beq
  G^{(\a)}_{\bs{\ell}_n;\bs{k}_N}  \; = \;  (-1)^{ (\ell_a k_a +\ell_b k_b)(N-2) } \cdot  \sinh\Big[ \i \pi \f{\om_2}{\om_1} k_{ab} \Big] \cdot  \sinh\Big[ \i \pi \f{\om_1}{\om_2} \ell_{ab} \Big] 
  \cdot  \mc{G}^{(\a;a,b)}_{\bs{\ell}_n;\bs{k}_N}\;, 
\enq
in which $\mc{G}^{(\a;a,b)}_{\bs{\ell}_n;\bs{k}_N} $ is symmetric under the permutation $(\ell_a,\ell_b) \hookrightarrow (\ell_b,\ell_a)$ or, independently, the permutation  $(k_a,k_b) \hookrightarrow (k_b,k_a)$. 
It takes the explicit form 
\bem
\mc{G}^{(\a;a,b)}_{\bs{\ell}_n;\bs{k}_N} \, = \,   (-1)^{ ( \ov{\bs{\ell}}_N + \ov{\bs{k}}_N ) (N-3) }
 \pl{ \substack{c<d \\ \not= a, b } }{N} \bigg\{ \sinh\Big[ \f{\pi}{\om_1} \big( v_{\a(c)\a(d)} + \i k_{cd} \om_2 \big) \Big] \cdot \sinh\Big[ \f{\pi}{\om_2} \big( v_{\a(c)\a(d)} + \i \ell_{cd} \om_1 \big) \Big] \bigg\}  \\
\times \pl{ \substack{d=1 \\ \not= a,b} }{ N } \Bigg\{ \pl{ k \in \{k_a,k_b\} }{} 
   \bigg\{  \f{\i}{2} \sinh\Big[ \f{\pi}{\om_1} \big( v_{\a(a)\a(d)} + \i  \om_2(k-k_d) \big) \Big]   \bigg\}
\cdot\pl{ \ell \in \{\ell_a,\ell_b\} }{}   \bigg\{  \f{\i}{2}  \sinh\Big[ \f{\pi}{\om_2} \big( v_{\a(a)\a(d)} + \i  \om_1 (\ell-\ell_d) \big) \Big] \bigg\}  \Bigg\} \\
\times \pl{ \substack{ d=1  \\ \not= \a(a) } }{N-1 } \Bigg\{  \varpi^{2}\Big(v_{d\a(a)}-\i \tfrac{\Om}{2}  \Big) 
\bigg\{ \pl{ k \in \{k_a,k_b\} }{} \pl{p=1}{k} \sinh\Big[ \f{\pi}{\om_1} \big( v_{d\a(a)} - \i p \om_2 \big) \Big] 
\cdot \pl{ \ell \in \{\ell_a,\ell_b\} }{}   \pl{p=1}{\ell}  \sinh\Big[ \f{\pi}{\om_2} \big( v_{d\a(a)} - \i p \om_1  \big) \Big] \bigg\}^{-1}
\Bigg\}   \\
\times  \pl{ \substack{ c=1 \\ c \not= a,b} }{N} \pl{ \substack{ d=1  \\ d \not= \a(c)}  }{N-1} \varpi\Big(  v_{d\a(c)} - \i\tfrac{\Om }{2}   - \i \ell_c \om_1 - \i k_c \om_2  \Big)  \; . 
\end{multline}
Likewise, one gets that 
\beq
  P^{(\a)}_{\bs{\ell}_n;\bs{k}_N}  \; = \;(-1)^{ (\ell_a k_a +\ell_b k_b)(N-1) } \cdot \mc{P}^{(\a;a,b)}_{\bs{\ell}_n;\bs{k}_N}  \;, 
\enq
where 
\bem
\mc{P}^{(\a;a,b)}_{\bs{\ell}_n;\bs{k}_N} \, = \, \ex{ 2\i\pi x \big( \ov{\bs{v}^{(\a)}} + \i \ov{\bs{\ell}}_N \om_1 + \i \ov{\bs{k}}_N \om_2    \big) } \cdot 
\Big( -2\i \Big)^{(k_a+k_b+\ell_a+\ell_b)(N-1)} \cdot 
\pl{ \substack{ c=1 \\ c \not= a,b} }{N} \pl{ d=1  }{N-1} \varpi\Big(   v_{\a(c)} - w_d - \i\tfrac{\Om }{2}   - \i \ell_c \om_1 - \i k_c \om_2  \Big) \\
\times \pl{ c=1   }{N-1 } \Bigg\{  \varpi^{2}\Big(v_{\a(a)}-w_c-\i \tfrac{\Om}{2}  \Big) 
\bigg\{ \pl{ k \in \{k_a,k_b\} }{} \pl{p=0}{k-1}  \sinh\Big[ \f{\pi}{\om_1} \big( v_{\a(a)}-w_c + \i p \om_2 \big) \Big]   
\cdot \pl{ \ell \in \{\ell_a,\ell_b\} }{}   \pl{p=0}{\ell-1}  \sinh\Big[ \f{\pi}{\om_2} \big( v_{\a(a)}-w_c + \i p \om_1  \big) \Big]   \Bigg\} \;. 
\end{multline}
Finally, one also gets  $\ga^{(\a)}_{\bs{\ell}_n;\bs{k}_N}  \; = \;  (-1)^{ (\ell_a k_a +\ell_b k_b)} \cdot \Ga^{(\a;a,b)}_{\bs{\ell}_n;\bs{k}_N}$, where 
\beq
\Ga^{(\a;a,b)}_{\bs{\ell}_n;\bs{k}_N}\; = \; \f{ 2^{N(N-1)}   }{ N! \cdot \big( \om_1 \om_2 \big)^{ \frac{N}{2} - 1 } } 
 \pl{c=1}{N} \Bigg\{  \big(  2\i \big)^{\ell_c + k_c}  \cdot 
\pl{p=1}{k_c}   \sinh\Big[ \i p \pi \tfrac{ \om_2 }{\om_1} \Big]   \cdot \pl{p=1}{\ell_c}  \sinh\Big[ \i p \pi \tfrac{\om_1 }{\om_2} \Big] \Bigg\}^{-1}
\cdot  \pl{ \substack{ c=1 \\ \not= a,b } }{N} \Big\{ (-1)^{k_c\ell_c} \Big\}  \; .  
\enq
Note that $\Ga^{(\a;a,b)}_{\bs{\ell}_n;\bs{k}_N}$ and $\mc{P}^{(\a;a,b)}_{\bs{\ell}_n;\bs{k}_N} $ enjoy the same symmetry properties as $\mc{G}^{(\a;a,b)}_{\bs{\ell}_n;\bs{k}_N}$.
Thus, in the end, one has that the residue expansion of the integral reorganises as 
\bem
 \underset{ \bs{\veps}_N\tend \bs{0}^+ }{ \e{Lim} } \Big\{ \msc{J}^{(N;N)}_{\bs{\veps}_N}\big( x\mid \bs{w}_{N-1}, \bs{v}_{N-1} \big) \Big\} \; = \; \sul{ \a \in \mc{M}_N }{} 
\sul{ \substack{ \ell_c, k_{c}  \in \mathbb{N} \\ c \not= a_{\a}, b_{\a} } }{ } \\
\times \sul{  \substack{ \ell_{ a_{\a}} ,  \ell_{ b_{\a}} \\ \in \mathbb{N} }  }{ }
\sul{  \substack{ k_{ a_{\a}} ,  k_{ b_{\a}} \\ \in \mathbb{N} }  }{ }
\Big( \mc{G}\cdot \mc{P} \cdot \Ga\Big)^{(\a;a_{\a},b_{\a})}_{\bs{\ell}_n;\bs{k}_N}  \cdot  \sinh\Big[ \i \pi \f{\om_2}{\om_1} k_{a_{\a}b_{\a}} \Big] \cdot  \sinh\Big[ \i \pi \f{\om_1}{\om_2} \ell_{a_{\a}b_{\a}} \Big]  \;. 
\label{ecriture dvpment integrale NN etape initialisation}
\end{multline}
Above, for a given $\a$, we denote by $ a_{\a}, b_{\a}$ any two distinct integers such that $ \a(a_{\a})=\a(b_{\a})$. Due to the symmetry properties of $\mc{G}, \mc{P}$ and $\Ga$,
 the summand in \eqref{ecriture dvpment integrale NN etape initialisation} is antisymmetric under $\ell_{a_{\a}} \leftrightarrow \ell_{b_{\a}}$ as well as under
$k_{a_{\a}} \leftrightarrow k_{b_{\a}}$. This entails that, in the sense of distributions,  the last line in \eqref{ecriture dvpment integrale NN etape initialisation} vanishes \textit{viz}. 
\beq
 \underset{ \bs{\veps}_N\tend \bs{0}^+ }{ \e{Lim} } \Big\{ \msc{J}^{(N;N)}_{\bs{\veps}_N}\big( x\mid \bs{w}_{N-1}, \bs{v}_{N-1} \big) \Big\} \; = \; 0 \;. 
\label{ecriture vanishing integrale NN}
\enq

\vspace{2mm}

{\bf $\bullet $ The fundamental induction equation}

\vspace{2mm}

Let $m\in \intn{0}{N-1}$ and assume the $\veps_a$'s to be small enough. Then, in $ \msc{J}^{(N;m)}_{\bs{\veps}_N}$ one deforms the $y_N$ integration to $\R-\i\a$ with $\a>0$ small enough and such that $\a > \e{max}\big\{ |\veps_a| \big\}$.
One crosses the poles at 
\beq
y_N=w_a-\i \zeta_N^{(m)}=w_a-\i\veps_{m+1} \;\; , \quad  \e{with} \qquad a=1,\dots, N-1 \; .
\enq
This yields
\bem
\msc{J}^{(N;m)}_{\bs{\veps}_N}\big( x\mid \bs{w}_{N-1}, \bs{v}_{N-1} \big) \; = \;
\Int{ \msc{C}^m }{} \dd^m y \Int{ \R^{N-m-1} }{} \pl{a=m+1}{N-1} \dd y_a \Int{\R-\i\a }{} \dd y_N \, \cdot \; \mc{I}_{N}^{(m)}\big( \bs{\veps}_N\mid x,\bs{y}_N ; \bs{w}_{N-1}, \bs{v}_{N-1}  \big)  \\
 \; + \;  \sul{k=1}{N-1} \Int{ \msc{C}^m }{} \dd^m y \Int{ \R^{N-m-1} }{} \pl{a=m+1}{N-1} \dd y_a  
\cdot \ex{2\i\pi (w_k-\i\veps_{m+1})x}   
 \f{ \pl{  \substack{ a=1 \\ \not= k }   }{N-1} \varpi\Big( w_{ka} -\i\frac{\Om }{2}\Big) \pl{a=1}{N-1} \varpi\Big(  v_a -w_k - \i\frac{\Om }{2}  +2 \i \veps_{m+1} \Big)  }
{ \pl{a=1}{N-1} \varpi\Big(  w_k-y_a -\i\frac{\Om }{2} + \i \zeta^{(m)}_{a}, y_a - w_k - \i\frac{\Om }{2}  + \i \zeta^{(m)}_{a} \Big)  }  \\
\times  \f{ 1 }{ N \sqrt{\om_1\om_2} }  \cdot \mc{I}_{N-1}^{(m)}\Big(\bs{\veps}_{N;[m]}\mid x,\bs{y}_{N-1} ; \bs{w}_{N-1}, \bs{v}_{N-1}  \Big)  \;. 
\label{ecriture J N m apres 1ere def ctr}
\end{multline}
Above, the vector $\bs{\veps}_{N;[m]} \in \R^{N-1}$ is defined by
\beq
\Big(\bs{\veps}_{N;[m]} \Big)_a \; = \; \veps_a \;\; , \;\; a=1,\dots, m \quad \e{and} \quad 
\Big(\bs{\veps}_{N;[m]} \Big)_a \, = \, \veps_{a+1} \; \; , \;  \; a=m+1,\dots, N-1 \;. 
\enq
The fact that the last line of \eqref{ecriture J N m apres 1ere def ctr} does indeed take the form as written  follows from the below identification  
\beq
\left\{ \ba{ccc}  \check{\zeta}_a^{(m)} & \equiv & - \Big(\bs{\veps}_{N;[m]} \Big)_a \; = \; - \veps_a \;\; , \;\; a=1,\dots, m  \vspace{2mm} \\ 
\check{\zeta}_a^{(m)} & \equiv & \Big( \wt{\bs{\veps}_{N;[m]}} \Big)_a^{(m)}   \, = \,   \veps_{N+m+1-a} \; \; , \;  \; a=m+1,\dots, N-1 \;. \ea \right. 
\label{definition des check zeta N m}
\enq
in which $\check{\zeta}_a^{(m)} $ is the sequence built up from $\bs{\veps}_{N;[m]} $. Hence, \eqref{definition des check zeta N m} ensures that $\check{\zeta}_a^{(m)} = \zeta_a^{(m)}$ for $a=1,\dots, N-1$.

Define $\op{P}_a$ as the operator $\Big( \op{P}_a\cdot \veps_N \Big)_b \, = \, (-1)^{\de_{ab}} \veps_b $. 
Let $\ov{\zeta}_a^{(m)}$ be the sequence built up from the vector $\op{P}_{m+1}\cdot \bs{\veps}_N$, namely 
\beq
\ov{\zeta}_a^{(m)} \, = \, - \veps_a \;\; , \;\; a=1,\dots, m \; \; ; \; \; \ov{\zeta}_N^{(m)} \, = \, - \veps_{m+1}  \quad \e{and} \quad 
\ov{\zeta}_a^{(m)} \, = \,   \veps_{N+m+1-a} \; \; , \;  \; a=m+1,\dots, N -1 \;. 
\enq
Thus, one has 
\beq
\ov{\zeta}_a^{(m)} \, = \,\zeta_a^{(m+1)}  \;\; , \;\; a=1,\dots, m \; \; ; \; \; \ov{\zeta}_N^{(m)} \, = \, \zeta_{m+1}^{(m+1)} \quad \e{and} \quad 
\ov{\zeta}_a^{(m)} \, = \,   \zeta_{a+1}^{(m+1)}  \; \; , \;  \; a=m+1,\dots, N -1 \;. 
\enq
Upon defining 
\beq
\ov{y}_a \, = \, y_a \;\; , \;\; a=1,\dots, m \; \; ; \; \; \ov{y}_{m+1} \, = \,y_N \quad \e{and} \quad 
\ov{ y }_{a+1} \, = \,   y_a  \; \; , \;  \; a=m+1,\dots, N -1 \;, 
\enq
one gets the below product identities 
\beq
 \pl{b=1}{N} \varpi\Big(  y_b-w -\i\tfrac{\Om }{2} +\i \ov{\zeta}^{(m)}_{b}\, , \,  v -y_b - \i\tfrac{\Om }{2}    + \i \ov{\zeta}^{(m)}_{b} \Big) \, = \, 
 \pl{b=1}{N} \varpi\Big( \ov{y}_b-w -\i\tfrac{\Om }{2} +\i \zeta^{(m+1)}_{b}\, , \,  v - \ov{y}_b - \i\tfrac{\Om }{2}    + \i \zeta^{(m+1)}_{b} \Big) 
\enq
and
\bem
\pl{a<b}{N} \varpi\Big(  y_{ba} -\i\tfrac{\Om }{2} +\i \ov{\zeta}^{(m)}_{b} + \i \ov{\zeta}^{(m)}_{a}, y_{ab} - \i\tfrac{\Om }{2} - \i \ov{\zeta}^{(m)}_{b} + \i \ov{\zeta}^{(m)}_{a} \Big) \\ 
\; = \; \pl{a<b}{N} \varpi\Big(  \ov{y}_{ba} -\i\tfrac{\Om }{2} +\i \zeta^{(m+1)}_{b} + \i \zeta^{(m+1)}_{a}, \ov{y}_{ab} - \i\tfrac{\Om }{2} - \i \zeta^{(m+1)}_{b} + \i \zeta^{(m+1)}_{a} \Big) \\
\times \pl{a=m+2}{N}  \varpi\Bigg( \ba{c } \ov{y}_{m+1a}- \i\tfrac{\Om }{2}  +\i \zeta^{(m+1)}_{m+1} + \i \zeta^{(m+1)}_{a}, \ov{y}_{a m+1}- \i\tfrac{\Om }{2}  - \i \zeta^{(m+1)}_{m+1} + \i \zeta^{(m+1)}_{a}   \vspace{2mm} \\ 
			       \ov{y}_{m+1a}- \i\tfrac{\Om }{2}  +\i \zeta^{(m+1)}_{m+1} - \i \zeta^{(m+1)}_{a}, \ov{y}_{a m+1}- \i\tfrac{\Om }{2} + \i \zeta^{(m+1)}_{m+1} + \i \zeta^{(m+1)}_{a}     \ea \Bigg) \;. 
\end{multline}
Thus, define
\bem
\de\! \msc{J}^{(N;m)}_{\bs{\veps}_N}\big( x\mid \bs{w}_{N-1}, \bs{v}_{N-1} \big) \; = \; 
\Int{ \msc{C}^m }{} \dd^m y  \Int{ \R^{N-m-1} }{} \pl{a=m+1}{N-1} \dd y_a \Int{\R-\i\a }{} \dd y_N  \\
\times \bigg\{ \mc{I}_{N}^{(m)}\Big(\bs{\veps}_N\mid x,\bs{y}_N ; \bs{w}_{N-1}, \bs{v}_{N-1}  \Big)  
\, - \, \mc{I}_{N}^{(m)}\Big( \op{P}_{m+1}\bs{\veps}_N\mid x,\bs{y}_N ; \bs{w}_{N-1}, \bs{v}_{N-1}  \Big)  \bigg\}  \; ,  
\label{ecriture integrale de difference}
\end{multline}
Then  in the counter-term integral, \textit{viz}. the one associated with $ \op{P}_{m+1}\bs{\veps}_N$ in \eqref{ecriture integrale de difference}, 
one deforms the $y_N$-integration curve $\R-\i\a$ to $\msc{C}$. One picks up poles at $w_k+\i\veps_{m+1}=w_k-\i \ov{\zeta}^{(m)}_N$ with $k=1,\dots, N-1$, and this yields 
\bem
\Int{ \msc{C}^m }{} \dd^m y \Int{ \R^{N-m-1} }{} \pl{a=m+1}{N-1} \dd y_a \Int{\R-\i\a }{} \dd y_N \, \cdot \; \mc{I}_{N}^{(m)}\Big( \bs{\veps}_N\mid x,\bs{y}_N ; \bs{w}_{N-1}, \bs{v}_{N-1}  \Big)  \\
\hspace{3cm} \; = \; 
\de\! \msc{J}^{(N;m)}_{\bs{\veps}_N}\big( x\mid \bs{w}_{N-1}, \bs{v}_{N-1} \big) \, + \, \wt{\msc{J}}^{\, (N;m+1)}_{\bs{\veps}_N}\big( x\mid \bs{w}_{N-1}, \bs{v}_{N-1} \big) \\
\; - \;  \sul{k=1}{N-1} \Int{ \msc{C}^m }{} \dd^m y \Int{ \R^{N-m-1} }{} \pl{a=m+1}{N-1} \dd y_a  
\; \ex{2\i\pi (w_k+\i\veps_{m+1})x}    \cdot 
 \f{ \pl{  \substack{ a=1 \\ \not= k }   }{N-1} \varpi\Big( w_{ka} -\i\frac{\Om }{2}\Big) \pl{a=1}{N-1} \varpi\Big(  v_a -w_k - \i\frac{\Om }{2}  - 2 \i \veps_{m+1} \Big)  }
{ \pl{a=1}{N-1} \varpi\Big(  w_k-y_a -\i\frac{\Om }{2} + \i \zeta^{(m)}_{a}, y_a - w_k - \i\frac{\Om }{2}  + \i \zeta^{(m)}_{a} \Big)  }  \\
\times  \f{ 1 }{ N \sqrt{\om_1\om_2} }  \cdot \mc{I}_{N-1}^{(m)}\Big(\bs{\veps}_{N;[m]}\mid x,\bs{y}_{N-1} ; \bs{w}_{N-1}, \bs{v}_{N-1}  \Big)  \;.
\end{multline}
Above, we have introduced 
\bem
 \wt{\msc{J}}^{\, (N;m+1)}_{\bs{\veps}_N}\big( x\mid \bs{w}_{N-1}, \bs{v}_{N-1} \big) \; = \; 
\Int{ \msc{C}^{m+1} }{} \dd^{m+1} y \Int{ \R^{N-m} }{} \pl{a=m+2}{N} \dd y_a  \, \cdot \; \mc{I}_{N}^{(m+1)}\Big( \bs{\veps}_N\mid x,\bs{y}_N ; \bs{w}_{N-1}, \bs{v}_{N-1}  \Big) \\
\times \pl{a=m+2}{N}  \varpi\Bigg( \ba{c } y_{m+1\, a}- \i\frac{\Om }{2}  +\i \zeta^{(m+1)}_{m+1} + \i \zeta^{(m+1)}_{a}, y_{a\,  m+1}- \i\frac{\Om }{2}  - \i \zeta^{(m+1)}_{m+1} + \i \zeta^{(m+1)}_{a}   \vspace{2mm} \\ 
			       y_{m+1\, a}- \i\frac{\Om }{2}  +\i \zeta^{(m+1)}_{m+1} - \i \zeta^{(m+1)}_{a}, y_{a \, m+1}- \i\frac{\Om }{2} + \i \zeta^{(m+1)}_{m+1} + \i \zeta^{(m+1)}_{a}     \ea \Bigg) \;. 
\end{multline}

Thus, all-in-all, one gets 
\bem
\Big( \msc{J}^{(N;m)}_{\bs{\veps}_N} \, - \, \de\! \msc{J}^{(N;m)}_{\bs{\veps}_N} \,- \, \wt{\msc{J}}^{\, (N;m+1)}_{\bs{\veps}_N}\Big) \big( x\mid \bs{w}_{N-1}, \bs{v}_{N-1} \big) \\
\, = \, 
 \sul{k=1}{N-1} \f{  \ex{2\i\pi  w_k x}  }{ N \sqrt{\om_1\om_2} } \Int{ \msc{C}^m }{} \dd^m y \Int{ \R^{N-m-1} }{} \hspace{-1mm}\pl{a=m+1}{N-1} \dd y_a \;\;
 \mc{I}_{N-1}^{(m)}\Big(\bs{\veps}_{N;[m]}\mid x,\bs{y}_{N-1} ; \bs{w}_{N-1;[k]}, \bs{v}_{N-1}  \Big) \\
 \times \f{ \pl{  \substack{ a=1 \\ \not= k }   }{N-1} \varpi\Big( w_{ka} -\i\frac{\Om }{2}\Big)  }
{ \pl{a=1}{N-1} \varpi\Big(  w_k-y_a -\i\frac{\Om }{2} + \i \zeta^{(m)}_{a}\Big)  }  
\cdot   \mc{D}_{\veps_{m+1}}\big(\bs{v}_{N-1},w_k)  \;.
\end{multline}
Here $\bs{w}_{N-1;[k]}$ is as defined through \eqref{definition vecteur avec coordonnees omises},  
\beq
 \mc{D}_{\veps}\big(\bs{v}_{N-1},w_k) \, = \, \ex{2\pi x \veps}  \pl{a=1}{N-1} \varpi\Big(  v_a -w_k - \i\tfrac{\Om }{2}  + 2 \i \veps \Big) \; - \; \big( \, \veps\, \leftrightarrow -\veps \big) \;. 
\enq
Upon using that 
\beq
\varpi\Big( \la-\i\frac{\Om}{2} \Big) \; = \; \f{\i}{2} \f{ \varpi \big( \la + \i \frac{\tau}{2} \big) }{ \sinh\Big[ \frac{\pi}{\om_1} \la \Big] }
\enq
and applying the pole expansion 
\beq
\ex{ \f{\pi}{\om_1} x \de_N  } \pl{a=1}{N-1} \Bigg\{ \f{ 1 }{ \sinh\big[ \frac{\pi}{\om_1}(v_a-x)\big]  }\Bigg\} \; = \; 
\sul{\ell=1}{N-1} \f{ \ex{ \f{\pi}{\om_1} v_{\ell} \de_N  }  }{  \sinh\big[ \frac{\pi}{\om_1}(v_a-x)\big]  } \cdot  \f{1}{ \pl{ \substack{a=1 \\ a \not= \ell } }{N-1}    \sinh\Big[ \frac{\pi}{\om_1} (v_{a}-v_{\ell}) \Big]   } 
\quad \e{with} \quad 
\de_N \,=\, \left\{ \ba{cc} 1 &   N \; \e{odd} \\ 
			      0 &   N \; \e{even}  \ea \right. 
\enq
one obtains the decomposition 
\beq
 \mc{D}_{\veps}\big(\bs{v}_{N-1},w_k) \, = \, \Big( \frac{\i}{2} \Big)^{N-1} \sul{\ell=1}{N-1} 
 \f{ \pl{a=1}{N-1} \varpi\Big(  v_a -w_k - \i\frac{\tau }{2}  + 2 \i \veps \Big) }
	{ \pl{ \substack{a=1 \\ a \not= \ell } }{N-1}    \sinh\Big[ \frac{\pi}{\om_1} (v_{a}-v_{\ell}) \Big]   } \cdot
	\f{ \ex{ \f{\pi}{\om_1} (v_{\ell}-w_k+2\i\veps) \de_N  } \cdot  \ex{2\pi\veps x}  }{   \sinh\Big[ \frac{\pi}{\om_1} \big(v_{\ell} -w_k   + 2 \i \veps \big)  \Big]  }
\; - \; \big( \, \veps\, \leftrightarrow -\veps \big) \;.
\enq
This decomposition entails that, in the sense of distributions, 
\beq
 \mc{D}_{\veps}\big(\bs{v}_{N-1},w_k) \, \limit{ \veps}{ 0^+ } \, \sqrt{\om_1\om_2} \sul{\ell=1}{N-1}  \de\big(  v_{\ell} \, - \, w_k \big)
  \pl{ \substack{a=1 \\ a \not= \ell }}{N-1} \varpi\Big(  v_a -v_{\ell} - \i\frac{\Om }{2}  \Big)   \;.
\enq
Thus, all-in-all
\bem
\lim_{\veps_{m+1}\tend 0^+} \Big( \msc{J}^{(N;m)}_{\bs{\veps}_N} \, - \, \de\! \msc{J}^{(N;m)}_{\bs{\veps}_N} \,- \, \wt{\msc{J}}^{\, (N;m+1)}_{\bs{\veps}_N}\Big) \big( x\mid \bs{w}_{N-1}, \bs{v}_{N-1} \big)   \\
\hspace{4cm} \; = \; \f{1}{N} \sul{k,\ell=1}{N-1}  \de\big(  v_{\ell} \, - \, w_k \big) \cdot \ex{2\i\pi w_k x }\cdot 
  \pl{ \substack{a=1 \\ a \not= \ell }}{N-1} \varpi\Big(  v_{a\ell} - \i\frac{\Om }{2}  \Big)  \\
\times    \pl{ \substack{a=1 \\ a \not= k }}{N-1} \varpi\Big(  w_{ka} - \i\frac{\Om }{2}  \Big) 
 \msc{J}^{(N;m)}_{\bs{\veps}_{N;[m]} }\big( x\mid \bs{w}_{N-1;[k]}, \bs{v}_{N-1;[\ell]} \big)  \;. 
\end{multline}
Finally, one observes that the difference structure of the integrand of $\de\! \msc{J}^{(N;m)}_{\bs{\veps}_N}$ entails that 
\beq
\lim_{\bs{\veps}_{N}\tend \bs{0}^+}  \Big\{ \de\! \msc{J}^{(N;m)}_{\bs{\veps}_N}\big( x\mid \bs{w}_{N-1}, \bs{v}_{N-1} \big) \Big\}  \; = \; 0 \;. 
\enq
Likewise, since the additional $\bs{\veps}_N$ dependence present in  $\wt{\msc{J}}^{\, (N;m+1)}_{\bs{\veps}_N}\big( x\mid \bs{w}_{N-1}, \bs{v}_{N-1} \big) $
arises in the regular part of the integrand and since the singularities arising in the $\bs{\veps}_N \tend \bs{0}^+$ limit of the integrand generate at most 
Sokhotsky-Plemejl distributions, one has 
\beq
\lim_{\bs{\veps}_{N}\tend \bs{0}^+} \Big\{ \wt{\msc{J}}^{\, (N;m+1)}_{\bs{\veps}_N}\big( x\mid \bs{w}_{N-1}, \bs{v}_{N-1} \big) \Big\} 
\; = \; \lim_{\bs{\veps}_{N}\tend \bs{0}^+} \Big\{ \msc{J}^{\, (N;m+1)}_{\bs{\veps}_N}\big( x\mid \bs{w}_{N-1}, \bs{v}_{N-1} \big) \Big\}\;. 
\enq
All of the above leads to the relation among the various integrals:
\bem
\lim_{\bs{\veps}_{N}\tend \bs{0}^+} \Big\{  \msc{J}^{(N;m)}_{\bs{\veps}_N} \big( x\mid \bs{w}_{N-1}, \bs{v}_{N-1} \big) \Big\} 
\,- \, \lim_{\bs{\veps}_{N}\tend \bs{0}^+} \Big\{  \msc{J}^{\, (N;m+1)}_{\bs{\veps}_N}  \big( x\mid \bs{w}_{N-1}, \bs{v}_{N-1} \big)  \Big\}  \\
\hspace{3.5cm} \; = \; \f{1}{N} \sul{k,\ell=1}{N-1}  \de\big(  v_{\ell} \, - \, w_k \big) \cdot \ex{2\i\pi w_k x }\cdot 
  \pl{ \substack{a=1 \\ a \not= \ell }}{N-1} \varpi\Big(  v_{a\ell} - \i\frac{\Om }{2}  \Big) \\
  \times    \pl{ \substack{a=1 \\ a \not= k }}{N-1} \varpi\Big(  w_{ka} - \i\frac{\Om }{2}  \Big) 
\lim_{\bs{\veps}_{N-1}\tend \bs{0}^+} \Big\{ \msc{J}^{(N;m)}_{\bs{\veps}_{N-1} }\big( x\mid \bs{w}_{N-1;[k]}, \bs{v}_{N-1;[\ell]} \big)  \Big\} \;, 
\label{ecriture fundamental induction equation}
\end{multline}
which holds for any $m\in \intn{0}{N-1}$.

\vspace{2mm}

{\bf $\bullet $ The induction sep}

\vspace{2mm}

Upon setting $m=N-1$ and then decreasing the value of $m$ up to $m=2$, one infers from \eqref{ecriture vanishing integrale NN} and  \eqref{ecriture fundamental induction equation} that 
\beq
\underset{ \bs{\veps}_N \tend \bs{0}^+ }{ \e{Lim} } \Big\{ \msc{J}^{(N;m)}_{\bs{\veps}_N} \big( x\mid \bs{w}_{N-1}, \bs{v}_{N-1} \big) \Big\}  \, = \, 0 \qquad \e{for} \; \e{any} \qquad 
m\in \intn{2}{N} \;. 
\enq
When $m=1$, the previous results and the induction hypothesis leads to 
\bem
\underset{ \bs{\veps}_N \tend \bs{0}^+ }{ \e{Lim} } \Big\{ \msc{J}^{(N;m)}_{\bs{\veps}_N} \big( x\mid \bs{w}_{N-1}, \bs{v}_{N-1} \big) \Big\}  \, = \,
\f{1}{N(N-1)}\sul{k,\ell=1}{N-1}  \de\big(  v_{\ell} \, - \, w_k \big) \cdot \ex{2\i\pi w_k x }\cdot 
  \pl{ \substack{a=1 \\ a \not= \ell }}{N-1} \varpi\Big(  v_{a\ell} - \i\frac{\Om }{2}  \Big)   \\
  \hspace{2.5cm} \times  \pl{ \substack{a=1 \\ a \not= k }}{N-1} \varpi\Big(  w_{ka} - \i\frac{\Om }{2}  \Big)  \cdot \pl{ \substack{a\not= b  \\ a,b \not= k }}{N-1} \varpi\Big(  w_{ab} - \i\frac{\Om }{2}  \Big) \cdot  \de(x) \cdot  \de_{\e{sym}}\Big( \bs{w}_{N-1;[k]} - \bs{v}_{N-1;[\ell]} \Big) \\
  \hspace{2.5cm} \, = \, \f{1}{N} \de(x)  \pl{  a\not= b  }{N-1} \varpi\Big(  w_{ab} - \i\frac{\Om }{2}  \Big) \cdot \sul{ k = 1 }{ N-1 } \f{1}{N-1} \sul{\ell=1}{N-1} \sul{ \substack{ \sg \in \mf{S}_{N-1} \\ \sg(k)=\ell} }{}
\pl{a=1}{N-1} \de\big(v_{\sg(a)}-w_a \big) \\
\, = \,  \f{1}{N} \de(x) \cdot \de_{\e{sym}}\big(\bs{v}_{N-1}-\bs{w}_{N-1} \big) \cdot  \pl{  a\not= b  }{N-1} \varpi\Big(  w_{ab} - \i\frac{\Om }{2}  \Big) \;. 
\end{multline}
An analogous reasoning establishes the induction hypothesis for $m=0$, hence entailing the claim. \qed

\subsection{The completeness}

We are finally in position to establish the completeness of the system of Eigenfunctions introduced in Proposition \ref{Proposition rep MellinBarnes}.

\begin{prop}

The family $\{ \psi^{(-)}_{\bs{y}_N} \}_{\bs{y}_N \in \R^{N} }$ forms a complete system on $L^{2}\big( \R^N, \dd^N x \big)$
in respect to an $L^{2}_{\e{sym}}\big( \R^N, \dd \mu_N(\bs{y}_N) \big)$ integration of its indices, namely
\beq
\f{1}{N!} \Int{\R^N}{} \Big( \psi_{\bs{y}_N}^{(-)} \big( \bs{x}_N^{\prime} \big) \Big)^* \cdot  \psi_{\bs{y}_N}^{(-)} \big( \bs{x}_N \big) \cdot \dd \mu_{N}(\bs{y}_{N})     \; = \; \pl{a=1}{N}\de(x_a-x^{\prime}_a) \;,
\enq
where 
\beq
 \dd \mu_{N}(\bs{y}_{N}) \; = \;  \mu_{N}(\bs{y}_{N}) \dd^N y 
\enq
and $ \mu_{N}(\bs{y}_{N})$ is as defined in \eqref{definition densite mesure Harish-Chandra}.

\vspace{2mm}

The family $\{ \psi^{(+)}_{y_0,\bs{y}_{N-1}} \}_{ y_0 \in \R, \bs{y}_{N-1} \in \R^{N-1} }$ forms a complete system on $L^{2}\big( \R^N, \dd^N x \big)$
in respect to an $L^{2}_{\cdot \times \e{sym}}\big( \R \times \R^{N-1}, \dd y_0 \otimes \dd \mu_{N-1}(\bs{y}_{N-1}) \big)$ integration of its indices, namely
\beq
\f{1}{(N-1)!} \Int{\R^N}{} \Big(  \psi^{(+)}_{y_0,\bs{y}_{N-1}}  \big( \bs{x}_N^{\prime} \big) \Big)^* \cdot   \psi^{(+)}_{y_0,\bs{y}_{N-1}} \big( \bs{x}_N \big) \cdot \dd \mu_{N}(\bs{y}_{N})     \; = \; \pl{a=1}{N}\de(x_a-x^{\prime}_a) \;. 
\enq

\end{prop}

\Proof 

Assume that completeness, in the above sense, holds for  $\{ \psi^{(-)}_{ \bs{y}_{N-1} } \}_{\bs{y}_{N-1} \in \R^{N-1} }$. 
Then, by using the Mellin-Barnes representation and the form of the regularisation one has to impose for the integral representation to make sense, one gets 
that (\textit{c.f.} \cite{KozUnitarityofSoVTransform} for a rigorous treatment of the various steps  in the sense of distributions) 
\bem
\f{1}{N!} \Int{\R^N}{} \Big( \psi_{\bs{y}_N}^{(-)} \big( \bs{x}_N^{\prime} \big) \Big)^* \cdot  \psi_{\bs{y}_N}^{(-)} \big( \bs{x}_N \big) \cdot \dd   \mu_{N}(\bs{y}_{N})    \\ 
\; = \; 
\Int{ \R^{N-1} }{}   \f{  \dd\mu_{N-1}(\bs{w}_{N-1}) }{ (N-1)! }    \Int{ \R^{N-1} }{}  \f{ \dd^{N-1}v  }{ (N-1)! } \cdot \Big( \psi_{\bs{v}_{N-1} }^{(-)} \big( \bs{x}_{N-1}^{\prime} \big) \Big)^* \cdot  \psi_{\bs{w}_{N-1}}^{(-)} \big( \bs{x}_{N-1} \big)
\cdot \lim_{ \bs{\veps}_N \tend \bs{0}^+ }\chi_N^{(\bs{\veps}_N)}\big( \bs{w}_{N-1}, \bs{v}_{N-1} ; x_N , x_N^{\prime} \big)
\label{ecriture PS completude via Mellin Barnes}
\end{multline}
where 
\beq
\chi_N^{(\bs{\veps}_N)}\big( \bs{w}_{N-1}, \bs{v}_{N-1} ; x_N, x_N^{\prime} \big) \; = \; \pl{a=1}{N-1} \bigg\{ \f{ \varpi (w_a+\kappa)  }{ \varpi(v_a+\kappa) } \bigg\}^{N-1}  \cdot 
   \f{ \psi^{(-)}_{\ov{\bs{v}}_{N-1}}( x_N^{\prime} ) }{ \psi^{(-)}_{\ov{\bs{w}}_{N-1}}( x_N ) } \cdot 
\f{ \mc{K}_N^{(\bs{\veps}_N)}\big( \bs{w}_{N-1}, \bs{v}_{N-1} ; x_N - x_N^{\prime} \big) }{ \pl{a\not=b}{N-1}\varpi\Big(v_{ab}-\i\tfrac{\Om}{2} \Big)   }  \;, 
\label{equation chi N en fct de K N}
\enq
while 
\bem
\mc{K}_N^{(\bs{\veps}_N)}\big( \bs{w}_{N-1}, \bs{v}_{N-1} ; x  \big) \; = \;  \Int{ \R^N }{}   \f{ \ex{   2 \i \pi  x \ov{\bs{y}}_N  }   }{N! (\om_1 \om_2)^{N-1} } 
\f{  \varpi\big( \ov{\bs{y}}_N-\ov{\bs{v}}_{N-1} -\kappa\big) }{  \varpi\big( \ov{\bs{y}}_N-\ov{\bs{w}}_{N-1}-\kappa \big)  }  \\
\times \f{ \pl{a=1}{N}\pl{b=1}{N-1} \varpi\Big(  y_a-w_b -\i\f{\Om }{2} + \i \veps_{N+1-a}\Big) \,  \varpi\Big( v_b -y_a-\i\f{\Om  }{2} + \i \veps_{N+1-a} \Big)  }
{ \pl{a<b}{N} \varpi\Big(  y_b-y_a -\i\f{\Om }{2} + \i \big( \veps_{N+1-b} + \veps_{N+1-a} \big) , y_a-y_b -\i\f{\Om }{2} + \i \big( \veps_{N+1-a} - \veps_{N+1-a} \big) \Big)   } \cdot \dd^N y \;. 
\end{multline}
The use of the integral representation,  
\beq
\f{  \varpi\big( \ov{\bs{y}}_N-\ov{\bs{v}}_{N-1} -\kappa\big) }{  \varpi\big( \ov{\bs{y}}_N-\ov{\bs{w}}_{N-1}-\kappa \big)  } 
\; = \; \Int{ \R }{} \dd t \Int{\R}{}\dd s \f{  \varpi\big( s-\ov{\bs{v}}_{N-1} -\kappa\big) }{  \varpi\big( s-\ov{\bs{w}}_{N-1}-\kappa \big)  } \ex{2\i\pi t(s-\ov{\bs{y}}_N)}
\enq
allows one to recast the previous expression as 
\beq
\mc{K}_N^{(\bs{\veps}_N)}\big( \bs{w}_{N-1}, \bs{v}_{N-1} ; x  \big) \; = \;  \Int{ \R }{} \dd t \Int{\R}{}\dd s \f{  \varpi\big( s-\ov{\bs{v}}_{N-1} -\kappa\big) }{  \varpi\big( s-\ov{\bs{w}}_{N-1}-\kappa \big)  } \ex{2\i\pi t s}
\msc{J}^{(N;0)}_{\bs{\veps}_N}\big( x-t\mid \bs{w}_{N-1}, \bs{v}_{N-1} \big) \;, 
\enq
with $\msc{J}^{(N;0)}_{\bs{\veps}_N}$ as defined through \eqref{definition integrale mathscr J N m-}-\eqref{ecriture expression explicite integrande modele}. Then, by virtue of Proposition \ref{Proposition integrale multiple pour delta sym}
and straightforward integrations,  
\beq
\lim_{\bs{\veps}_N\tend \bs{0}^+} \mc{K}_N^{(\bs{\veps}_N)}\big( \bs{w}_{N-1}, \bs{v}_{N-1} ; x  \big) \; = \;
 \pl{a\not=b}{N-1}\varpi\Big(v_{ab}-\i\tfrac{\Om}{2} \Big)  \cdot \de\big( x \big) \cdot \de_{\e{sym}}\Big(\bs{w}_{N-1}-\bs{v}_{N-1} \Big)   \;. 
\enq
Thus, upon inserting this result in \eqref{equation chi N en fct de K N} and then \eqref{ecriture PS completude via Mellin Barnes}, one obtains:
\bem
\f{1}{N!} \Int{\R^N}{} \Big( \psi_{\bs{y}_N}^{(-)} \big( \bs{x}_N^{\prime} \big) \Big)^* \cdot  \psi_{\bs{y}_N}^{(-)} \big( \bs{x}_N \big) \cdot  \dd \mu_{N}(\bs{y}_{N}) \\
= \f{ \de\big(x_N-x_N^{\prime} \big)  }{(N-1)!} \cdot 
\Int{ \R^{N-1} }{} \Big( \psi_{\bs{y}_{N-1}}^{(-)} \big( \bs{x}_{N-1}^{\prime} \big) \Big)^* \cdot  \psi_{\bs{y}_{N-1}}^{(-)} \big( \bs{x}_{N-1} \big) \cdot  \dd \mu_{{N-1}}(\bs{y}_{{N-1}})  
\,  = \,   \pl{a=1}{N} \de\big(x_a-x_a^{\prime} \big) \, \;. 
\end{multline}

 \vspace{2mm} The proof of completeness in the case of the system $\{ \psi^{(+)}_{y_0,\bs{y}_{N-1}} \}_{ y_0 \in \R, \bs{y}_{N-1} \in \R^{N-1} }$ goes along similar lines. \qed

\section{Complete and orthogonal system of Eigenfunctions of $\big[\op{T}_{N}(\la)\big]_{ab}$}
\label{Section Complete and orthogonal system}

In this section we summarise the results established in the previous two sections.

\begin{theorem}
\label{Theorem fct propres entrees matrice monodromie} 
 
Each of the four operator entries of the monodromy matrix $\op{T}_{N}(\la)$ \eqref{ecriture matrice de monodromie XXZ} admits a complete and orthogonal system of Eigenfunctions,  
$\Big\{ \Phi^{(\op{E}\,)}_{\bs{y}_N} \Big\}_{\bs{y}_N\in \R^N}$  for $\op{E}\in \{\op{A}, \op{D} \}$ and 
 $\Big\{ \Phi^{(\op{E}\,)}_{y_0,\bs{y}_{N-1}} \Big\}_{y_0\in \R, \bs{y}_{N-1}\in \R^{N-1}}$, for for $\op{E}\in \{\op{B}, \op{C} \}$. 
Let $\dd \mu_N(\bs{y}_N)=\mu_N(\bs{y}_N) \dd^N y$ with $\mu_N(\bs{y}_N)$ as defined in \eqref{definition densite mesure Harish-Chandra}.
 \begin{itemize}
  
  \item The system $\Big\{ \Phi_{\bs{y}_N}^{(\op{E}\,)} \Big\}_{ \bs{y}_N \in \R^N}$,  $\op{E}\in \{\op{A}, \op{D} \}$,  satisfies:
\beq
\f{1}{N!} \Int{ \R^N }{} \Big( \Phi_{ \bs{y}_N}^{(\op{E}\,)}  \big( \bs{x}_N^{\prime} \big) \Big)^* \cdot  \Phi_{\bs{y}_N}^{(\op{E}\,)} \big( \bs{x}_N \big) \cdot  \dd \mu_{N}(\bs{y}_{N}) \; = \; \pl{a=1}{N} \de\big( x_a - x^{\prime}_a \big)
\enq
along with
\beq
\Int{\R^N}{} \Big( \Phi_{ \bs{y}_N^{\prime} }^{(\op{E}\,)} \big( \bs{x}_N \big) \Big)^* \cdot  \Phi_{\bs{y}_N}^{(\op{E}\,)} \big( \bs{x}_N \big) \cdot  \dd^N x \; = \;  \f{1}{\mu_N(\bs{y}_N) } \cdot  \de_{\e{sym}}\big( \bs{y}_N - \bs{y}^{\prime}_N \big) \;. 
\enq
Furthermore, the generalised Eigenvalue equation takes the form
\beq
\op{E}_N(\la) \cdot  \Phi_{\bs{y}_N}^{(\op{E}\,)}  \big( \bs{x}_N \big) \; = \; \pl{a=1}{N} \Big\{ -2\i \sinh\big[\tfrac{\pi}{\om_1}(\la-y_a)  \big] \Big\}\cdot  \Phi_{\bs{y}_N}^{(\op{E}\,)}  \big( \bs{x}_N \big) \;. 
\enq

   \item The system $\Big\{ \Phi_{y_0, \bs{y}_{N-1} }^{(\op{E}\,)}  \Big\}_{  y_0\in \R  \, , \,  \bs{y}_{N-1} \in \R^{N-1}  }$,  $\op{E}\in \{\op{B}, \op{C} \}$ satisfies:
\beq
\f{1}{(N-1)!} \Int{\R^N}{} \Big( \Phi_{y_0^{\prime}, \bs{y}_{N-1}^{\prime} }^{(\op{E}\,)} \big( \bs{x}_N^{\prime} \big) \Big)^* \cdot  \Phi_{\bs{y}_N}^{(\op{E}\,)}  \big( \bs{x}_N \big) \cdot \dd y_0 \cdot \dd \mu_{N-1}(\bs{y}_{N-1}) 
\; = \; \pl{a=1}{N} \de\big( x_a - x^{\prime}_a \big)
\enq
along with
\beq
\Int{\R^N}{} \Big( \Phi_{y_0^{\prime}, \bs{y}_{N-1}^{\prime} }^{(\op{E}\,)}   \big( \bs{x}_N \big) \Big)^* \cdot  \Phi_{y_0, \bs{y}_{N-1} }^{(\op{E}\,)}  \big( \bs{x}_N \big) \cdot  \dd^N x \; = \; 
\f{1}{\mu_{N-1}(\bs{y}_{N-1}) } \cdot \de(y_0-y_0^{\prime}) \cdot  \de_{\e{sym}}\big( \bs{y}_{N-1} - \bs{y}^{\prime}_{N-1} \big) \;. 
\enq
Finally, the generalised Eigenvalue equation takes the form
\beq
\op{E}_N(\la) \cdot  \Phi_{y_0, \bs{y}_{N-1} }^{(\op{E}\,)}  \big( \bs{x}_N \big) \; = \; \ex{2\pi \om_2 y_0} \pl{a=1}{N-1} \Big\{ -2\i \sinh\big[\tfrac{\pi}{\om_1}(\la-y_a)  \big] \Big\}\cdot  \Phi_{y_0, \bs{y}_{N-1} }^{(\op{E}\,)}  \big( \bs{x}_N \big) \;. 
\enq

 \end{itemize}

The generalised Eigenfunctions admit Gauss-Givental and Mellin-Barnes integral representations:
\beq
\Phi_{\bs{y}_N}^{(\op{A})}  \big( \bs{x}_N \big) \, = \, \psi^{(-)}_{\bs{y}_N}\big( \bs{x}_N \big) \, = \, c_{\op{A}} \, \vp^{(-)}_{\bs{y}_N}\big( \bs{x}_N \big)
\qquad  and  \qquad 
\Phi_{y_0, \bs{y}_{N-1} }^{(\op{B})}  \big( \bs{x}_N \big) \, = \, \psi^{(+)}_{y_0, \bs{y}_{N-1} }\big( \bs{x}_N \big) \, = \, c_{\op{B}} \,  \vp^{(+)}_{y_0, \bs{y}_{N-1} }\big( \bs{x}_N \big)
\label{ecriture proportionalite MB et GG reps}
\enq
Above, $c_{\op{A}}$, resp. $c_{\op{B}}$, are $\bs{y}_N$, resp. $y_0, \bs{y}_{N-1}$, independent constants equal to $\pm 1$. 
 
\end{theorem}

We conjecture that, in fact, the proportionality constants in \eqref{ecriture proportionalite MB et GG reps} equal $1$.

\Proof 

We shall only establish the properties of the generalised Eigenfunctions of the operator $\op{A}_N(\la)$,
as the case of the $\op{B}_N(\la)$ operator can be dealt with similarly. 
Furthermore, the case of the operators $\op{D}_N(\la)$ and $\op{C}_N(\la)$ is a direct consequence of 
the results relative to the operators $\op{A}_N(\la)$ and $\op{B}_N(\la)$. 

Indeed, observe that upon introducing the operator $\Om_a$ such that 
\beq
\Om_a \op{x}_a \Om_a \, = \, - \op{x}_a \qquad \e{and} \qquad  \Om_a \op{p}_a \Om_a \, = \, - \op{p}_a \;, 
\enq
one has the relation 
\beq
\Om_n \cdot \op{L}_n^{(\kappa)}(\la)\cdot \Om_n \, = \, \sg^{x} \cdot \op{L}_n^{(\kappa)}(\la) \cdot \sg^x   \qquad i.e. \qquad 
\pl{a=1}{N}\Om_a \cdot \op{T}_N(\la)\cdot \pl{a=1}{N}\Om_a \, = \, \sg^{x} \cdot \op{T}_N(\la)\cdot \sg^x \;. 
\label{definition operateurs parite Omega}
\enq
This entails that  
\beq
\pl{a=1}{N}\Om_a \cdot \op{D}_N(\la) \cdot \pl{a=1}{N}\Om_a =  \op{A}_N(\la) \qquad \e{and} \qquad  \pl{a=1}{N}\Om_a \cdot \op{C}_N(\la) \cdot \pl{a=1}{N}\Om_a =  \op{B}_N(\la) \; .
\label{ecriture relation AB et DC}
\enq
Thus, if $\Big\{ \Phi^{(\op{A})}_{\bs{y}_N} \Big\}_{\bs{y}_N\in \R^N}$, resp.  $\Big\{ \Phi^{(\op{B})}_{y_0,\bs{y}_{N-1}} \Big\}_{y_0\in \R, \bs{y}_{N-1}\in \R^{N-1}}$, 
is the complete and  orthogonal system of generalised Eigenfunctions of the $\op{A}_N(\la)$, resp. $\op{B}_N(\la)$, operator
then 
\beq
\bigg\{ \pl{a=1}{N}\Om_a \cdot \Phi^{(\op{A})}_{\bs{y}_N} \bigg\}_{\bs{y}_N\in \R^N}\; , \qquad \e{resp}.  \qquad 
\bigg\{ \pl{a=1}{N}\Om_a \cdot  \Phi^{(\op{B})}_{y_0,\bs{y}_{N-1}} \bigg\}_{y_0\in \R, \bs{y}_{N-1}\in \R^{N-1}} \;, 
\enq
is the complete and  orthogonal system of generalised  Eigenfunctions of the $\op{D}_N(\la)$, resp. $\op{C}_N(\la)$, 
operator. Furthermore, the form of the generalised Eigenvalues is a direct consequence of the conjugation relation between the operators \eqref{ecriture relation AB et DC}
and the form of the generalised Eigenvalues of the operators $\op{A}_N(\la)$ and $\op{B}_N(\la)$.

\subsubsection*{$\bullet$ Proprotionality of $\vp^{(-)}_{\bs{y}_N}$ and $\psi^{(-)}_{\bs{y}_N}$ }

We first show that, for any fixed $\bs{y}_N \in \R^N$, the functions $\vp^{(-)}_{\bs{y}_N}$ and $\psi^{(-)}_{\bs{y}_N}$ are are non-identically zero. 
Given $f\in L^2_{\e{sym}}\big( \R^N, \dd \mu_{N}(\bs{y}_N) \big)$, smooth and compactly supported, define
\beq
\mc{U}_N[f](\bs{x}_N) \; = \; \f{1}{N!} \Int{ \R^N }{} \vp^{(-)}_{\bs{y}_N}(\bs{x}_N) f(\bs{y}_N) \dd \mu_{N}(\bs{y}_N) \;. 
\enq
Then, the orthogonality relation satisfied by the functions $ \vp^{(-)}_{\bs{y}_N}(\bs{x}_N)$ ensures that, for any such $f$, it holds
\beq
\Int{ \R^N }{} \dd^N x \,  \Big( \vp^{(-)}_{\bs{y}_N}(\bs{x}_N) \Big)^* \cdot \mc{U}_N[f](\bs{x}_N)  \; = \; f(\bs{y}_N) \;. 
\enq
Thus,  $\vp^{(-)}_{\bs{y}_N}$ cannot be identically zero. 

Regarding to $\psi^{(-)}_{\bs{y}_N}$, as in \cite{KozIPForTodaAndDualEqns}, one may compute the $\bs{x}_N\tend \infty$, $x_{a+1}-x_a \tend +\infty$ of $\psi^{(-)}_{\bs{y}_N}(\bs{x}_N)$
staring from its Mellin-Barnes integral representation by pushing the various integration contours slightly in the upper-half plane. 
This shows that the function is non-vanishing in this asymptotic regime, and hence is non-identically zero.

\vspace{3mm}

Since 
\begin{itemize}
\item $\op{A}_N(\la) \psi^{(-)}_{\bs{y}_N} = a(\la\mid \bs{y}_N) \psi^{(-)}_{\bs{y}_N}$ with $a(\la\mid \bs{y}_N)  \, = \, \pl{a=1}{N} \Big\{-2\i \sinh\big[\tfrac{\pi}{\om_1}(\la-y_a)\big] \Big\} $, 
\item $\psi^{(-)}_{\bs{y}_N}$ is non-identically zero,
\item the system $\big\{ \psi^{(-)}_{\bs{y}_N} \big\}_{\bs{y}_N \in \R^N }$ forms a complete system, 
\end{itemize}
each generalised Eigenvalue $a(\la\mid \bs{y}_N) $ of $\op{A}_N(\la)$ appears with exactly multiplicity $1$. Since it also holds $\op{A}_N(\la) \vp^{(-)}_{\bs{y}_N} = a(\la\mid \bs{y}_N) \vp^{(-)}_{\bs{y}_N}$,
with $\vp^{(-)}_{\bs{y}_N}$ non-identically zero, it follows that there exists a constant $c_N(\bs{y}_N,\kappa) \in \Cx^*$ such that  
\beq
\psi^{(-)}_{\bs{y}_N} \, = \,  c_N(\bs{y}_N,\kappa) \vp^{(-)}_{\bs{y}_N} \;. 
\enq
The proprotionality constant may, in principle, depend on $\bs{y}_N$ and the representation parameter $\kappa$. 
Below, we establish various properties that ought to be satisfied by these constants. This will strongly constrain
its value.

\subsubsection*{$\bullet$ $c_N(\bs{y}_N,\kappa)$ is unimodular}

We first establish that the proportionality constant is unimodular. The orthogonality relation for $\vp^{(-)}_{\bs{y}_N}$ leads to the relation 
\beq
\Int{\R^N}{}  \Big( \psi_{\bs{y}_N^{\prime}}^{(-)}(\bs{x}_N) \Big)^*\cdot \psi_{\bs{y}_N}^{(-)}(\bs{x}_N) \cdot \dd^N x \; = \; 
\f{ |c_N(\bs{y}_N,\kappa)|^2  }{ \mu_{N}\big(\bs{y}_N\big) } \cdot \de_{\e{sym}}\big( \bs{y}_N - \bs{y}_N^{\prime}  \big)
\enq
Integrating both sides of this equation versus 
\beq
\f{\dd\mu_N(\bs{y}^{\prime}_N)}{N!}\otimes \f{ \dd\mu_N(\bs{y}_N) }{ N! }  \psi_{\bs{y}_N^{\prime}}^{(-)}(\bs{x}_N^{\prime})  \Big( \psi_{\bs{y}_N}^{(-)}(\bs{x}_N^{\prime\prime}) \Big)^*
\enq
and using completeness of the $\psi_{\bs{y}_N^{\prime}}^{(-)}$ one is lead to 
\beq
\pl{a=1}{N} \de\big(x_a^{\prime}-x_a^{\prime\prime}\big) \; = \; |c_N(\bs{y}_N,\kappa)|^2\pl{a=1}{N} \de\big(x_a^{\prime}-x_a^{\prime\prime}\big) \;.  
\enq
Thus, indeed, one has that $|c_N(\bs{y}_N,\kappa)|=1$.

\subsubsection*{$\bullet$ $c_N(\bs{y}_N,\kappa)$ is invariant under $\kappa$-reflections}

It follows from \eqref{equation entrelacement rep kappa et moins kappa} that 
\beq
\pl{a=1}{N} D_{-\kappa}\big( \op{p}_a \big)  \cdot  \op{T}_N(\la;\kappa) \, = \,  \op{T}_N(\la;-\kappa)  \cdot \pl{a=1}{N} D_{-\kappa}\big( \op{p}_a \big)
\enq
where we have explicitly stressed the dependence of the monodromy matrix on the parameter $\kappa$. In the following, it will be convenient to make also explicit the dependence of the functions 
$\psi^{(-)}_{\bs{y}_N},  \vp^{(-)}_{\bs{y}_N}$ on $\kappa$, \textit{viz}. $\psi^{(-)}_{\bs{y}_N}(\bs{x}_N;\kappa),  \vp^{(-)}_{\bs{y}_N}(\bs{x}_N;\kappa)$. 
The intertwining of the monodromy matrix suggests that analogous relations should exist between the generalised Eigenfunctions. 
These will be established below.

The Mellin-Barnes integral representation induction immediately leads to 
\beq
\psi^{(-)}_{\bs{y}_N}(\bs{x}_N;\kappa) \; = \; \pl{s=1}{N-1} \Bigg\{ \Int{ (\R-\i \a_s)^{N-s} }{} \f{ \dd^{N-s}w^{(s)} }{ (N-s)!}  \Bigg\}
\pl{ s=1 }{N} \psi^{(-)}_{ \ov{\bs{w}}_{N-s+1}^{(s-1)}- \ov{\bs{w}}_{N-s}^{(s)}}(x_{N-s+1}) \pl{s=1}{N}\Big\{ \Phi\big( \bs{w}_{N-s}^{(s)} \mid \bs{w}_{N-s+1}^{(s-1)} \big) \Big\}
\enq
which can be reorganised as
\beq
\psi^{(-)}_{\bs{y}_N}(\bs{x}_N;\kappa) \; = \; \sg\big( \bs{y}_N;\kappa \big)\pl{s=1}{N-1} \Bigg\{ \Int{ (\R-\i \a_s)^{N-s} }{} \f{ \dd^{N-s}w^{(s)} }{ (N-s)!}  \Bigg\}
\pl{ s=1 }{N} \psi^{(-)}_{ \ov{\bs{w}}_{N-s+1}^{(s-1)}- \ov{\bs{w}}_{N-s}^{(s)}}(x_{N-s+1}) \cdot \xi_{\kappa}\Big( \big\{ \bs{w}_{N-s}^{(s)} \big\}_{s=1}^{N} \Big)
\label{ecriture MB compacte totale pour psi moins}
\enq
where 
\beq
 \xi_{\kappa}\Big( \big\{ \bs{w}_{N-s}^{(s)} \big\}_{s=1}^{N} \Big) \; = \; \f{  \pl{s=1}{N-1}\pl{a=1}{N-s} \varpi\big( w_a^{(s)}-\kappa, w_a^{(s)}+\kappa \big)  }
 { \pl{s=1}{N}  \varpi\big( \ov{\bs{w}}_{N-s+1}^{(s-1)}- \ov{\bs{w}}_{N-s}^{(s)} -\kappa \big) } 
\cdot  
\pl{s=1}{N-1} \Bigg\{ \f{ \pl{a=1}{N-s} \pl{b=1}{N-s+1}  \varpi\big( w_b^{(s-1)} -  w_a^{(s)}- \i \tfrac{\Om}{2} \big)   }{  \big( \om_1 \om_2 \big)^{N-s} \cdot  \pl{a \not= b}{N-s}  \varpi\big( w_b^{(s)} -  w_a^{(s)}- \i \tfrac{\Om}{2} \big)  } 
\Bigg\} \;, 
\enq
and 
\beq
\sg\big( \bs{y}_N;\kappa \big) \, = \, \pl{a=1}{N} \f{ \varpi(y_a-\kappa) }{ \big[ \varpi(y_a+\kappa) \big]^{N-1} } \;. 
\enq
The relation 
\beq
 \Big( D_{-\kappa}\big( \op{p}  \big) \cdot \psi^{(-)}_{y}\Big)(x) \; = \; D_{-\kappa}(y) \, \psi^{(-)}_{y}(x)
\enq
and the fact that 
\beq
 \xi_{\kappa}\Big( \big\{ \bs{w}_{N-s}^{(s)} \big\}_{s=1}^{N} \Big) \; = \;  
 \pl{s=1}{N}\f{  \varpi\big( \ov{\bs{w}}_{N-s+1}^{(s-1)}- \ov{\bs{w}}_{N-s}^{(s)} + \kappa \big) }{  \varpi\big( \ov{\bs{w}}_{N-s+1}^{(s-1)}- \ov{\bs{w}}_{N-s}^{(s)} -\kappa \big) } 
 \cdot \xi_{-\kappa}\Big( \big\{ \bs{w}_{N-s}^{(s)} \big\}_{s=1}^{N} \Big)  
\enq
then immediately leads to 
\beq
\pl{a=1}{N} D_{-\kappa}\big( \op{p}_a \big)  \cdot  \psi^{(-)}_{\bs{y}_N}(\bs{x}_N;\kappa) \, = \, \pl{a=1}{N}D_{-\kappa}^N(y_a) \cdot \psi^{(-)}_{\bs{y}_N}(\bs{x}_N;-\kappa) \; .  
\enq

The same property holds for $\vp^{(-)}_{\bs{y}_N}$, namely 
\beq
\pl{a=1}{N} D_{-\kappa}\big( \op{p}_a \big)  \cdot  \vp^{(-)}_{\bs{y}_N}(\bs{x}_N;\kappa) \, = \, \pl{a=1}{N}D_{-\kappa}^N(y_a) \cdot \vp^{(-)}_{\bs{y}_N}(\bs{x}_N;-\kappa) \; .  
\label{ecriture action operateur entrelacement rep de spin}
\enq
Indeed, this identity is a direct consequence of an inductive application of Lemma \ref{Proposition action operateur spin flip sur operateur Lambda}
on the level of the recursive construction of $\vp^{(-)}_{\bs{y}_N}(\bs{x}_N;\kappa)$. 

This entails that $c_N(\bs{y}_N,\kappa)=c_N(\bs{y}_N,-\kappa)$.

\subsubsection*{$\bullet$ Complex conjugation of $c_N(\bs{y}_N,\kappa)$}

Finally, the $\La^{(N)}_{y,\eps}$ operator can be represented as
\beq
\Big( \La^{(N)}_{y,\eps}\cdot f \Big)(\bs{x}_N) \, = \, \Int{ \R^{N-1} }{} \La^{(N)}_{y,\eps}\big(\bs{x}_N,\bs{x}^{\prime}_{N-1}; \kappa) f (\bs{x}^{\prime}_{N-1})  \cdot \dd^{N-1}x^{\prime}
\enq
where the integral kernel takes the form 
\bem
 \La^{(N)}_{y,\eps}\big(\bs{x}_N,\bs{x}^{\prime}_{N-1}; \kappa) \; = \; \f{ \ex{2\i\pi y_- x_1 }  \ex{2\i\pi y_+^{\star} x_N }  }{ \Big( \sqrt{\om_1\om_2} \mc{A}(y_+) \Big)^{N-1} } \\
\times \pl{a=1}{N-1} \bigg\{ D_{y_--y_+}\big( \om_1 \om_2 x_{aa+1} \big)  D_{y_-^{\star}}\big( \om_1 \om_2 (x_{a}-x^{\prime}_a) \big) D_{y_+}\big( \om_1 \om_2 (x_{a+1}-x^{\prime}_a) \big)\bigg\} \;. 
\end{multline}
It is then immediate to check that 
\beq
 \Big( \La^{(N)}_{y,\eps}\big(\bs{x}_N,\bs{x}^{\prime}_{N-1}; \kappa)  \Big)^{*} \; = \;  \La^{(N)}_{-y,\eps}\big(\bs{x}_N,\bs{x}^{\prime}_{N-1}; -\kappa)  \;. 
\enq
This relation then leads to 
\beq
 \Big(\vp^{(-)}_{\bs{y}_N}(\bs{x}_N;\kappa) \Big)^{*} \, = \,   \vp^{(-)}_{-\bs{y}_N}(\bs{x}_N;-\kappa) \;. 
\enq
\vspace{2mm}

\noindent Now starting from the Mellin-Barnes representation \eqref{ecriture MB compacte totale pour psi moins} and upon using that 
for 
\beq
\bs{w}^{(s)}_{N-s } \in \R^{N-s} \quad  \e{and} \quad  \bs{\a}_s=\a_s(1,\dots, 1)\in \R^{N-s} \quad \e{with} \quad  \a_s \quad \e{real} \; ,
\enq
it holds 
\beq
 \Big( \xi_{\kappa}\Big( \big\{ -\bs{w}_{N-s}^{(s)} -\i \bs{\a}_s \big\}_{s=1}^{N} \Big) \Big)^{*} \; = \; \xi_{-\kappa}\Big( \big\{  \bs{w}_{N-s}^{(s)} -\i \bs{\a}_s \big\}_{s=1}^{N} \Big) \;, 
\enq
one entails that, as well 
\beq
 \Big(\psi^{(-)}_{\bs{y}_N}(\bs{x}_N;\kappa) \Big)^{*} \, = \,   \psi^{(-)}_{-\bs{y}_N}(\bs{x}_N;-\kappa) \;. 
\enq
This entails that $ \Big( c_N(\bs{y}_N,\kappa) \Big)^*=c_N(-\bs{y}_N,-\kappa)$. Hence, by invoking the $\kappa$-reflection property, one infers that the constant $c_N$ behaves under complex conjugation
as
\beq
\Big( c_N(\bs{y}_N,\kappa) \Big)^* = c_N(-\bs{y}_N, \kappa) \; . 
\enq

\subsubsection*{$\bullet$ $c_N(\bs{y}_N,\kappa)$ is $\bs{y}_N$-independent}

It remains to prove that the constant does not depend on $\bs{y}_N$. As follows from Propositions \ref{Proposition action operateur B} and \ref{Proposition commutativite des Lambda et echange Lambda dagger Lambda},
$\vp^{(-)}_{\bs{y}_N}$ satisfies identically the same equations \eqref{ecriture action B}-\eqref{ecriture action C} as $\psi^{(-)}_{\bs{y}_N}$. 
By projecting these on a given variable $\bs{y}_N$, this yields that 
\beq
 c_N(\bs{y}_N,\kappa) \, = \,  c_N(\bs{y}_N+\i \om_2 \bs{e}_k,\kappa) 
\enq
where $\bs{e}_k$ is the $k^{\e{th}}$ unit vector in $\R^N$. The same equation holds for $\om_1\leftrightarrow \om_2$ since both 
$\vp^{(-)}_{\bs{y}_N}$ and $\psi^{(-)}_{\bs{y}_N}$ satisfy the dual equations to  \eqref{ecriture action B}-\eqref{ecriture action C} as well. 
 
Thus, all-in-all, we get that $\big(c_N(\bs{y}_N,\kappa) \big)^*= c_N(\bs{y}_N,\kappa) $, what along with $| c_N(\bs{y}_N,\kappa) |=1$ implies that 
$ c_N(\bs{y}_N,\kappa) =\pm 1$. \qed

\section{The Sinh-Gordon model}
\label{Section Eigenfunctions for Sinh Gordon}

The Lax matrix of the lattice Sinh-Gordon model takes the form \cite{BytskoTeschnerRmatrixForModularDouble}:
\beq
\op{L}_n^{(\e{SG})}(\la) \, = \, \sg^x \op{L}_n^{(\kappa)}(\la) \;. 
\label{ecriture lien Lax XXZ et Lax SG}
\enq
The  monodromy matrix of the $N$-site lattice Sinh-Gordon model thus takes the form  
\beq
\op{T}_N^{(\e{SG})}(\la) \; = \; \op{L}_1^{(\e{SG})}(\la)\cdots \op{L}_N^{(\e{SG})}(\la) \; = \; \left(\ba{cc} \op{A}_N^{(\e{SG})}(\la)   & \op{B}_N^{(\e{SG})}(\la)   \\ 
								  \op{C}_N^{(\e{SG})}(\la)   &   \op{D}_N^{(\e{SG})}(\la) \ea  \right) \;. 
\enq
The analysis of the system of Eigenfunctions of the operators $\op{A}_N$ and $\op{B}_N$ that was carried out in the previous sections allows us 
to access to the system of  Eigenfunctions of the operator  $ \op{B}_N^{(\e{SG})}(\la) $ and characterise its spectrum. 
In this manner we prove the conjecture raised in \cite{BytskoTeschnerRmatrixForModularDouble} relatively to the spectrum of this operator. 
In order to state the result, it is convenient to recall the operators $\Om_a$ introduced in \eqref{definition operateurs parite Omega} which enjoy the exchange relations:
\beq
\Om_a \op{x}_a \Om_a \, = \, - \op{x}_a \qquad \e{and} \qquad  \Om_a \op{p}_a \Om_a \, = \, - \op{p}_a \;. 
\enq

\begin{theorem}
 The operator $ \op{B}_N^{(\e{SG})}(\la) $ admits a complete and orthogonal system of Eigenfunctions:
 \begin{itemize}
  
  \item for $N$ odd, this systems is $\{ \Phi_{\bs{y}_N} \}_{ \bs{y}_N \in \R^N}$ and it satisfies :
\beq
\f{1}{N!} \Int{ \R^N }{} \Big( \Phi_{ \bs{y}_N} \big( \bs{x}_N^{\prime} \big) \Big)^* \cdot  \Phi_{\bs{y}_N} \big( \bs{x}_N \big) \cdot  \dd \mu_{N}(\bs{y}_{N}) \; = \; \pl{a=1}{N} \de\big( x_a - x^{\prime}_a \big)
\enq
along with
\beq
\Int{\R^N}{} \Big( \Phi_{ \bs{y}_N^{\prime} } \big( \bs{x}_N \big) \Big)^* \cdot  \Phi_{\bs{y}_N} \big( \bs{x}_N \big) \cdot  \dd^N x \; = \;  \f{1}{\mu_N(\bs{y}_N) } \cdot  \de_{\e{sym}}\big( \bs{y}_N - \bs{y}^{\prime}_N \big) \;. 
\enq
Furthermore, the generalised Eigenvalue equation takes the form
\beq
\op{B}_N^{(\e{SG})}(\la) \cdot  \Phi_{\bs{y}_N} \big( \bs{x}_N \big) \; = \; \pl{a=1}{N} \Big\{ -2\i \sinh\big[\tfrac{\pi}{\om_1}(\la-y_a)  \big] \Big\}\cdot  \Phi_{\bs{y}_N} \big( \bs{x}_N \big) \;. 
\enq

   \item for $N$ even, this systems is $\{ \Phi_{y_0, \bs{y}_{N-1} } \}_{  y_0\in \R  \, , \,  \bs{y}_{N-1} \in \R^{N-1}  }$ and it satisfies :
\beq
\f{1}{(N-1)!} \Int{\R^N}{} \Big( \Phi_{y_0^{\prime}, \bs{y}_{N-1}^{\prime} }\big( \bs{x}_N^{\prime} \big) \Big)^* \cdot  \Phi_{\bs{y}_N} \big( \bs{x}_N \big) \cdot \dd y_0 \cdot \dd \mu_{N-1}(\bs{y}_{N-1}) 
\; = \; \pl{a=1}{N} \de\big( x_a - x^{\prime}_a \big)
\enq
along with
\beq
\Int{\R^N}{} \Big( \Phi_{y_0^{\prime}, \bs{y}_{N-1}^{\prime} }  \big( \bs{x}_N \big) \Big)^* \cdot  \Phi_{y_0, \bs{y}_{N-1} } \big( \bs{x}_N \big) \cdot  \dd^N x \; = \; 
\f{1}{\mu_{N-1}(\bs{y}_{N-1}) } \cdot \de(y_0-y_0^{\prime}) \cdot  \de_{\e{sym}}\big( \bs{y}_{N-1} - \bs{y}^{\prime}_{N-1} \big) \;. 
\enq
Finally, the generalised Eigenvalue equation takes the form
\beq
\op{B}_N^{(\e{SG})}(\la) \cdot  \Phi_{y_0, \bs{y}_{N-1} } \big( \bs{x}_N \big) \; = \; \ex{2\pi \om_2 y_0} \pl{a=1}{N-1} \Big\{ -2\i \sinh\big[\tfrac{\pi}{\om_1}(\la-y_a)  \big] \Big\}\cdot  \Phi_{y_0, \bs{y}_{N-1} } \big( \bs{x}_N \big) \;. 
\enq

 \end{itemize}

\end{theorem}

\Proof

The relation between the Lax matrices \eqref{ecriture lien Lax XXZ et Lax SG} implies that the monodromy matrix of the Sinh-Gordon model is related to the one of the modular XXZ chain as
\beq
\op{T}^{(\e{SG})}_N(\la) \; = \; 
\left\{  \ba{cc}		 \pl{a=1}{\tf{N}{2}}\Om_{2a-1} \, \cdot \, \op{T}_N(\la) \, \cdot \,   \pl{a=1}{\tf{N}{2}}\Om_{2a-1}  & \e{if} \;  N \; \e{is} \; \e{even}  \vspace{5mm}  \\
	 \pl{a=1}{\tf{(N+1)}{2}}\Om_{2a-1} \, \cdot \,  \op{T}_N(\la) \,  \sg^x \, \cdot \,  \pl{a=1}{\tf{(N+1)}{2}}\Om_{2a-1}  & \e{if} \;  N \; \e{is} \; \e{odd}  \ea \right. 
\label{Ecriture lien matrice de monodromie XXZ et SG}
\enq
Indeed, this relation is a simple consequence of the local identity 
\beq
\sg^x \, \op{L}_n^{(\kappa)}(\la) \, \sg^x \, = \, \Om_n \, \op{L}_n^{(\kappa)}(\la) \, \Om_n \;. 
\enq
The representation \eqref{Ecriture lien matrice de monodromie XXZ et SG} then ensures that the system of generalised Eigenfunctions of the operator $\op{B}_{N}^{(\e{SG})}(\la) $ is given by 
\beq
\left\{  \ba{cc}		 \pl{a=1}{ \tf{N}{2} }\Om_{2a-1} \, \cdot \, \Phi^{ (\op{A}) }_{ \bs{y}_N}  & \e{if} \;  N \; \e{is} \; \e{even}  \vspace{5mm}  \\
	 \pl{a=1}{ \tf{(N+1)}{2} }\Om_{2a-1} \, \cdot \, \Phi^{(\op{B})}_{ y_0, \bs{y}_{N-1}}   & \e{if} \;  N \; \e{is} \; \e{odd}  \ea \right. \;, 
\enq
where $\Phi^{(\op{A})}_{\bs{y}_N} $, $\Phi^{(\op{B})}_{y_0, \bs{y}_{N-1}}  $ are as introduced in Theorem \ref{Theorem fct propres entrees matrice monodromie}. 
The rest follows from the stated properties of the functions  $\Phi^{(\op{A})}_{\bs{y}_N} $, $\Phi^{(\op{B})}_{y_0, \bs{y}_{N-1}} $ in that theorem. \qed

\section*{Conclusion}

 In this work we have constructed the Eigenfunctions of the entries of the $N$-site monodromy matrix of the modular XXZ magnet. 
We have established that each system associated with a given entry forms a complete and orthogonal system.  
The proof of the orthogonality was achieved by means of handlings on the level of the Gauss-Givental representation for these Eigenfunctions. 
The proof of the completeness was carried out on the level of the Mellin-Barnes representation. We stress that we have proposed a new and simple
method for proving the completeness. The technique we developed is general and applicable to a wide variety of quantum integrable models
solvable by the separation of variables method. 
As a by product of our analysis, we have proved the conjectures raised by Bytsko-Teschner on the spectrum of the $B$-operator
for the lattice Sinh-Gordon model.

\section*{Acknowledgments}


The authors are indebted to J.M.-Maillet for stimulating discussions.
The work of S.D. is supported by the RFBR
grant no. 17-01-00283a.
K.K.K. acknowledges support from CNRS and ENS de Lyon.
A. M. acknowledges support from the DFG grant MO 1801/1-3.
S.D. would like to thank the Laboratoire de physique of ENS de Lyon for hospitality
during his visit there where this work was initiated.

\appendix

\section{Main notations}
\label{Appendice notation}

\begin{itemize}
 
 \item $N$-dimensional vectors are denoted as $\bs{x}_N=(x_1,\dots, x_N)$;
 
 \item $N-1$ dimensional vectors built from an $N$-dimensional vector $\bs{x}_N$ with the removed $m^{\e{th}}$ coordinate are denoted as $\bs{x}_{N;[m]}$
 and read
\beq
\bs{x}_{N;[m]} \, = \,  \big( x_1,\dots, x_{m-1},x_{m+1},\dots, x_N  \big) \;. 
\label{definition vecteur avec coordonnees omises}
\enq
\item Given an $N$-dimensional vector $\bs{x}_N$, we denote 
\beq
\ov{\bs{x}}_N=\sul{a=1}{N}x_a \; .
\label{definition moyenne coord vecteur}
\enq

\item Ratios of products of one variable functions appearing with multi-component entries are denoted using the hypergeomertic notations, \textit{e.g.} 
\beq
f\left(\ba{cc}  a_1,\dots, a_n \\ b_1,\dots, b_m  \ea \right) \, = \,  \f{ \pl{k=1}{n} f(a_k)  }{  \pl{k=1}{m} f(b_k)    } \;. 
\label{definition notation produit hypergeometrique}
\enq
\item Given indexed symbols $x_{a},x_{b}$, we denote $x_{ab}=x_a-x_b$.  

\item Given $y \in \Cx$, $y^*$ stands for its complex conjugate and $y^{\star}=-y-\i\tfrac{\Om}{2}$.

\end{itemize}

\section{Properties of the auxiliary special functions}
\label{Appendice fct speciale}

\subsection{The quantum dilogarithm and the $D_{\a}$ functions}

The quantum dilogarithm $\varpi$ is a meromorphic function on $\Cx$ which admits the integral representation
\beq
\varpi(\la) \; = \;  \exp\Bigg\{  \pm \f{ \i \pi }{ 2 \om_1 \om_2 } \cdot \Big( \la^2 \,+\, \f{ \om^2_1+\om^2_2 }{ 12 }   \Big) 
\; - \; \i \Int{ \R  \pm  \i 0^+}{}  \f{ \dd t }{ 4 t } \f{ \ex{-2\i \la t}  }{ \sinh\big(t\om_1 \big) \cdot \sinh\big(t\om_2 \big)  }  \Bigg\} \;, 
\label{definition quantum dilog}
\enq
valid for $|\Im(\la)| \, < \, \tf{ \Om }{2}$. 

This function is self-dual and satisfies to the first-order finite difference equations
\beq
\varpi\big(\la+\i\om_2\big) \, = \, 2\i  \sinh\Big[ \f{\pi}{\om_1}\big(\la+\i\f{\tau}{2} \big) \Big] \cdot \varpi(\la) \quad \e{and} \quad 
\varpi\big(\la+\i\om_1\big) \, = \, 2\i  \sinh\Big[ \f{\pi}{\om_2}\big(\la - \i\f{\tau}{2} \big) \Big] \cdot \varpi(\la)  \;. 
\label{ecriture eqn diff finite dilog}
\enq
From there one entails that 
\bem
\varpi\Big(\la-\i \tfrac{ \Om }{ 2 } + \i \ell \om_1 + \i k \om_2 \Big) \, = \, (-1)^{k\ell} \big( -2\i \big)^{\ell + k}  \cdot 
\pl{p=0}{k-1}   \sinh\Big[ \f{\pi}{\om_1}\big(\la + \i p \om_2 \big) \Big]   \\
\times \pl{p=0}{\ell-1}  \sinh\Big[ \f{\pi}{\om_2}\big(\la + \i p \om_1 \big) \Big]   \cdot 
\varpi\Big(\la-\i \tfrac{ \Om }{ 2 } \Big) 
\label{ecriture recurrence generale direct dilogarithme}
\end{multline}
and symmetrically, 
\bem
\varpi\Big(\la-\i \tfrac{ \Om }{ 2 } - \i \ell \om_1 - \i k \om_2 \Big) \, = \, (-1)^{k\ell} \Big( \f{\i}{2}  \Big)^{\ell + k}  \cdot 
\pl{p=1}{k} \Big\{ \sinh\Big[ \f{\pi}{\om_1}\big(\la - \i p \om_2 \big) \Big] \Big\}^{-1}  \\ 
\times \pl{p=1}{\ell } \Big\{ \sinh\Big[ \f{\pi}{\om_2}\big(\la - \i p \om_1 \big) \Big] \Big\}^{-1} \cdot 
\varpi\Big(\la-\i \tfrac{ \Om }{ 2 } \Big)  \;. 
\label{ecriture recurrence generale inverse dilogarithme}
\end{multline}
The quantum dilogarithm has only simple poles and zeroes. These are located at
\beq
\varpi(x)=0 \quad \e{iff} \quad x \in  \i\f{\Om }{2} + \i\mathbb{N} \om_1+ \i\mathbb{N} \om_2 \qquad \e{and} \qquad 
\varpi^{-1}(x)=0 \quad \e{iff} \quad x \in  -\i\f{\Om }{2} - \i\mathbb{N} \om_1 - \i\mathbb{N} \om_2 \;. 
\enq
$\varpi$ satisfies to the inversion identity $\varpi(\la)\varpi(-\la)=1$ and $\big(\varpi(\la^*)\big)^*=\varpi^{-1}(\la)$. 
One can also establish that 
\beq
\e{Res}\Big(\varpi\big( \la - \i \tfrac{\Om}{2}\big) \cdot  \dd \la, \la=0 \Big) \, = \, \f{\i}{2\pi} \sqrt{\om_1 \om_2}  \qquad \e{and} \qquad 
\varpi( \tfrac{\i}{2} \tau) \; = \; \sqrt{ \f{\om_2}{\om_1 } } \;. 
\enq
The above entails that, for $(k,\ell)\in \mathbb{N}^2$,
\beq
\e{Res}\Big(\varpi\big( \la - \i \tfrac{\Om}{2}\big) \cdot  \dd \la, \la=-\i \ell \om_1 -\i k \om_2  \Big) \, = \, \f{\i }{2  \pi} \sqrt{\om_1 \om_2}  
(-1)^{k\ell} \Big( \f{ 1 }{2 \i } \Big)^{\ell + k}  \cdot \bigg\{
\pl{p=1}{k}   \sinh\Big[ \i p \pi \f{ \om_2 }{\om_1} \Big]   \cdot \pl{p=1}{\ell}  \sinh\Big[ \i p \pi \f{\om_1 }{\om_2} \Big] \bigg\}^{-1} \;. 
\label{ecriture formule residu general du dilogarithme}
\enq

\vspace{3mm}

\noindent The function $D_{\a}$ is defined by the below ratio of dilogarithms
\beq
D_{\a}(\la) \; = \; \f{\varpi(\la+\a) }{ \varpi(\la-\a) } \qquad \e{so}\; \e{that} \qquad  D_{\a}(\la) \, = \, \Big( D_{-\a^*}(\la^*) \Big)^*   \;. 
\label{definition D alpha et pte conj complexe}
\enq

The function $D_{\a}$ is a meromorphic function on $\Cx$ that admits a holomorphic determination of the logarithm on 
\beq
 \mc{S}_{ \f{\Om}{2}-|\Im(\a)| }(\R) \; = \; \Big\{ \la \in \Cx \, : \, \big| \Im(\la) \big| \; < \; \f{1}{2} \Big( \om_1 \, + \, \om_2 \Big) - \,  \big| \Im(\a) \big|  \Big\}
\enq
given by 
\beq
\ln D_{\a}(\la) \; = \; \mp \f{ 2 \i \pi }{ \om_1 \om_2 } \a \la 
\; + \; \i \Int{ \R  \pm  \i 0^+}{}  \f{ \dd t }{ 2 t } \f{ \ex{2\i \la t} \sin (2\a t )  }{ \sinh\big(t\om_1 \big) \cdot \sinh\big(t\om_2 \big) } \;. 
\enq

\noindent The function $D_{\a}(\la)$ satisfies to the properties
\begin{itemize}
\item $D_{\a}(\la)$ is self-dual, namely invariant under the exchange $\om_1\leftrightarrow \om_2$;
\item for $\om_2>\om_1$ it admits the asymptotic behaviours
\beq
%
%
%
D_{\a}(\la)  =   \ex{ \mp \f{ 2 \i \pi}{ \om_1\om_2} \cdot  \la \a  }\cdot \big( 1 \; + \; \e{O}\Big( \ex{ \mp \f{ 2\pi}{ \om_2 } \la } \sinh\Big[ \f{2\pi}{ \om_2}  \a \Big] \Big)  \bigg)
	\quad \e{when} \quad  \la \tend \infty\; , \; \;  |\arg(\pm \la)| \, < \, \f{\pi}{2} ;
\enq
\item $D_{\a}(\la)$ satisfies to the difference equation 
\beq
\f{D_{\a}(\la+\i\om_2)}{D_{\a}(\la) } \; = \;  \f{ \cosh\Big[ \f{\pi}{\om_1}\big(\la+\i\f{\om_2}{2}+\a \big) \Big]  }{  \cosh\Big[ \f{\pi}{\om_1}\big(\la+\i\f{\om_2}{2}-\a \big) \Big]  } 
\enq
as well as its dual $\om_1\leftrightarrow \om_2$;
\item $D_{\a}(\la)$ enjoys the transmutation properties
\beq
D_{\a}\Big( \la \mp \i \tfrac{\om_2}{2} \Big) \, = \, \f{  D_{\a- \i \frac{\om_2}{2} }\big( \la  \big)  }{   2 \cosh\Big[\frac{\pi}{\om_1}(\la \pm \a) \Big] } \;; 
\label{ecriture ptes transmutation fct D alpha}
\enq
\item $D_{\a}(\la)$ has simple zeroes at 
\beq
\pm \Big\{ -\a \, + \, \i\f{\Om }{2}\, + \,   \i n \om_1   \, + \,  \i m \om_2\; : \; (n,m) \in \mathbb{N}^2 \Big\} \; ; 
\enq
\item $D_{\a}(\la)$ has simple poles at 
\beq
\pm \Big\{ \a \, + \, \i \f{ \Om }{2} \, + \,  \i n \om_1   \, + \,  \i m \om_2\; : \; (n,m) \in \mathbb{N}^2 \Big\} \;. 
\enq

 \end{itemize}

\subsection{Integral identities}

The function $D_{\a}$ admits the Fourier transform
\beq
\mc{F}[D_{\a}](t) \; = \; \sqrt{\om_1\om_2} \cdot  \mc{A}(\a) \cdot D_{\a^{\star}}\Big( \f{\om_1\om_2}{2\pi} t \Big)
\label{ecriture TF de fct D}
\enq
with
\beq
\mc{F}[f](t)\;= \; \Int{ \R }{}  \ex{it x} f(x) \cdot \dd x \qquad \e{for} \quad f \in L^{1}(\R,\dd x) \;. 
\enq
Here we remind that
\beq
\mc{A}(\a) \; = \; \varpi\big( 2 \a + \i\f{ \Om }{2} \big) \qquad \e{and} \qquad \a^{\star}=-\a-\i\f{ \Om }{2} \;. 
\enq
\eqref{ecriture TF de fct D} enatils that 
\beq
\lim_{\a\tend 0} \Big\{ \mc{A}(\a) D_{\a^{\star}}(t)  \Big\} \, = \, \de(t) \;. 
\label{ecriture realisation fct delta via fct D alpha}
\enq

The $D_{\a}$ functions satisfy to the three term integral identity \cite{Kashaev3termIntegralRelationDfcts}
\bem
\Int{ \R }{}  D_{\a}\big( \om_1 \om_2 (x-u) \big) \cdot  D_{\be}\big( \om_1 \om_2 (x-v)\big) \cdot D_{\ga}\big( \om_1 \om_2 (x-w) \big)\cdot  \dd x  \\
\; = \;  \f{ \mc{A}\big(\a,\be,\ga \big) }{ \sqrt{ \om_1 \om_2  } }  D_{\a^{\star}}\big( \om_1 \om_2(w-v)\big)\cdot  D_{\be^{\star}}\big( \om_1 \om_2(u-w)\big) \cdot  D_{\ga^{\star}}\big( \om_1 \om_2(u-v) \big)
\label{ecriture identite itle 3 termes}
\end{multline}
provided that $\a+\be+\ga=-\i \Om$ and \cite{BytskoTeschnerSinhGordonFunctionalBA}
\bem
\Int{ \R }{}  D_{\a}\big( \om_1 \om_2 (x-u)\big)\cdot  D_{\be}\big( \om_1 \om_2 (x-v)\big)\cdot  D_{\ga}\big( \om_1 \om_2 (x-w)\big)\cdot  D_{\de}\big( \om_1 \om_2 (x-z)\big)\cdot   \dd x \\
\; = \;\mc{A}\big(\a,\be,\ga,\de\big)  \cdot D_{\a + \be + \i \f{ \Om }{ 2 } } \bigg( \ba{c}  \om_1 \om_2(u - v) \\  \om_1 \om_2 (w - z) \ea  \bigg) \\
\times \Int{ \R }{}  D_{\a^{\star}}\big( \om_1 \om_2 (x-v)\big)\cdot  D_{\be^{\star}}\big( \om_1 \om_2 (x-u) \big)\cdot  D_{\ga^{\star}}\big( \om_1 \om_2 (x-z)\big)\cdot  D_{\de^{\star}}\big( \om_1 \om_2 (x-w)\big)\cdot   \dd x
\label{ecriture identite integrale 3 fonction D et exposant}
\end{multline}
provided that $\a+\be+\ga+\de \, = \, -\i \Om$.

Sending one of the integration variables to infinity provides one with the auxiliary identities
\beq
\Int{ \R }{}  D_{\a}\big( \om_1 \om_2(x-u) \big) \cdot \ex{ \pm   2 \i \pi \be x} \cdot  D_{\ga}\big( \om_1 \om_2 (x-w) \big)\cdot  \dd x 
\; = \;   \f{ \mc{A}\big(\a,\be,\ga \big) }{ \sqrt{ \om_1 \om_2  } }  \cdot  \ex{ \pm  2\i \pi  \big(\a^{\star} w + \ga^{\star} u \big) }   D_{\be^{\star}}\big( \om_1 \om_2 (u-w) \big)
\label{ecriture identite integrale 2 fonction D et exposant}
\enq
provided that that $\a+\be+\ga=-\i \Om$ and 
\bem
\Int{ \R }{}  D_{\a}\big( \om_1 \om_2 (x-u) \big) \cdot D_{\be}\big( \om_1 \om_2 (x-v)\big) \cdot D_{\ga}\big( \om_1 \om_2 (x-w) \big) \cdot \ex{ \pm    2 \i \pi  \de x}  \cdot  \dd x   \\
\; = \; \mc{A}\big(\a,\be,\ga,\de\big) \ex{ \pm    2 \i \pi \big(\a+\be+\i\f{\Om}{2}\big) w }   D_{\a + \be + \i \f{ \Om }{ 2 } } \big( u - v  \big)  \\
\times \Int{ \R }{}  D_{\a^{\star}}\big( \om_1 \om_2 (x-v) \big) \cdot D_{\be^{\star}}\big( \om_1 \om_2 (x-u) \big) \cdot  D_{\de^{\star}}\big( \om_1 \om_2 (x-w) \big) \cdot \ex{ \pm   2 \i \pi  \ga^{\star} x}  \cdot  \dd x
\label{ecriture identite integrale 3 fonction D et exposant}
\end{multline}
provided that $\a+\be+\ga+\de \, = \, -\i \Om$.

The three term integral relation can be recast in an operator form as
\beq
D_u(\op{p})\cdot D_{u+v}(\om_1\om_2\op{x}) \cdot D_v(\op{p}) \; = \; D_v(\om_1\om_2\op{x}) \cdot D_{u+v}(\op{p}) \cdot D_u(\om_1\om_2\op{x}) 
\label{ecriture identite etoile triangle}
\enq
whereas its degenerate form can be recast as
\beq
D_{\a}(\op{p}) \cdot \ex{\pm 2\i \pi \be \op{x} }  \cdot D_{\ga}(\op{p}) \; = \;\ex{\pm 2\i \pi \ga \op{x} }  \cdot D_{\be }(\op{p}) \cdot \ex{\pm 2\i \pi \a \op{x} }
\label{ecriture degenerescence exponentielle etoile triangle}
\enq
where $\a,\be,\ga$ fulfill the constraint $\be=\a+\ga$. 

Let $y_{\pm}$, $t_{\pm}$ be parameters as in \eqref{definition parametres y pm}. Then, the integral identity involving four $D$ functions can be recast in the operator form as
\bem
D_{ y_-}\big(\op{p}_2\big)\cdot D_{ y_+}\big(\om_1\om_2\op{x}_{23}\big)\cdot D_{ t_-}\big(\om_1\om_2\op{x}_{12}\big) \cdot D_{ t_+}\big(\op{p}_2\big) \\
\; = \;  D_{y_-^{\star} + t_- + \i \f{ \Om }{ 2 } } \big(\om_1\om_2\op{x}_{12}\big) \cdot D_{ t_-}\big(\op{p}_2\big)\cdot D_{ t_+}\big(\om_1\om_2\op{x}_{23}\big)\cdot D_{ y_-}\big(\om_1\om_2\op{x}_{12}\big) \cdot D_{ y_+}\big(\op{p}_2\big)
\cdot   D_{y_-^{\star} + t_- + \i \f{ \Om }{ 2 } }^{ - 1 } \big(\om_1\om_2\op{x}_{32}\big) \;. 
\label{ecriture identite echange 4 op D}
\end{multline}
Likewise, its exponential degenerate form can be recast as 
\bem
\ex{2\i\pi y_- \op{x}_1} \cdot D_{ y_-}\big(\op{p}_1\big)\cdot D_{ y_+}\big(\om_1\om_2\op{x}_{12}\big) \cdot \ex{2\i\pi t_- \op{x}_1}  \cdot D_{ t_+}\big(\op{p}_1\big) \\
\; = \;  \ex{2\i\pi t_- \op{x}_1} \cdot  D_{ t_-}\big(\op{p}_1\big)\cdot D_{ t_+}\big(\om_1\om_2\op{x}_{12}\big) \cdot \ex{2\i\pi y_- \op{x}_1}  \cdot D_{ y_+}\big(\op{p}_1\big)
\cdot   D_{y_-^{\star} + t_- + \i \f{ \Om }{ 2 } }^{ - 1 } \big(\om_1\om_2\op{x}_{21}\big) \;. 
\label{ecriture identite echange 4 op D degeneree gauche}
\end{multline}
Finally, the exponential degenerate form of the four D function integral \eqref{ecriture identite integrale 3 fonction D et exposant} can be also recast as 
\bem
\mc{A}(t_+) \,  D_{ y_-}\big(\op{p}_2\big)\cdot D_{ y_+}\big(\om_1\om_2\op{x}_{23}\big)\cdot \ex{ \pm 2\i\pi (t_+^{\star} \op{x}_2+y_+^{\star}\op{x}_3)}  \cdot  D_{ t_-}\big(\om_1\om_2\op{x}_{12}\big)   \\
\; = \; \mc{A}(y_+) \, D_{y_-^{\star} + t_- + \i \f{ \Om }{ 2 } } \big(\om_1\om_2\op{x}_{12}\big)    D_{ t_-}\big(\op{p}_2\big)\cdot D_{ t_+}\big(\om_1\om_2\op{x}_{23}\big)
\cdot \ex{ \pm 2\i\pi (y_+^{\star} \op{x}_2+t_+^{\star}\op{x}_3)}  \cdot  D_{ y_-}\big(\om_1\om_2\op{x}_{12}\big) \;. 
\label{ecriture identite echange 4 op D degeneree droite}
\end{multline}

\end{document}